\documentclass[a4paper,11pt]{article}

\usepackage[T1]{fontenc}
\usepackage[english]{babel}
\usepackage[utf8]{inputenc}
\usepackage[normalem]{ulem}
\usepackage{comment}

\usepackage[vmargin=2.4cm,hmargin=2.4cm,includeheadfoot]{geometry}
\usepackage[skip=0.25\baselineskip,indent,parfill]{parskip}

\usepackage{tocloft} 
\usepackage{color}
\usepackage{subcaption}
\usepackage{tikz}
\usepackage{pgfplots}
\usetikzlibrary{arrows.meta}
\usetikzlibrary{tikzmark}
\usetikzlibrary{shapes.multipart}
\usetikzlibrary{decorations.markings}
\usetikzlibrary{pgfplots.fillbetween}
\pgfplotsset{compat=1.18}

\definecolor{darkblue}{rgb}{0.1,0.1,.7}
\usepackage[colorlinks, linkcolor=darkblue, citecolor=darkblue, urlcolor=darkblue, linktocpage, pagebackref]{hyperref}

\renewcommand*{\backref}[1]{}
\renewcommand*{\backrefalt}[4]{%
  \ifcase #1 %
  \or
    {\scriptsize\,[p.~#2]}%
  \else
    {\scriptsize\,[pp.~#2]}%
  \fi
}

\usepackage[dont-mess-around]{fnpct}

\usepackage[numbers,sort&compress]{natbib}

\makeatletter
\patchcmd\NAT@citexnum{\let\NAT@last@num\NAT@num}{\MakeLinkTarget[cite]{}\Hy@backout{\@citeb\@extra@b@citeb}\let\NAT@last@num\NAT@num}{}{\fail}
\makeatother

\usepackage{booktabs}

\usepackage{amsmath}
\usepackage{amssymb}
\usepackage{mathrsfs}
\usepackage{amsthm}
\usepackage{mathtools}
\usepackage{enumitem}
\usepackage{braket}

\usepackage{multirow}
\usepackage{nicematrix}

\numberwithin{equation}{section}

\let\originalleft\left
    \let\originalright\right
\renewcommand{\left}{\mathopen{}\mathclose\bgroup\originalleft}
    \renewcommand{\right}{\aftergroup\egroup\originalright}

\newcommand{\Z}{\mathbb{Z}}
\newcommand{\R}{\mathbb{R}}
\newcommand{\id}{\text{\usefont{U}{bbold}{m}{n}1}}

\newcommand{\be}{\begin{equation}} \newcommand{\ee}{\end{equation}}

\renewcommand{\tilde}{\widetilde}

\newcommand{\floor}[1]{\left\lfloor #1 \right\rfloor}

\newcommand{\ped}[1]{\textormath{\textsubscript{#1}}{_{\mathrm{#1}}}}
\newcommand{\ap}[1]{\textormath{\textsuperscript{#1}}{^{\mathrm{#1}}}}
\newcommand{\nn}{\nonumber}

\DeclareMathOperator{\tr}{tr}

\usepackage{xifthen}
\newcommand{\dif}[1][]{\mathord{\ifthenelse{\isempty{#1}}{\mathrm{d}}{\mathrm{d}^{#1}}}}

\begin{document}

\begin{titlepage}
    \vspace*{1em}
    \begin{center}
        {\LARGE \bfseries A Bootstrap Study of Confinement in AdS} \\[3em]
        {\bf
            Lorenzo Di Pietro$^{a,b}$, Stefanos R.\ Kousvos$^{c}$, Marco Meineri$^{c}$,\\ Alessandro Piazza$^{b,d}$, Marco Serone$^{b,d}$, Alessandro Vichi$^{e}$ 
        } \\[2em]
         \( {}^{a} \) {\itshape Department of Physics, University of Trieste, Strada Costiera 11, 34151 Trieste, Italy} \\
         \( {}^{b} \) {\itshape INFN, Section of Trieste, Via Valerio 2, I-34127 Trieste, Italy} \\
        \( {}^{c} \) {\itshape Department of Physics, University of Torino and INFN, Section of Torino,\\[-1pt] Via P.\ Giuria 1, 10125 Torino, Italy} \\
        \( {}^{d} \) {\itshape SISSA, Via Bonomea 265, I-34136 Trieste, Italy} \\
        \( {}^{e} \) {\itshape Department of Physics, University of Pisa and INFN, \\Largo Pontecorvo 3, I-56127 Pisa, Italy} \\[2em]
        {\small{\texttt{ldipietro@units.it, stefanosrobert.kousvos@unito.it, marco.meineri@unito.it, \\ apiazza@sissa.it, serone@sissa.it,  alessandro.vichi@unipi.it}}}
    \end{center}
    \vspace{3em}
    \begin{abstract}

Yang–Mills theory in $\mathrm{AdS}_{4}$ with Dirichlet boundary conditions is expected to undergo a transition as the AdS radius varies, since the boundary data is incompatible with confinement in flat space. 
Various mechanisms have been proposed for the disappearance of the Dirichlet boundary condition. 
From the boundary viewpoint, the associated $3d$ CFT is a deformation of a generalised free theory of non-Abelian conserved currents, with the deformation governed by the bulk gauge coupling.
We test these scenarios by deriving non-perturbative constraints from the numerical conformal bootstrap of the four-point function of non-Abelian conserved currents.
We rule out the scenario in which the boundary current decouples. 
Bounds on the lightest scalar operators disfavour a bulk Higgs mechanism and instead point to a transition driven by a scalar singlet becoming marginal. 
We also obtain bounds on other scalar operators and on the current central charge, and we refine character-based techniques incorporating parity and charge-conjugation symmetry to determine the operator spectrum of the $3d$ Generalised Free Vector theory. 
These results may be of independent interest beyond Yang–Mills theory in AdS.

    \end{abstract}
\end{titlepage}

\tableofcontents


\section{Introduction and summary of results}\label{sec:intro}

A quantitative understanding of confinement and the emergence of a mass gap in four-dimensional non-Abelian gauge theories remains one of the hardest and most enduring open problems in Quantum Field Theory (QFT)~\cite{Jaffe:2000ne,Greensite:2011zz}.
Despite decades of effort, progress has been limited by the absence of small parameters away from the limit of a large number of colours, as well as by the intrinsically strong coupling that governs the infrared (IR) regime.

As initially proposed in~\cite{CALLAN1990366,Aharony:2012jf}, the problem can be addressed from a new angle by studying pure Yang–Mills (YM) theory on a curved four-dimensional background, namely Euclidean Anti-de Sitter space ($\mathrm{AdS}_{4}$). 
While this reformulation does not solve the issue of strong coupling in any obvious way, it introduces a natural infrared regulator without sacrificing spacetime symmetries.
Excitations with wavelengths larger than the AdS radius $L$ are gapped already at the classical level, and the running of the coupling $ g\ped{YM}(\mu)$ effectively freezes below the scale $\mu \sim 1/L$, where curvature effect dominates over quantum fluctuations. 
Unlike a finite box, however, AdS preserves the full $SO(4,1)$ symmetry group, which coincides with the conformal group in $3d$ flat space. 

As a result, asymptotically free gauge theories in AdS are controlled by a single dimensionless parameter, which may be taken as $g\ped{YM}(1/L)$ or, equivalently, $L \Lambda\ped{YM} $, where $\Lambda\ped{YM}$ denotes the dynamically generated scale of the flat-space theory. 
The physics interpolates between two regimes.
\begin{itemize}
    \item \emph{Weak coupling} ($L \Lambda\ped{YM} \ll 1$): the theory can be described by the classical Yang–Mills action with perturbative corrections. 
    The bulk spectrum consists of weakly interacting gluonic excitations with a mass gap of order $1/L$.
    \item \emph{Strong coupling} ($ L \Lambda\ped{YM}  \gg 1$): the gauge interactions become non-perturbative before curvature effects set in, leading to confinement. 
    Physical states are colour-singlet bound states with a mass gap of order $\Lambda\ped{YM}$.
\end{itemize}

Varying the dimensionless parameter $L \Lambda\ped{YM}$ interpolates between these limits and is expected to induce a confinement–deconfinement transition at a critical value of the radius $L_{*}\sim1/\Lambda\ped{YM}$. 
Such a transition has been conjectured and explored in~\cite{Aharony:2012jf,Copetti:2023sya}, and more recently studied perturbatively in~\cite{Ciccone:2024guw,Ciccone:2025dqx}. 
Its existence is also required by consistency with the flat-space limit $L\to\infty$, where pure YM theory on $\mathbb{R}^4$ confines.

Last but not least, this framework allows us to use the powerful machinery of conformal bootstrap to investigate the boundary Conformal Field Theory (CFT).\footnote{
    Through this work, we will use the expression ``boundary CFT'' to refer to the boundary theory. 
    Formally, this is a set of correlation functions that behave like correlators of a conformally invariant theory, even though there is no local stress tensor among the boundary operators. 
    In any case, the axioms of a conformal theory, such as unitarity, crossing, and the existence of a convergent OPE, are all we need to make use of bootstrap techniques.
}
Since the formulation of the AdS/CFT correspondence~\cite{Maldacena:1997re,Witten:1998qj,Gubser:1998bc}, CFTs have been used to study theories in AdS, initially as a tool to understand gravity and string theory, but later even to study ordinary QFT in a static AdS background~\cite{Heemskerk:2009pn,Penedones:2010ue,Paulos:2016fap,Mazac:2016qev,Komatsu:2020sag,Antunes:2021abs}. 
Among the tools that CFTs offer, one particularly promising one is the numerical conformal bootstrap. 
Since its renaissance~\cite{Rattazzi:2008pe,Rychkov:2009ij}, the bootstrap approach has led to many interesting results, reviewed for instance in~\cite{Poland:2018epd}  (see~\cite{Rychkov:2023wsd} for more recent updates). 
Initially formulated only for correlation functions of scalars, this technique was soon generalised to fermions in three~\cite{Iliesiu:2015qra,Iliesiu:2017nrv,Erramilli:2022kgp,Mitchell:2024hix} and four dimensions~\cite{Karateev:2019pvw}, and finally to spin-1 currents~\cite{Dymarsky:2017xzb,Reehorst:2019pzi,He:2023ewx,Bartlett-Tisdall:2023ghh,Bartlett-Tisdall:2024mbx} and the stress tensor~\cite{Dymarsky:2017yzx,Chang:2024whx} in three dimensions. 
The latter led to a world-record determination of the critical exponents in the $3d$ Ising model, improving on the previous bootstrap record of~\cite{Kos:2016ysd}.

In this work, we make use of the numerical conformal bootstrap to study the correlation functions of conserved currents in the boundary CFT. 
The goal of this paper is to elucidate the mechanism underlying the (de)confinement transition in $\mathrm{AdS}_{4}$ and to investigate how the appropriate boundary conditions for the gauge fields evolve as one moves between the two regimes. 
In particular, we study the disappearance of the Dirichlet boundary condition, identified in~\cite{Aharony:2012jf} as the smoking gun for confinement, and we find non-perturbative constraints on the possible realisations of this phenomenon.

\subsection{A family of boundary theories}

The bulk gauge coupling, or equivalently 
\begin{equation}
    L \Lambda\ped{YM} \sim \exp\left( - \frac{8\pi^2}{\beta_0 g\ped{YM}^2}\right) \, ,
\end{equation}
corresponds to a one-parameter family of deformations of the boundary CFTs.
As reviewed in the previous section, when we vary $L \Lambda\ped{YM}$, we interpolate between the two following regimes.

\paragraph{Small radius (\texorpdfstring{$L \Lambda\ped{YM} \ll 1$}{L LamYM << 1})} In this regime, computations can be carried out perturbatively using the UV Lagrangian. 
In particular, one can use a classical analysis to understand the possible AdS-invariant boundary conditions for the bulk gauge field $A^{a}_{\mu}$, which transforms in the adjoint representation of the gauge group $G\ped{YM}$.
The standard possibilities are the Dirichlet (D) boundary condition (bc)
\begin{equation}\label{eq:Dirichlet-bc}
    A^a_\mu(z,\vec{x}) \underset{z\to 0}{\sim} z J_\mu^a(\vec{x}) \, ,
\end{equation}
and the Neumann (N) boundary condition
\begin{equation}
    A^a_\mu(z,\vec{x}) \underset{z\to 0}{\sim} a_\mu^a(\vec{x}) \, .
\end{equation} 
Here we are using Poincaré coordinates $(z,\vec{x})$ with $z>0$, and $\mu$ denotes an index parallel to the boundary. The metric is taken to be
\begin{equation}
    \dif{s}^{2} = \frac{L^2}{z^2}\Big(\dif{z}^2 + \dif{\vec{x}}^{\, 2}\Big) \, .
\end{equation}
Choosing Dirichlet as a bc, the theory has a $G\ped{YM}$ global symmetry. 
This symmetry acts locally on the boundary, meaning that among the boundary operators there is an associated conserved current $J^a_\mu$, with protected scaling dimension $\Delta=2$ (i.e.\ $L \Lambda\ped{YM}$-independent) in the adjoint representation of $G\ped{YM}$. 
In fact, in the limit $g\ped{YM}\to 0$, the boundary CFT approaches the Generalised Free Vector (GFV) theory built out of this conserved current. 
Choosing Neumann as a bc, $G\ped{YM}$ is now gauged also at the boundary. 
In practice, this means that the boundary CFT has no continuous global symmetry.\footnote{
 Moreover, the only discrete global symmetry is charge conjugation as far as point-like operators are concerned. 
    Since we restrict to point-like operators, our analysis is not sensitive to the global structure of the gauge group. 
    Boundary operators can also arise from the line operators of the bulk theory, which carry information about the global structure of the gauge theory~\cite{Aharony:2013hda}. 
    For instance, the bulk line operators can give rise to boundary line operators, or they can be open and end at boundary points.
    Some discrete global symmetry, whose precise form depends on the choice of global structure, can act on this broader class of operators, even with Neumann boundary condition. \label{foot:disc}
}
In the limit $g\ped{YM}\to 0$, all the local boundary operators are gauge invariant combinations of a boundary field-strength operator $f^a_{\mu\nu}$. 
There is no boundary operator of protected scaling dimension. 
Note that in both Dirichlet and Neumann boundary condition, for $g\ped{YM}\to 0$, all Lorentz-scalar operators are irrelevant. 
In other words, there is no additional bc that we can define by perturbing Dirichlet or Neumann boundary conditions with a relevant operator. 
The boundary theories are stable as far as such perturbations are concerned.

\paragraph{Large radius (\texorpdfstring{$L \Lambda\ped{YM} \gg 1$}{L LamYM >> 1})} In this regime, the spectrum of boundary operators reproduces the spectrum of asymptotic scattering states of the theory in flat space (both single- and multi-particle states). 
Since a unique vacuum with a definite spectrum is expected to exist in flat space, there should be a single possibility as far as boundary conditions are concerned.
Given that a mass gap is also expected in flat space, all scaling dimensions should grow linearly with $L$, $\Delta_i\sim M_i L$, where $M_i$ are the masses of the corresponding asymptotic states in the flat space theory.
Furthermore, due to the presumed confinement, there should be no colour symmetry $G\ped{YM}$ labelling the states. 
More generally, glueball asymptotic states should not carry charges under any continuous global symmetry.

\paragraph{Intermediate regime (\texorpdfstring{$L \Lambda\ped{YM} \sim 1$}{L LamYM ~ 1})} As the radius of AdS increases, we depart from the perturbative description, and we enter the strongly coupled regime. 
The arguments reviewed above suggest that the Dirichlet bc cannot be valid all the way to infinite $L$: something must occur to prevent this from happening. 
The confinement-deconfinement transition is then conjectured to happen at some critical radius \( L_{*}\). 
For \( L \lesssim L_{*} \), the Dirichlet bc is stable and the boundary CFT has a \( G\ped{YM} \) global symmetry, having in its spectrum a non-Abelian conserved current \( J^{a}_{\mu} \). 
For \( L \gtrsim  L_{*} \), the Dirichlet bc no longer exists as a stable bc and the continuous boundary global symmetry is lost. 
 
\subsection{Summary of confinement scenarios}

Various scenarios for the disappearance of the Dirichlet bc have been proposed. They can be phrased in terms of features of the data of the boundary CFT as we approach \( L_{*} \) in a continuous way from below ($L\leq L_*$). 
\begin{itemize}
    \item \textbf{Decoupling}: the current \( J^{a}_{\mu} \) decouples from the spectrum by becoming a null operator. 
    In terms of CFT data, this means that the current two-point function coefficient \( C_{J} = C_{J}(L \Lambda\ped{YM})\) satisfies 
    \(C_{J} \to 0 \) as \( L \to L_{*} \).
    
    \item \textbf{Merging}: a scalar \( S \) singlet under \( G\ped{YM} \) becomes marginal, triggering a merging and annihilation of the boundary CFT with another putative CFT.
    This is a generic mechanism discussed in~\cite{Kaplan:2009kr,Gorbenko:2018ncu,Gorbenko:2018dtm}, and considered in the context of QFT in AdS in~\cite{Hogervorst:2021spa,Lauria:2023uca,Copetti:2023sya}. 
    In terms of CFT data this means that the dimension of $S$ satisfies \( \Delta_{S} = \Delta_{S}(L \Lambda\ped{YM}) \to 3 \) as \( L \to L_{*} \). 
    
    \item \textbf{Higgsing}: a scalar \( \mathcal{O} \) charged under \( G\ped{YM} \) becomes marginal, allowing for a new boundary condition for YM in \( \mathrm{AdS}_{4} \) where the boundary global symmetry is explicitly broken. 
    In terms of CFT data this means that \( \Delta_{\mathcal{O}} = \Delta_{\mathcal{O}}(L \Lambda\ped{YM}) \to 3 \) as \( L \to L_{*} \).
\end{itemize}
Additional details on these scenarios are discussed in Section~\ref{sec:scenarios}.

The above scenarios have been analysed in~\cite{Ciccone:2024guw} through the lens of perturbation theory, valid in the regime $L \Lambda\ped{YM} \ll 1$. 
On the one hand, this provides a first quantitative approach going beyond the qualitative explorations of~\cite{Aharony:2012jf} and a first evidence in favour of merging. 
On the other hand, the results obtained there were compatible with all the scenarios above and were based on a one-loop analysis.
The bootstrap analysis presented in this work shares the same goal of a quantitative approach to the (de)confinement transition of YM in AdS, this time on a non-perturbative footing.

\subsection{Results}\label{sec:intro-results}

The boundary CFT of YM in \( \mathrm{AdS}_{4} \) with Dirichlet bc contains a non-Abelian conserved current \( J^{a}_{\mu} \), in the adjoint representation of the gauge group.
For definiteness, the gauge group will be taken to be $G\ped{YM} = SU(N)$ from now on. We set the $\theta$-angle to zero, so the YM theory
preserves parity and charge conjugation symmetry. The most important observables involved in the above scenarios are accessible in the Operator Product Expansion (OPE) of such a current, and hence in the four-point function \( \braket{JJJJ} \). 
This motivates us to consider the setup of the non-Abelian current bootstrap developed in~\cite{He:2023ewx}, and reviewed in Section~\ref{sec:bootstrap-setup}. 

As a first application, we considered a lower bound on the current central charge $C_J$. 
This quantity is normally defined as the normalisation of the current two-point function, in the specific choice of normalisation where $J$ obeys the Ward identities of the global symmetry~\cite{Osborn:1993cr}.
In the bootstrap setup, after the appropriate rescaling, the current central charge appears in 
the coefficient of the $J$ exchange in the $J\times J$ OPE. 
This OPE contains two possible tensor structures, and the precise mixture is parametrised by a further coefficient $\alpha_{JJJ}$~\cite{Li:2015itl}. 
A lower bound on $C_J$ as a function of $\alpha_{JJJ}$ is reported in Figures~\ref{fig:SU2_CJ-bound_pd-11-19-23} and~\ref{fig:SU3_CJ-bound_pd-11-19} respectively for $SU(2)$ and $SU(3)$.
It is strictly positive, preventing the current central charge from vanishing in any unitary CFT. 
A lower bound on $C_J$ as a function of $N$ is reported in Figure~\ref{fig:CJmin}, which again shows that it is always strictly above zero.
In this case, the bound is obtained by setting $\alpha_{JJJ}$ at the value $\alpha_{JJJ}^*$, which minimises $C_J$.

The striking consequence of these bounds is that \emph{the ``decoupling'' scenario discussed in the previous section cannot be reached continuously}: the boundary theory must violate some assumption (unitarity or conformal symmetry) before reaching the limit of vanishing $C_J$.

Even though a map between $C_J$ and $g\ped{YM}$ valid at strong coupling is not known, 
it is expected that, at least within some natural set of scheme choices for the definition of $g\ped{YM}$, they are inversely proportional to each other.
This is indeed the case perturbatively.
If this relation persists at strong coupling, our results show that the coupling cannot diverge, since $C_J > 0$. 

Next, we search for evidence in favour of or against the ``merging'' and ``Higgsing'' scenarios. 
Our strategy is to show that, in YM-like theories, either a singlet scalar (for merging) or a charged scalar (for Higgsing) operator is allowed to cross marginality. 
Here we face a subtle problem, namely, defining what it means for a conformal theory to be YM-like -- in particular, at strong coupling, where the transition is expected to happen. 
At this stage, we cannot precisely characterise what a CFT describing the boundary of YM in the bulk should look like. 
Our intuitive picture is that, at weak coupling, the CFT spectrum should be perturbatively close to that of a GFV. 
Then, as the bulk coupling increases, the spectrum will continuously be deformed away from the GFV.
This motivates us to study in detail the GFV spectrum. We perform such task using character techniques. Taking into account charge conjugation symmetries of operators is not entirely trivial and requires some more formalism, which we extensively develop in Appendix~\ref{app:characters-GFV}.

The above intuition is verified in Figure~\ref{fig:SU3_Pe-S_Pe-adjS_allowed-region_strong-gaps_pd-11-19}, where it is shown that with extreme gap assumptions, the allowed region shrinks to an island around the GFV point. 
More realistic assumptions are obtained by parametrising the deformation away from the GFV spectrum schematically as follows:
\begin{equation}\label{eq:GFTspectrum}
    \big|\Delta - \Delta\ped{GFV}\big| \leq \frac{\delta}{\ell^n} \, ,
\end{equation}
for the leading spin $\ell > 0$ operators in the spectrum -- see Section~\ref{sec:bootstrap-dimensions-GFV} for precise details. 
We then explore allowed values for the dimensions of the lightest scalar operators in the $J\times J$ OPE.\footnote{
    We assume that there is one such isolated operator below a certain threshold. 
    See Section~\ref{sec:bootstrap-dimensions-GFV} for more details.
}
As long as $\delta \ll 1$, we are in the perturbative regime and we find that (as expected) no scalar operator is allowed to be relevant. 
As $\delta$ increases, the space of theories consistent with the spectral assumption~\eqref{eq:GFTspectrum} becomes larger, with YM at strong coupling being part of it. 
Figures~\ref{fig:JJJJ_SU2_Pe-S_Pe-SSb_delta-gaps_min-compare_pd19} and~\ref{fig:JJJJ_SU3_Pe-S_Pe-adjS_delta-gaps_min-compare_pd19} for $SU(2)$ and $SU(3)$ show that, as $\delta$ increases, a scalar singlet is allowed to cross marginality before scalar operators in any other non-singlet representation. 
We interpret this as \emph{evidence in favour of the merging scenario}.

Similar evidence can be obtained by examining the extremal spectrum arising from the optimisation of various quantities. 
As shown in Figures~\ref{fig:JJJJ_SU2_Pe-S_Pe-SSb_delta-gaps_pd19} and~\ref{fig:JJJJ_SU3_Pe-S_Pe-adjS_Pe-SSb_delta-gaps_pd19}, we obtained bounding boxes for the scaling dimensions of singlet and non-singlet scalars, as a function of $\delta$. Although we don't have evidence that these extremal spectra are necessarily related to YM theory, among all CFTs consistent with the assumption~\eqref{eq:GFTspectrum}, those that minimise the dimension of a non-singlet scalar operator necessarily contain a lighter singlet scalar in the spectrum, which crosses marginality first.

To summarise, our analysis shows that there exist two values $\delta_{*}$ and $\delta_{**}$ such that, for $\delta < \delta_{*}$, no scalar operator with a spectrum of the form~\eqref{eq:GFTspectrum} is allowed to be relevant. 
For $\delta_{*} < \delta < \delta_{**}$, only a scalar singlet can become marginal. 
That doesn't necessarily imply that this will be the case in YM, leading to the merging scenario, but only that this possibility can occur before anything else. 
This picture confirms the perturbative result of~\cite{Ciccone:2024guw}, which showed that the scalar singlet acquires the largest negative anomalous dimension at leading order. 
Finally, for $\delta > \delta_{**}$, non-scalar singlets are allowed to be relevant as well.
Beyond this point, we are not able to make any conclusive statements.
Nevertheless, it appears that the merging scenario is the most plausible.

\subsection{Structure of the work}

Before diving into the bootstrap setup, in Section~\ref{sec:scenarios} we provide further details on the possible scenarios for the (de)confinement transition. 
This will help us outline concrete goals for the bootstrap analysis and interpret the results at the end.
In Section~\ref{sec:spectrum} we review the main ingredients of the boundary theory, including the spectrum of the CFT in the limit $g\ped{YM}\rightarrow 0$ and perturbative results from earlier works.

Next, in Section~\ref{sec:bootstrap-setup}, we review the setup of the non-Abelian current bootstrap. 
As a first application, we obtain a universal lower bound on $C_J$ in Section~\ref{sec:bootstrap-CJ}, valid for any unitary $SU(N)$ symmetric CFT in $3d$. 
We move on to carve out the space of allowed scaling dimensions for the lowest-lying scalar operators in Section~\ref{sec:bootstrap-dimensions}. 
In Section~\ref{generalbounds}, we first inspect universal bounds that apply to any unitary CFT.
Next, we show in Sections~\ref{sec:bootstrap-dimesions-feasibility} and~\ref{sec:bootstrap-dimensions-feasibility} that by imposing strong or moderate gaps on the second-lightest and some spinning sectors, we can obtain island- or peninsula-like allowed regions for the smallest dimension of certain isolated scalar operators.
Finally, we implement a parametric set of more physically motivated gap assumptions designed to pinpoint YM-like theories and inspect their consequences in Section~\ref{sec:bootstrap-dimensions-GFV}. 
We do this for $SU(2)$ and $SU(3)$ theories.

Several appendices complement this work. 
In Appendix~\ref{app:AHM} we explore a QFT toy model in the bulk where the Higgsing transition can be analysed in detail. 
This serves as an illustration of the more general mechanism discussed in the main body of the paper. 
Appendix~\ref{app:characters-GFV} is a stand-alone review of character techniques for extracting the spectrum of a free CFT.
It highlights the role of the discrete symmetries $P$ (parity) and $C$ (charge conjugation), and obtains the necessary characters and spectrum for our theory of interest (the mean field theory of a conserved $SU(N)$ current).
The detailed results can be found in the ancillary file \texttt{GFV-spectrum.nb}.
Appendix~\ref{sec:projectors} contains technical details about $SU(N)$ projectors.
Appendix~\ref{app:numerics} collects the precise gap assumptions used in this work, as well as technical details on the numerical implementation.

\section{The confinement-deconfinement transition}
\label{sec:scenarios}

In this section, we review and clarify the proposed scenarios for the transition between the small-radius and large-radius phases of non-Abelian gauge theories in AdS.

The puzzle is to find a way to connect these two regimes coherently at intermediate values of $L \Lambda\ped{YM}$. 
A necessary condition to connect them is that the Dirichlet bc should disappear at some non zero critical value of $L \Lambda\ped{YM}$.
Such a critical value must exist because  there is no scalar operator at $\Delta=3$ in the GFV.
On the other hand, the boundary condition Neumann could very well be the one that approaches continuously the single confined theory in the large radius limit.\footnote{
    See~\cite{Gabai:2025hwf} for a recent exploration in this direction from the point of view of flux tubes.
} 
The latter possibility seems very natural, but it is not guaranteed to be the case. 
Let us outline some possible scenarios~\cite{Aharony:2012jf, Copetti:2023sya} for the behaviour at intermediate $L \Lambda\ped{YM}$.

\paragraph{Decoupling} 
The coefficient $C_J$ of the two-point function of the $SU(N)$ currents vanishes at a critical value of the radius. 
Because of the Cauchy-Schwarz inequality, the null state $\ket{J_\mu^a}$ must have vanishing overlap with any finite norm state in a unitary CFT. 
One may object that a state with the same quantum numbers still exists, and is created by the unit-normalised operator $J_\mu^a/\sqrt{C_J}$.
In fact, this scenario involves a radical change in the spectrum: all the local operators that are charged under $SU(N)$ should be quotiented out. 
From the bulk point of view, this is an expected feature of a confining transition: all the multi-gluon states should disappear. 
From the point of view of the CFT data, the fate of these scaling operators is sealed by the three-point function $\braket{J_\mu^a O^* O}$, where $O$ and $O^*$ are in conjugate representations of the $\mathfrak{su}(N)$ algebra and have non-vanishing two-point function. 
This correlator is fixed by the $\mathfrak{su}(N)$ representation and is, in particular, of order one. 
Hence, the Cauchy-Schwarz inequality requires that the state created by acting with the two charged local operators at separated points becomes non-normalizable as $C_J \to 0$.
While single operator insertions can be normalised to one, this statement, corresponding to the divergence of a four-point function, is physical. 
The origin of the divergence in the four-point function $\braket{O O^* O O^*}$ is, of course, visible in the OPE, and related to the discussion above: the conformal block decomposition contains the current, with an OPE coefficient proportional to $1/C_J$. 
Other blocks cannot compensate for the corresponding pole, as one can see, for instance, by evaluating the correlator in a reflection positive configuration.

The singlet sector of the theory is unaffected by these divergences and can continue to exist, at least on the plane, at smaller AdS radii. 
In the language of~\cite{Aharony:2012jf}, this is a continuous transition. 
Nevertheless, crossing and unitarity do constrain the behaviour of the charged sector as $C_J$ gets smaller, and one of the aims of this paper is to show, via the conformal bootstrap, that the unit normalisation of the two-point function $\braket{O O^*}$ and the growth of the four-point function are in tension with each other.

\begin{figure}[t!]
    \centering
    \begin{tikzpicture}[x=0.45\linewidth,y=0.3\linewidth]
        \coordinate (M) at (0,0.45);

        \draw[-{Latex},line width=0.8pt] (-1,0) -- (1,0) node[below,yshift=-2] {\( L \Lambda \)};

        \draw[color=green!50!black,line width=1.2pt] (-0.9,0.3) node[below,yshift=2] {\( \mathrm{D} \)} to (-0.3,0.3) to [out=0,in=-90] (M);
        \draw[color=red!80!black,line width=1.2pt] (-0.4,0.6) node[above, yshift=2] {\( \mathrm{D}^{\star} \)} to (-0.3,0.6) to [out=0,in=90] (M);

        \draw[color=red!80!black,line width=1.2pt,dashed] (-0.6,0.6) node[left] {\( ? \)} -- (-0.4,0.6);
        \draw[color=cyan!80!black,line width=1.2pt] (-0.9,0.15) node[below,yshift=-2] {\( \mathrm{N} \)} to (0.9,0.15);
        \draw[dashed, line width=0.5pt] (M) -- (0,0) node[below,yshift=-2] {\( (L \Lambda)\ped{M} \)};
        \filldraw (M) circle (0.05cm);

        \node[above left, align=right] at (1,0.6) {flat space};
    \end{tikzpicture}
    \caption{Sketch of the boundary conditions available in terms of real boundary CFTs as the coupling $g\ped{YM}$ grows towards flat space. Here we illustrate the simplest possible scenario, namely merging with transition to Neumann. At some critical value, the Dirichlet boundary condition is assumed to merge with a hitherto unknown boundary condition dubbed $\mathrm{D}^{\star}$ and annihilate into the complex plane. Thus, sufficiently close to flat space, one naively expects only the Neumann boundary condition to remain. Note that the vertical axis does not represent a definite quantity; it is merely used to juxtapose the different options. 
    }\label{fig:merg}
\end{figure}
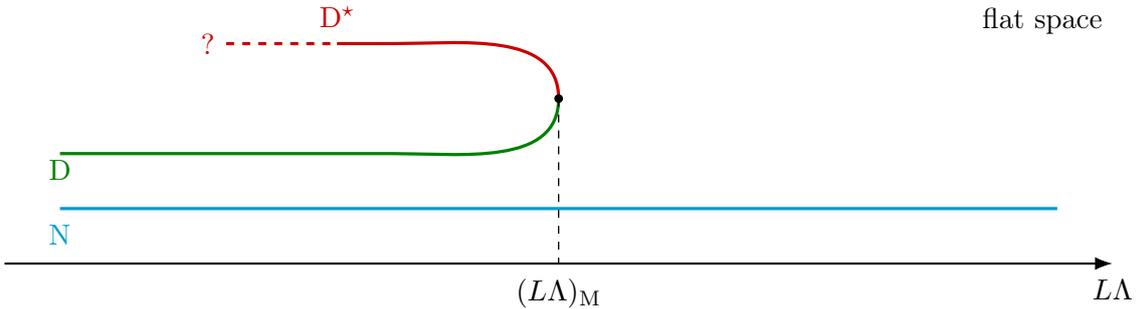

\paragraph{Merging with transition to Neumann}
One of the initially irrelevant scalar singlet operators of Dirichlet bc becomes marginal at some critical value $(L \Lambda\ped{YM})\ped{M}$, before any operator charged under $G\ped{YM}$. 
As discussed in~\cite{Lauria:2023uca, Ciccone:2024guw}, this generically implies that the Dirichlet bc merges with a second $G\ped{YM}$-symmetric bc, called Dirichlet${}^\star$ ($\mathrm{D}^{\star}$), and for $L\Lambda\ped{YM} > (L \Lambda\ped{YM})\ped{M}$ both Dirichlet and $\mathrm{D}^{\star}$ are \emph{complex CFTs}~\cite{Gorbenko:2018ncu, Gorbenko:2018dtm}. 
In this context, this means that they remain AdS-invariant boundary conditions (i.e.\ boundary conditions that preserve the whole isometry group) only if one deforms the action by a complex coupling proportional to the operator that crossed marginality, thus breaking
unitarity.
The consequence of this for the physical theory, constrained to real values of the coupling, is that necessarily the boundary condition breaks the AdS isometries, i.e.\ the conformal group at the boundary. 
The breaking takes the form of a boundary RG, triggered by the running of the coupling associated with the singlet operator. 
In the deep IR of this RG flow, barring the exotic possibility of a boundary condition that preserves only scale transformations and not the full conformal group, the theory must settle in a new AdS-invariant boundary condition.
The most natural possibility is that it settles in N, which then continuously extrapolates to the large radius boundary condition. 
This is the simplest possible scenario. It is illustrated in Figure~\ref{fig:merg}.

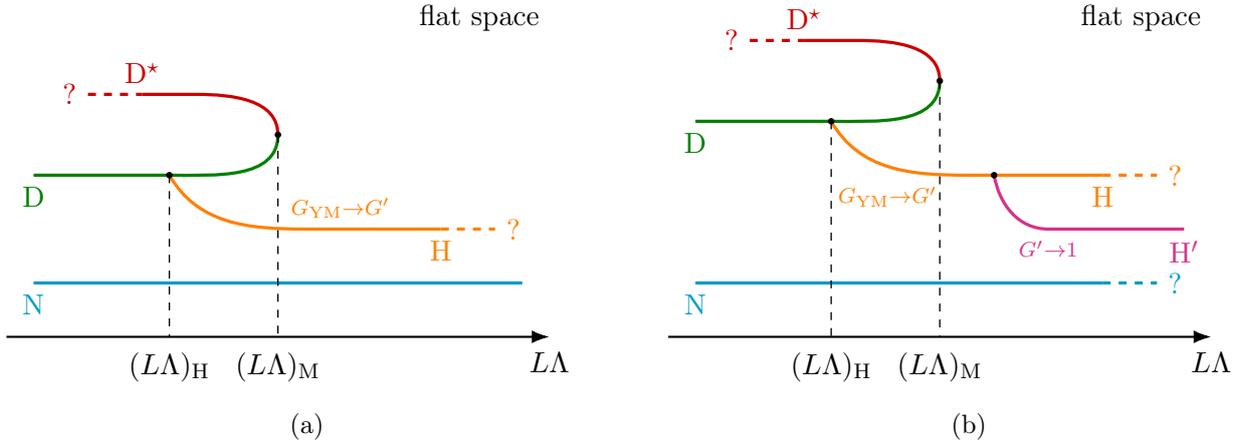
\begin{figure}[t]
    \centering
    \begin{subfigure}[b]{0.49\linewidth}
        \begin{tikzpicture}[x={0.45\linewidth},y={0.45\linewidth}]
            \coordinate (M) at (0,0.75);
            \coordinate (H) at (-0.4,0.6);
            \draw[-{Latex},line width=0.8pt] (-1,0) -- (1,0) node[below,yshift=-2] {\( L \Lambda \)};

            \draw[color=green!50!black,line width=1.2pt] (-0.9,0.6) node[below] {\( \mathrm{D} \)} to (H) to [out=0,in=-90] (M);
            \draw[color=red!80!black,line width=1.2pt] (-0.5,0.9) node[above] {\( \mathrm{D}^{\star} \)} to (-0.3,0.9) to [out=0,in=90] (M);

            \draw[color=red!80!black,line width=1.2pt,dashed] (-0.7,0.9) node[left] {\( ? \)} -- (-0.5,0.9);
            \draw[color=cyan!80!black,line width=1.2pt] (-0.9,0.2) node[below] {\( \mathrm{N} \)} to (0.9,0.2);

            \draw[color=orange,line width=1.2pt] (H) to [out=-60,in=-180] (0.1,0.4) node[above right,xshift=-2ex] {\( \scriptstyle G\ped{YM} \to G' \)} to (0.6,0.4) node[below] {\( \mathrm{H} \)};

            \draw[color=orange,line width=1.2pt,dashed] (0.6,0.4) -- (0.8,0.4) node[right] {\( ? \)};
            \draw[dashed, line width=0.5pt] (M) -- (0,0) node[below,yshift=-2] {\( (L \Lambda)\ped{M} \)};
            \draw[dashed, line width=0.5pt] (H) -- (-0.4,0) node[below,yshift=-2] {\( (L \Lambda)\ped{H} \)};
            \filldraw (M) circle (0.01);
            \filldraw (H) circle (0.01);

            \node[above left,text width=3cm,align=right] at (1,1.1) {flat space};
        \end{tikzpicture}
        \caption{}
    \end{subfigure}
    \hfill
    \begin{subfigure}[b]{0.49\linewidth}
        \begin{tikzpicture}[x={0.45\linewidth},y={0.45\linewidth}]
            \coordinate (M) at (0,0.95);
            \coordinate (H) at (-0.4,0.8);
            \coordinate (Hp) at (0.2,0.6);

            \draw[-{Latex},line width=0.8pt] (-1,0) -- (1,0) node[below,yshift=-2] {\( L \Lambda \)};

            \draw[color=green!50!black,line width=1.2pt] (-0.9,0.8) node[below] {\( \mathrm{D} \)} to (H) to [out=0,in=-90] (M);
            \draw[color=red!80!black,line width=1.2pt] (-0.5,1.1) node[above] {\( \mathrm{D}^{\star} \)} to (-0.3,1.1) to [out=0,in=90] (M);

            \draw[color=red!80!black,line width=1.2pt,dashed] (-0.7,1.1) node[left] {\( ? \)} -- (-0.5,1.1);
            \draw[color=cyan!80!black,line width=1.2pt] (-0.9,0.2) node[below] {\( \mathrm{N} \)} to (0.6,0.2);

            \draw[color=orange,line width=1.2pt] (H) to [out=-60,in=-180] (0.1,0.6) node[below left,xshift=-1.5ex] {\( \scriptstyle G\ped{YM} \to G' \)} to (0.6,0.6) node[below] {\( \mathrm{H} \)};

            \draw[color=orange,line width=1.2pt,dashed] (0.6,0.6) -- (0.8,0.6) node[right] {\( ? \)};
            \draw[color=cyan!80!black,line width=1.2pt,dashed] (0.6,0.2) -- (0.8,0.2) node[right] {\( ? \)};

            \draw[color=magenta!90!black,line width=1.2pt] (Hp) to [out=-80,in=-180] (0.4,0.4) node[below] {\( \scriptstyle G' \to 1 \)} -- (0.9,0.4) node[below] {\( \mathrm{H}' \)};

            \draw[dashed, line width=0.5pt] (M) -- (0,0) node[below,yshift=-2] {\( (L \Lambda)\ped{M} \)};
            \draw[dashed, line width=0.5pt] (H) -- (-0.4,0) node[below,yshift=-2] {\( (L \Lambda)\ped{H} \)};
            \filldraw (M) circle (0.01);
            \filldraw (H) circle (0.01);
            \filldraw (Hp) circle (0.01);

            \node[above left,text width=3cm,align=right] at (1,1.1) {flat space};
        \end{tikzpicture}
        \caption{}
    \end{subfigure}
    \hspace*{-1.5em}
    \caption{
    Two sketches of possible Higgsing scenarios. In both panels, at $(L\Lambda)\ped{H}< (L \Lambda)\ped{M}$ a charged operator under $G\ped{YM}$ becomes relevant, guaranteeing the existence of a boundary condition with smaller global symmetry. 
This new boundary condition may itself contain other charged operators which eventually also become relevant, leading to a cascade of symmetry breaking until no symmetry is left.  In panel (a), the Neumann bc extrapolates to flat space. However, since H emerges before the merging transition, the latter has one more possible endpoint. In panel (b), a completely Higgsed boundary condition H$'$, rather than N, extrapolates to flat space.
\label{fig:Higgs}}
\end{figure}

\paragraph{\textbf{Higgsing}}
One of the initially irrelevant Lorentz scalar operators of Dirichlet bc, charged under $G\ped{YM}$, becomes marginal at some critical value $(L \Lambda\ped{YM})\ped{H}$, before any singlet operator does. 
We expand more on this scenario because it was only mentioned in~\cite{Copetti:2023sya, Ciccone:2024guw} briefly. 
Let us denote the operator becoming marginal as $\mathcal{O}^I$, where $I$ is an index in the (non-trivial) representation $R$ of $G\ped{YM}$ this operator belongs to.
When this happens, we gain the possibility of defining a new boundary condition: we simply perturb the Dirichlet bc by the relevant coupling $g_I$
\begin{equation}\label{explicitsymmetrybreaking}
    S\ped{YM} \to S\ped{YM} + \int_{z=0} \dif[3]{\vec{x}} \, g_I \, \mathcal{O}^I(\vec{x}) \, ,
\end{equation}
and let the boundary RG flow settle at a new AdS-invariant boundary condition.\footnote{
    In~\cite{Copetti:2023sya, Ciccone:2024guw}, the operator $\mathcal{O}^I$ triggering Higgsing was assumed to be in the adjoint representation of $G\ped{YM}$. 
    However, this does not have to be the case; any non-trivial representation could do the job. 
    Clearly, point-like operators of the Dirichlet boundary condition can only be in representations obtained by products of the adjoint representation with itself. 
    One reason why the adjoint representation is special is that, according to the perturbative calculations for $SU(N)$ YM in~\cite{Ciccone:2024guw}, it is the charged operator with the smallest (irrelevant) scaling dimension at one-loop order.
} This RG flow is short when $L \Lambda\ped{YM}$ is larger than $(L \Lambda\ped{YM})\ped{H}$ but parametrically close to it. 
In this limit, the spectrum of the new boundary condition can be obtained as a perturbation of Dirichlet bc. 
Since the perturbing operator is charged,  the global symmetry of the new boundary condition is explicitly broken to a subgroup of $G\ped{YM}$. 
The current conservation is violated by the deformation as follows
\begin{equation}\label{brokencurrent}
    \partial^\mu J^a_\mu \propto g^*_I(T^a)^I_{~J} \mathcal{O}^J \, ,
\end{equation}
where $(T^a)^I_{~J}$ denotes the $G\ped{YM}$ generator in the representation $R$, and $g^*_I$ is the value of the coupling at the IR fixed point of the boundary RG flow. 
We see that the broken currents are those associated with the generators for which $g^*_I(T^a)^I_{~J} \neq 0 $. 
From the point of view of the bulk, this breaking is interpreted as Higgsing of the gauge degrees of freedom. 
The resulting anomalous dimension of the current operator corresponds to a non-zero mass of the gluons. 
For this reason, we call this a Higgsing bc and denote it with H. To illustrate this characterisation of Higgsing in AdS in a simpler weakly-coupled example, in Appendix~\ref{app:AHM} we discuss the abelian Higgs model in AdS, in particular deriving equations analogous to~\eqref{explicitsymmetrybreaking} and~\eqref{brokencurrent}.

It is important to stress that the emergence of this new Higgsing bc \emph{does not make Dirichlet disappear}.
Indeed, as depicted in Figure~\ref{fig:Higgs}, while the two continuous families of CFTs defined by Dirichlet and H join at a point, they do so in a different way compared to the merging of Dirichlet and $\mathrm{D}^{\star}$ discussed above. The two associated lines do not intersect smoothly, but rather cross and continue past the intersection point.\footnote{Depending on the form of the leading non-linear terms in the $\beta$ function of $g_I$, the line associated with Higgsing might exist as a unitary and AdS-invariant boundary condition on both sides of the intersection with D, or only on one side. The latter case is realised in the perturbative example discussed in the Appendix~\ref{app:AHM}. Note that in the discussion around~\eqref{brokencurrent} we are implicitly assuming that the sign of the leading non-linear term leads to a real IR fixed point parametrically close to the intersection point.} This is allowed because the smooth merger and annihilation can only happen between families of CFTs sharing the same symmetry, while Dirichlet and H, by definition, have different symmetries.
Similar behaviours arise, for instance, between the free and interacting CFTs near Wilson-Fisher and $d=2+\epsilon$ fixed points, if one treats the spacetime dimension as a continuous parameter.\footnote{
    Examples of continuous families of CFTs with a charged operator that crosses marginality can also be found in the context of local CFTs.
    For example, one can consider the $O(N)$-symmetric scalar CFT in $3d$, where the role of  $L \Lambda\ped{YM}$ is played by the number of fields $N$. 
    In this family of theories, a rank-$4$ traceless symmetric operator approaches marginality for $N\ped{H}\gtrsim 2.99$~\cite{Chester:2020iyt}. 
    This implies that, for $N>N\ped{H}$, the $O(N)$ model is unstable with respect to this perturbation, which, however, needs to be turned on explicitly. 
} 
This scenario recently appeared in~\cite{DeCesare:2025ukl}, which defined the two families as ``continuously but not analytically connected''.

One still needs to find a mechanism for Dirichlet to disappear. 
One possibility is that it still happens through a merger and annihilation, only at a larger value of the coupling $(L \Lambda\ped{YM})\ped{M} >(L \Lambda\ped{YM})\ped{H}$. 
In fact, rather than providing a full and clear picture of the connection between the regimes of small and large AdS, this scenario suggests that things could be far more involved than in the simple merger picture above.
For instance, we also need to consider what happens to the H boundary condition in the limit of a large radius. 
Note that H will be charged under an unbroken subgroup $G^\prime$ of $G\ped{YM}$, where $G^\prime$ is the group that leaves invariant the deforming operator in~\eqref{explicitsymmetrybreaking}. 
This implies that H shouldn't extrapolate continuously to the large radius boundary condition either. 
One possibility, illustrated in panel (a) of Figure~\ref{fig:Higgs}, is that it is still Neumann that is connected to the flat space observables. 
Another possibility is that a further Higgsing takes place in H, breaking $G'$ completely and giving rise to a fully Higgsed boundary condition $\mathrm{H}'$ with no continuous global symmetry. 
As illustrated in panel (b), $\mathrm{H}'$ could, in principle, be a valid alternative to Neumann as the boundary condition that is continuously connected to the flat space observables, compatibly with colour confinement and the mass gap.\footnote{
    A natural question is whether such a putative H$'$ bc is in fact the same as N.
    Depending on the global structure of the theory, one might be able to use discrete symmetries and boundary operators originating from bulk line operators to distinguish between H$'$ and N, see the comments in footnote~\ref{foot:disc}. 
    A comprehensive discussion of these global aspects and of bulk line operators goes beyond the scope of this paper.
}

From the discussion above, it is clear that the more intricate Higgsing scenario can be ruled out if one can show that Dirichlet disappears \emph{before} any charged scalar operator becomes relevant. 
This observation motivates part of our investigations below.

\section{Boundary spectrum of YM in AdS}\label{sec:spectrum}

At zero coupling \( g\ped{YM} = 0 \), the \( d = 3 \) boundary CFT is given by a generalised Free Vector Theory (GFV): the Mean Field Theory constructed from \( J^{a}_{\mu} \), which is a spin \( \ell = 1 \) operator in the adjoint representation of \( SU(N) \) with dimension \( \Delta = d-1 = 2 \). 
As we turn on bulk interactions \( g\ped{YM} \neq 0 \), the spectrum of the boundary CFT becomes a deformation of the GFV, where all composite operators built out of \( J^{a}_{\mu} \) acquire anomalous dimensions. 
The current itself and the vacuum do not obtain an anomalous dimension. 
We will assume that the spectrum of primaries in the boundary CFT of YM in \( \mathrm{AdS}_{4} \) is continuously connected to the spectrum of the GFV. 
Contrary to what typically happens for perturbative IR fixed points of flat space QFTs, the bulk equations of motion are not expected to imply discontinuous changes in the boundary spectrum of primaries between the free and the interacting theory. 
From the boundary point of view, this is easy to see since, other than the current itself (which remains conserved), there are no operators located at the unitarity bound at zero coupling. 

\subsection{Symmetries}\label{sec:spectrum-symmetries}
The classical Lagrangian of pure YM theory in $\mathrm{AdS}_{4}$ with gauge group $SU(N)$ is
\begin{equation}\label{eq:YM-action}
    S\ped{YM} = -\frac{1}{2g\ped{YM}^{2}} \int \dif{x} \tr\!\big[F_{MN} F^{MN}\big] \, ,
\end{equation}
where the field strength for an $\mathfrak{su}(N)$ connection $A_{M} = A_{M}^{a} T^{a}$ is
\begin{equation}\label{eq:Fmunu}
    F_{MN } = F_{MN}^{a} T^{a} \, , \qquad 
    F_{MN}^{a} = \nabla_{M} A_{N}^{a} - \nabla_{N} A_{M}^{a} + f^{abc} A_{M}^{b} A_{N}^{c} \, ,
\end{equation}
and $g\ped{YM}$ is the YM coupling constant. 
In what follows, we will use capital Latin letters to denote $AdS_4$ indices and Greek letters to denote CFT${}_3$ indexes. 
Lower case Latin letter will be reserved for $\mathfrak{su}(N)$ adjoint indices. 
Here $\{T^{a}, a = 1, \ldots, N^2-1\}$ are anti-hermitian $\mathfrak{su}(N)$ generators with 
\begin{equation}\label{eq:sungenerators}
    \tr\!\big[T^{a} T^{b}\big] = -\frac{1}{2} \delta^{ab} \, , \qquad
    \big[T^{a}, T^{b}\big] = f^{abc} T^{c} \, ,
\end{equation}
where the trace is in the fundamental representation, and the structure constants $f^{abc}$ are real.
The integral in~\eqref{eq:YM-action} is short-hand for the integral over $\mathrm{AdS}$ 
\begin{equation}\label{eq:AdS-integral}
    \int \dif{x} \, f(x) \equiv \int_{\mathrm{AdS}_{4}} \dif[4]{x} \, \sqrt{g} f(x) \, ,
\end{equation}
and $\nabla_{M}$ in~\eqref{eq:Fmunu} is the covariant derivative in AdS.

When Dirichlet bc~\eqref{eq:Dirichlet-bc} are imposed, boundary correlators of local operators are invariant under the faithful symmetry
\begin{equation}\label{eq:YM-AdS-symmetry}
    G = G\ped{space} \times G\ped{global} = \Big(SO(4,1)_{0} \rtimes \Z_{2}^{\mathcal{P}}\Big) \times \Big(PSU(N) \rtimes \Z_{2}^{\mathcal{C}}\Big) \, , \qquad (N \geq 3) \, ,
\end{equation}
while for $N = 2$ we have
\begin{equation}\label{eq:YM-AdS-symmetry-N2}
    G = \Big(SO(4,1)_{0} \rtimes \Z_{2}^{\mathcal{P}}\Big) \times SO(3) \, , \qquad (N = 2) \, .
\end{equation}
Let us explain the origin of each factor.
\begin{itemize}
    \item \textbf{Conformal symmetry}: $SO(4,1)_{0}$ denotes the connected component of the Euclidean conformal group of $\R^{3}$. 
    This coincides with the isometries of Euclidean $\mathrm{AdS}_{4}$ that are connected to the identity. 
    It is a well-known fact that boundary correlators of QFTs in rigid $\mathrm{AdS}$ space are invariant under this group.
    \item \textbf{Parity}: $\Z_{2}^{\mathcal{P}}$ denotes the $\Z_2$ group generated by the identity and parity transformations on the conformal boundary. 
    In Poincaré coordinates $(z, \vec{x})$, where $\mathrm{AdS}_{4}$ is Weyl equivalent to $\R_{+} \times \R^{3}$, a parity transformation is generated by a single coordinate reflection on $\R^{3}$, or equivalently by a total reflection $(z, \vec{x}) \to (z, -\vec{x})$.
    With the latter definition, the gauge connection transforms as $A^{M}(z, \vec{x}) \to {\mathcal{P}^{M}}_{N} A^{N}(z, -\vec{x})$ with ${\mathcal{P}^{z}}_{z} = 1, {\mathcal{P}^{\mu}}_{\nu} = -\delta^{\mu}_{\nu}$ and zero otherwise. 
    The action~\eqref{eq:YM-action} is manifestly invariant under this transformation. 
    Parity does not commute with translations and special-conformal transformations, hence the product is semi-direct in~\eqref{eq:YM-AdS-symmetry}.\footnote{
        For example, a level one descendant picks up an additional $-1$ factor compared to its corresponding primary.
    }
    However, it still commutes with the maximally compact subgroup $SO(1,1) \times SO(3)$, generated by dilatation and Euclidean rotations. 
    This makes it easy to take parity into account.
    Its action on a local primary operator with scaling dimension $\Delta$ and spin $\ell$, independently of its $PSU(N) \rtimes \Z_{2}^{\mathcal{C}}$ representation, is 
    \begin{equation}\label{eq:parityDef}
        \mathcal{O}_{\Delta,\ell}(\vec{x}) 
        \ \xrightarrow{\hspace{0.5em}\mathcal{P}\hspace{0.5em}} \
        p (-1)^{\ell} \mathcal{O}_{\Delta,\ell}(-\vec{x}) \, ,
    \end{equation}
    where $p=\pm$ denotes the intrinsic parity of the operator. 

    \item \textbf{$\boldsymbol{SU(N)}$ symmetry}: even though $SU(N)$ transformations are gauged in the bulk of $\mathrm{AdS}$, they are genuine symmetries of the Dirichlet boundary CFT. 
    Local boundary operators are obtained as products of currents, in irreducible representations arising from the tensor product of adjoint representations.
    Since the adjoint representation has zero $N$-ality, being invariant under transformations in the centre of $SU(N)$, all boundary operators will also be so.
    Thus the faithful symmetry is $PSU(N) = SU(N) / \Z_{N}$. 
    For $N = 2$, we have $PSU(2) = SO(3)$.

    \item \textbf{Charge conjugation}: $\Z_{2}^{\mathcal{C}}$ is the $\Z_2$ group generated by the identity and charge conjugation $\mathcal{C}$. 
    We define the action of $\mathcal{C}$ on the $\mathfrak{su}(N)$ connection as
    \begin{equation}\label{eq:charge-conjugation-Amu}
        A_{M}(x) = A_{M}^{a}(x) T^{a} 
        \ \xrightarrow[]{\hspace{0.5em}\mathcal{C}\hspace{0.5em}} \ 
        A_{M}(x)^{*} = \left(- C^{ab} A_{M}^{b}(x)\right) T^{a} \, ,
    \end{equation}
    where ${}^{*}$ denotes complex conjugation, and we used that $(T^{a})^{*} = - (T^{a})^{t}$ for anti-hermitian generators, while also defining the matrix
    \begin{equation}\label{eq:Cab}
        (T^{a})^{t} = C^{ab} T^{b}\, , \qquad 
        C^{ab} = - 2 \tr\!\big[(T^{a})^{t} \, T^{b}\big] \, .
    \end{equation}
    Under this transformation the field strength~\eqref{eq:Fmunu} transforms as\footnote{
        The structure constants $f^{abc} = - 2 \tr\left[[T^{a}, T^{b}] T^{c}\right]$ satisfy the following relation
        \(
            C^{a a'} C^{b b'} C^{c c'}f^{a'b'c'} 
            = - 2 \tr\left[[(T^{a})^{t}, (T^{b})^{t}] (T^{c})^{t}\right] 
            = -f^{abc} \,,
        \)
        which implies
        \(
            f^{abc} A_{M}^{b} A_{N}^{c} \xrightarrow{\mathcal{C}} f^{abc} C^{b b'} C^{c c'}  A_{M}^{b'} A_{N}^{c'} = - C^{a a'} f^{a' b' c'} A_{M}^{b'} A_{N}^{c'} \, .
        \)
        \label{foot:fabc-charge-conj}
    }
    \begin{equation}\label{eq:F-charge-conjugation-transform}
        F_{MN}(x) = F_{MN}^{a}(x) T^{a} 
        \ \xrightarrow[]{\hspace{0.5em}\mathcal{C}\hspace{0.5em}} \ 
        F_{MN}(x)^{*} = \left(- C^{ab} F_{MN}^{b}(x)\right) T^{a} \, .
    \end{equation}
    It is easy to check that this is a symmetry of~\eqref{eq:YM-action}.  
    Since~\eqref{eq:charge-conjugation-Amu} does not commute with $SU(N)$ transformations, the product in~\eqref{eq:YM-AdS-symmetry} is semi-direct. 
    For $N = 2$, complex conjugation can be realised by an $SU(2)$ transformation, i.e.\ there is $g_{0}$ such that $g^{*} = g_{0} g g_{0}^{-1}$ for any $g \in SU(2)$, explaining why we don't have a $\Z_{2}^{\mathcal{C}}$ factor in~\eqref{eq:YM-AdS-symmetry-N2}. 

    The boundary correlators inherit this symmetry.
    The action of $\mathcal{C}$ on a generic boundary primary operator $\mathcal{O}$ is
    \begin{equation}\label{eq:CCDef}
        \mathcal{O}(\vec{x}) 
        \ \xrightarrow{\hspace{0.5em}\mathcal{C}\hspace{0.5em}} \ 
        {\mathfrak{c}} \, \mathcal{O}(\vec{x}) ^{*} \, ,
    \end{equation}
    valid for any scaling dimension, spin, and $SU(N)$ representation. 
    The factor $\mathfrak{c}$ is a possibly present intrinsic charge parity, which we will discuss shortly. 
    Charge conjugation $\mathcal{C}$ acts on an irreducible representation $R_\lambda$ with highest-weight Dynkin labels $\lambda = (\lambda_1,\ldots,\lambda_{N-1})$ as 
    \begin{equation}\label{eq:DynkinCharge}
        \lambda 
        \ \xrightarrow[]{\hspace{0.5em}\mathcal{C}\hspace{0.5em}} \  
        \lambda^* = (\lambda_{N-1},\ldots,\lambda_{1}).
    \end{equation} 
    If $\lambda\neq \lambda^*$, irreducible representations of $PSU(N) \rtimes \Z_{2}^{\mathcal{C}}$ are given by $R_{\lambda}^c = R_\lambda \oplus R_{\lambda^{*}}$, where $\mathcal{C}$ acts by swapping the two subspaces. 
    In this case, the constant $\mathfrak{c}$ in~\eqref{eq:CCDef} is immaterial and can be taken to be one. 
    For brevity, we will denote by $\lambda_c$ the representation $R_{\lambda}^c$.
    If $\lambda=\lambda^*$, the representation is self-conjugate, and in this case we can have a non-trivial intrinsic charge conjugation $\mathfrak{c}=\pm 1$ in~\eqref{eq:CCDef}. 
    The corresponding representations of $PSU(N) \rtimes \Z_{2}^{\mathcal{C}}$ are $R_\lambda^\pm$ and we denote them simply as $\lambda_\pm$.\footnote{
        These exhaust all the finite-dimensional unitary irreducible representations, see e.g.~\cite[Prop. 1.5]{Adams2015}.
    } 

    Whenever we do not need to distinguish between the two cases, we will just write $\lambda_{(c,\pm)}$.
\end{itemize}
See Appendix~\ref{app:characters-GFV} for further details on the action of parity and charge conjugation.

Irreducible representations of $G = G\ped{space} \times G\ped{global}$ of the boundary primary fields will then be characterised by the quantum numbers \(\left(\Delta,\ell,p,\lambda_{(c,\pm)}\right)\).
For example, the local conserved $SU(3)$ current has
\begin{equation}\label{eq:J-quantum-numbers}
    J^{a}_{\mu} \quad \colon \quad \Big(2,1,+,(1,1)_{-}\Big).
\end{equation}
Details on how we fix the intrinsic sign under $\Z_{2}^{\mathcal{C}}$ are given below. 
For bootstrap applications, it will be of particular importance to know the spectrum of operators in the \( J \times J \) OPE that are allowed by symmetries. 
This will be reviewed in Section~\ref{sec:bootstrap-setup}, but let us anticipate here the first restriction imposed by the global symmetry to keep an eye out for the relevant representations in the following section. 

We denote by $\mathbf{adj}$ the adjoint representation of $SU(N)$.
For $N \geq 4$, the tensor product of two $\mathbf{adj}$ decomposes as
\begin{equation}\label{eq:SUN-adj-decompose}
    \mathbf{adj} \otimes \mathbf{adj} \cong
    \mathbf{S} \oplus 2 \, \mathbf{adj} \oplus \mathbf{S\bar{S}} \oplus \mathbf{A\bar{A}} \oplus \mathbf{S\bar{A}} \oplus \mathbf{A\bar{S}} \, ,
\end{equation}
where the Young diagrams and Dynkin labels of the above representations are reported in Table~\ref{tab:rep-names}.\footnote{
    The \( \mathbf{S\bar{S}} \), \( \mathbf{S\bar{A}} \), \( \mathbf{A\bar{S}} \) and \( \mathbf{A\bar{A}} \) representations where denoted in~\cite{Ciccone:2024guw} as \( R_{1} \), \( \overline{R_{2}} \), \( R_{2} \) and \( R_{3} \), respectively.
}
All these representations have vanishing $N$-ality and lift to representations of $PSU(N)$. 
For $N = 3$, the decomposition~\eqref{eq:SUN-adj-decompose} still applies but the $\mathbf{A\bar{A}}$ representation does not appear. 
For $N = 2$, we have 
\begin{equation}\label{eq:SU2-adj-decompose}
    \mathbf{adj}\otimes\mathbf{adj} \cong \mathbf{S}\oplus \mathbf{adj} \oplus \mathbf{S\bar{S}} \,,
\end{equation}
where $\mathbf{S}$, $\mathbf{adj}$ and $\mathbf{S\bar{S}}$ representations are identified with the singlet, the vector, and the traceless symmetric representations of $SO(3)$, respectively.

\begin{table}[t!]
    \centering
    \begin{tabular}{cllcc}
      \toprule
      Name & \multicolumn{2}{c}{Dynkin lables} & Young diagram & Dimension \\ \midrule
      \( \mathbf{S} \) & \( (0,\ldots,0) \) & & \( [0] \) & \( 1 \) \\
      \( \mathbf{adj} \) & \( (1,0,\ldots,0,1) \) & [\( (2) \) for \( N=2 \)] & \( [N-1,1] \) & \( N^{2} - 1 \) \\
      \( \mathbf{S\bar{S}} \) & \( (2,0,\ldots,0,2) \) & [\( (4) \) for \( N=2 \)] & \( [N-1,N-1,1,1] \) & \( \frac{1}{4}(N-1)N^{2}(N+3) \)\\
      \( \mathbf{S\bar{A}} \) & \( (2,0,\ldots,0,1,0) \) & [\( (3,0) \) for \( N = 3\)] & \( [N-2,1,1] \) & \( \frac{1}{4}(N^{2}-4)(N^{2}-1) \)  \\
      \( \mathbf{A\bar{S}} \) & \( (0,1,0,\ldots,0,2) \) & [\( (0,3) \) for \( N = 3\)] & \( [N-1,N-1,2] \) & \( \frac{1}{4}(N^{2}-4)(N^{2}-1) \) \\
      \( \mathbf{A\bar{A}} \) & \( (0,1,0,\ldots,0,1,0) \) & [\((0,2,0)\) for \( N = 4 \)] & \( [N-2,2] \)  & \( \frac{1}{4}(N-3)N^{2}(N+1) \) \\ \bottomrule
    \end{tabular}
    \caption{Irreducible \( SU(N) \) representations appearing in the tensor product of two adjoint representations. Note that $\mathbf{A\bar{A}}$ does not appear for $N = 2,3$, while $\mathbf{S\bar{A}}$ and $\mathbf{A\bar{S}}$ do not appear for $N=2$. A Young diagram $[l_{1},\ldots,l_{k}]$ has $l_{i}$ boxes in the $i$-th column.}\label{tab:rep-names}
\end{table}

When taking into account the transformation property under charge conjugation, for $N \geq 3$ the two $\mathbf{adj}$ in the right-hand side of~\eqref{eq:SUN-adj-decompose} have opposite charge conjugation parity and induce two inequivalent representations of $PSU(N) \rtimes \Z_{2}^{\mathcal{C}}$, denoted by $\mathbf{adj}_{\pm}$. Moreover, the $\mathbf{S\bar{A}}$ and $\mathbf{A\bar{S}}$ representations are grouped into a representation denoted by $\mathbf{S\bar{A}}_{c}$. The tensor product~\eqref{eq:SUN-adj-decompose} is then ``refined'' as follows for two $\mathbf{adj}_{\pm}$ representations:
\begin{equation}\label{eq:SUNZ2-adj-decompose}
    \mathbf{adj}_{\pm} \otimes \mathbf{adj}_{\pm} \cong
    \mathbf{S}_{+} \oplus \mathbf{adj}_{+} \oplus \mathbf{adj}_{-} \oplus \mathbf{S\bar{S}}_{+} \oplus \mathbf{A\bar{A}}_{+} \oplus \mathbf{S\bar{A}}_{c} \, .
\end{equation}
The signs under charge conjugation on the right-hand side can be determined using the characters of Appendix~\ref{app:characters-charge-conj}.

We further split representations in~\eqref{eq:SUNZ2-adj-decompose} and~\eqref{eq:SU2-adj-decompose} in terms of their behaviour under the swapping of the two tensor product factors, i.e.\ in the symmetric square ($\mathrm{Sym}^{2}$) and alternating square ($\mathrm{Alt}^{2}$) components. 
This is useful, since, as we will review in Section~\ref{sec:bootstrap-setup}, their symmetry properties under swapping determine further selection rules in the $J \times J$ OPE. 
The decomposition for $N \geq 3$ reads\footnote{
    Again, for $N=3$ the representation $\mathbf{A\bar{A}}$ does not appear.
}
\begin{equation}\label{eq:SUNZ2-adj-decompose-sym-alt}
    \begin{aligned}
        N \text{ odd} \ \colon \qquad &
    \begin{aligned}
        \mathsf{R}_{s} = \mathrm{Sym}^{2} \, (\mathbf{adj}_{\pm}) &\cong \mathbf{S}_{+} \oplus \mathbf{adj}_{+} \oplus \mathbf{S\bar{S}}_{+} \oplus \mathbf{A\bar{A}}_{+} \, , \\
        \mathsf{R}_{a} = \mathrm{Alt}^{2} \, (\mathbf{adj}_{\pm}) &\cong \mathbf{adj}_{-} \oplus \mathbf{S\bar{A}}_{c} \, , \\
    \end{aligned}
    \\[1em]
    N \text{ even} \ \colon \qquad &
    \begin{aligned}
        \mathsf{R}_{s} = \mathrm{Sym}^{2} \, (\mathbf{adj}_{\pm}) &\cong \mathbf{S}_{+} \oplus \mathbf{adj}_{-} \oplus \mathbf{S\bar{S}}_{+} \oplus \mathbf{A\bar{A}}_{+} \, , \\
        \mathsf{R}_{a} = \mathrm{Alt}^{2} \, (\mathbf{adj}_{\pm}) &\cong \mathbf{adj}_{+} \oplus \mathbf{S\bar{A}}_{c} \, , \\
    \end{aligned}
    \end{aligned}
\end{equation}
and for $N=2$
\begin{equation}\label{eq:SU2-adj-decompose-sym-alt}
    N = 2 \ \colon \qquad
    \begin{aligned}
        \mathsf{R}_{s} = \mathrm{Sym}^{2}(\mathbf{adj}) &\cong \mathbf{S} \oplus \mathbf{S\bar{S}} \, , \\
        \mathsf{R}_{a} = \mathrm{Alt}^{2}(\mathbf{adj}) &\cong \mathbf{adj} \, .
    \end{aligned}
\end{equation}

We observe, in particular, that $\mathbf{adj}_{-}$ (resp.\  $\mathbf{adj}_{+}$) appears in $\mathsf{R}_{a}$ for $N$ odd (reps.\ $N$ even). 
For Yang-Mills in $\mathrm{AdS}$ we require $A_{\mu}^{a} \underset{z\to 0}{\sim} z J^{a}_{\mu}$ to transform with the same sign as $f^{abc} A_{\mu}^{a} A_{\nu}^{c} \underset{z\to 0}{\sim} z^{2} f^{abc} J^{a}_{\mu} J^{b}_{\nu}$, in order for $F_{\mu\nu}^{a}$ to transform covariantly as in~\eqref{eq:F-charge-conjugation-transform}.
Since $f^{abc} A_{\mu}^{a} A_{\nu}^{c}$ is precisely the projection in the alternating square adjoint subspace, this implies that $A_{\mu}^{a}$, and thus $J_{\mu}^{a}$, has to be identified with the $\mathbf{adj}_{-}$ representation of $PSU(N) \rtimes \Z_{2}^{\mathcal{C}}$ for $N$ odd and with $\mathbf{adj}_{+}$ for $N$ even. 
This explains the origin of the ``$-$'' sign in~\eqref{eq:J-quantum-numbers}.

\subsection{Free theory spectrum}\label{sec:spectrum-GFV}

The spectrum of the GFV generated from a non-Abelian \( SU(N) \) conserved current can be obtained with character techniques, as was done in Appendix A of~\cite{Ciccone:2024guw}, and further refined in Appendix~\ref{app:characters-GFV} for the purposes of the present work.
In this section, we discuss some features of the spectrum that are relevant for the scenarios discussed in Section~\ref{sec:intro} and for the gap assumptions used in Sections~\ref{sec:bootstrap-dimesions-feasibility} and~\ref{sec:bootstrap-dimensions-GFV}.
The spectrum of primary operators of the GFV is obtained by \( J^{a}_{\mu} \) and products thereof, with appropriate insertions of derivatives and projection into irreducible representations of \( SU(N) \). 
Moreover, one can contract these operators with the \( \epsilon^{\mu\nu\rho} \) tensor to form operators that have intrinsic parity \( p = - \). 

We start by discussing the double trace operators of the form $[JJ]$: these are the lowest-dimensional operators, besides the identity and the conserved current itself. They are also the only operators for which anomalous dimensions have been computed, as we will review in Section~\ref{sec:spectrum-perturbative}.
In terms of the twist
\begin{equation}
    \Delta-\ell = \tau = 2 + m \, ,
\end{equation} 
parity-even and parity-odd operators have even and odd integer values of $m\geq 0$, respectively.
Consider first parity-even operators. 
They are of the form\footnote{
    The position of the derivative insertions it to be understood as schematic, for example by \( J^{a}_{\mu} \partial_{\mu_{1}}\cdots \partial_{\mu_{\ell-2}} J_{\nu}^{b} \) we mean that among all possible linear combinations of operators with 2 \( J \)-s and \( \ell-2 \) derivatives, we have to select the ones that correspond to primaries.
}
\begin{equation}\label{eq:ops-parity-even-generic2}
    \begin{aligned}
        (\Delta,\ell,p) &= (\ell+4+2n,\ell, +) & \colon \quad 
        & J^{a}_{\mu} \partial_{\mu_{1}}\cdots \partial_{\mu_{\ell}} \Box^{n} J_{\mu}^{b} \,, \qquad 
        & \ell\geq 0\,, n\geq 0 \,, \\
        (\Delta,\ell,p) &= (\ell+2+2n,\ell, +) & \colon \quad 
        & J^{a}_{\mu} \partial_{\mu_{1}}\cdots \partial_{\mu_{\ell-2}} \Box^{n} J_{\nu}^{b} \,, \qquad 
        & \ell\geq 2\,, n\geq 0 \,,
    \end{aligned}
\end{equation}
and are in representations $\mathsf{R}_{s}$ for $\ell$ even and $\mathsf{R}_{a}$ for $\ell$ odd. 
The first Regge trajectory $m=0$ starts from $\ell\geq 2$ and is non-degenerate, as are the operators with $\ell = 0 $ and $\ell = 1$ for any $m\geq 2$.
Operators with $\ell\geq 2$ and $m\geq 2$ are instead doubly degenerate.

Double trace parity-odd operators have the same schematic form~\eqref{eq:ops-parity-even-generic2}, provided we replace one of the two currents with its dual
\( \tilde{J}^{\mu,a} = \epsilon^{\mu\nu\rho} \partial_{\nu}J^{a}_{\rho} \).
We then have
\begin{equation}\label{eq:ops-parity-odd-generic2}
    \begin{aligned}
        (\Delta,\ell,p) &= (\ell+5+2n,\ell, -) & \colon \quad 
        & \tilde J^{a}_{\mu} \partial_{\mu_{1}}\cdots \partial_{\mu_{\ell}} \Box^{n} J_{\mu}^{b} \,, \qquad 
        & \ell\geq 0\,, n\geq 0 \,, \\
        (\Delta,\ell,p) &= (\ell+3+2n,\ell, -) & \colon \quad & 
        \tilde J^{a}_{\mu} \partial_{\mu_{1}}\cdots \partial_{\mu_{\ell-2}} \Box^{n} J_{\nu}^{b} \,, \qquad 
        & \ell\geq 2\,, n\geq 0 \,.
      \end{aligned}
\end{equation}
Operators with $\ell=0$ and $\ell=1$ are non-degenerate for any $m\geq 3$ and are respectively in representations $\mathsf{R}_{s}$ and $\mathsf{R}_{a}$. 
Operators with $\ell\geq 2$ and any $m \geq 1$ are also non-degenerate, but they appear in both $\mathsf{R}_{s}$ and $\mathsf{R}_{a}$ representations.

Determining the spectrum (quantum numbers and degeneracies) of $k$-trace primary operators by explicit construction becomes increasingly hard as $k$ increases, since one has to disentangle primaries from descendants and decompose into irreducible representations products of more and more adjoint currents.
For this purpose, in Appendix~\ref{app:characters-GFV} we use character techniques that allow us to determine the spectrum of the GFV systematically -- including the intrinsic charge conjugation for self-conjugate representations -- without the need to construct the operators explicitly.
We refer the reader to Tables~\ref{tab:spectrum-SU2} and~\ref{tab:spectrum-SU3} and to the ancillary file \texttt{GFV-spectrum.nb} for a systematic collection of the lowest lying GFV operators for the cases of $SU(2)$, $SU(3)$ and $SU(4)$.

For illustration purposes, we report below the schematic form of the lowest-dimensional triple trace operators appearing in the GFV spectrum.
The first parity-even $[J^3]$ operators read
\begin{equation}
    \begin{aligned}
        (\Delta, \ell, p) &= (7, 0, +) & \quad \colon \quad & J_{\mu}^{a} J_{\nu}^{b} \partial^{\mu}J_{\nu}^{c} \, , \\
        (\Delta, \ell, p) & = (6, 1, +) & \quad \colon \quad &  J^{a}_{\mu} J^{b}_{\mu} J^{c}_{\rho} \, , \\
        (\Delta, \ell, p) &= (7, 2, +) & \quad \colon \quad &   J^{a}_{\mu} J^{b}_{\nu} \partial^{\mu} J^{c}_{\rho} \, , \\
        (\Delta,\ell,p) &= (3 + \ell, \ell, +) & \quad \colon \quad  & J^{a}_{\mu} J^{b}_{\nu} \partial_{\mu_{1}}\cdots \partial_{\mu_{\ell-3}} J^{c}_{\nu} \, , && \quad \ell\geq 3 \, .
    \end{aligned}
\end{equation}
While for the parity-odd sector, these are
\begin{equation}
    \begin{aligned}
        (\Delta, \ell, p) &= (6,0,-) & \quad \colon \quad & \epsilon^{\mu\nu\rho} J_{\mu}^{a} J_{\nu}^{b} J_{\rho}^{c} \, , \\
        (\Delta, \ell, p) & = (7, 1, -) & \quad \colon \quad &  J^{a}_{\mu} J^{b}_{\mu} \tilde{J}^{c}_{\rho} \, , \\
        (\Delta, \ell, p) &= (6, 2, -) & \quad \colon \quad & \epsilon^{\mu\nu\rho}J_{\mu}^{a} J_{\nu}^{b} J_{\sigma}^{c} \, , \\
        (\Delta,\ell,p) &= (4 + \ell,\ell,-) & \quad \colon \quad &  J^{a}_{\mu} J^{b}_{\nu} \partial_{\mu_{1}}\cdots \partial_{\mu_{\ell-3}} \tilde{J}^{b}_{\nu} \, , && \quad \ell \geq 3 \, .
    \end{aligned}
\end{equation}
These operators are generically in irreducible representations appearing in $\mathbf{adj}^{\otimes 3}$, which include both $\mathsf{R}_{s}$ and $\mathsf{R}_{a}$ representations in~\eqref{eq:SUNZ2-adj-decompose-sym-alt}.

\subsection{Perturbative results}\label{sec:spectrum-perturbative}

In this subsection, we collect the known results about anomalous dimensions of double-trace operators of the form $[JJ]$ in the GFV spectrum.
We define the anomalous dimensions $\vec{\gamma}$ as 
\begin{equation}
    \vec{\Delta}_{[JJ]_m} = 2 + \ell + m + \vec{\gamma}_{[JJ]_m} \, ,
\end{equation}
where $\vec{\gamma}$ takes into account the different irreducible representations which can appear. 
For $N\geq 4$, $\vec{\gamma} = (\gamma_{\mathbf{S}_{+}}, \gamma_{\mathbf{ adj}_{+}}, \gamma_{\mathbf{S\bar{S}}_{+}}, \gamma_{\mathbf{A\bar{A}}_{+}}, \gamma_{\mathbf{adj}_{-}},\gamma_{\mathbf{S\bar{A}}_{c}})$.\footnote{Recall from section~\ref{sec:spectrum-GFV} that not always all representations appear for given quantum numbers of $[JJ]$. In particular, for parity even scalars, we can only have the representations $\mathbf{S}_{+}$, $\mathbf{adj}_{+}$, $\mathbf{S\bar{S}}_{+}$, $\mathbf{A\bar{A}}_{+}$.}  
The two-point function of double trace operators with spin $\ell$ has the following schematic group-theoretical structure at NLO:\footnote{In terms of Witten diagrams in the bulk, NLO means two loop level.}
\begin{equation}\label{eq:gammaRed}
    \braket{[JJ]^{a_1b_1} [JJ]^{a_2 b_2}}\big\vert_{\text{1-loop}} \propto \left(T\pm (-1)^\ell U\right)^{a_1 b_1 a_2 b_2} \, ,
\end{equation}
where 
\begin{equation}\label{eq:gammaTU}
    T^{a_1 b_1 a_2 b_2} = f^{a_1a_2 e} f^{b_1b_2 e} \, , \qquad 
    U^{a_1 b_1 a_2 b_2} =  f^{a_1b_2 e} f^{b_1a_2 e} \, .
\end{equation}
Under the exchange $a_1\leftrightarrow b_1$ or $a_2\leftrightarrow b_2$, $U\leftrightarrow T$, and $T+U$ and $T-U$ are respectively symmetric and antisymmetric. 
Depending on the spin $\ell$, then, only symmetric $(R_s$) or antisymmetric $(R_a$) representations can acquire a non-trivial anomalous dimension at NLO. 
The group theoretical coefficients $\vec{v}$ of anomalous dimensions for operators with the same spacetime quantum numbers but in different representations of $SU(N)$ can be easily obtained by projecting~\eqref{eq:gammaRed} on the various representations using the results in Appendix~\ref{sec:projectors}.
We get
\begin{equation}
    \vec{v}_{s} = (2N,N,-2,2) \, , \qquad
    \vec{v}_{a} = (N,0) \, ,
\end{equation}
where $\vec{v} = \vec{v}_{s} \oplus \vec{v}_{a}$, and the subscripts $s,a$ refer respectively to the representations $R_s$ and $R_a$ in~\eqref{eq:SUNZ2-adj-decompose-sym-alt}. 
The same decomposition applies at leading order in a $1/\ell$ expansion~\cite{Li:2015itl}.

The anomalous dimensions for the first three Regge trajectories $m=0,1,2$ have been computed in~\cite[Eqs.\ 5.16--5.18]{Li:2015itl} at leading order in a $1/\ell$ expansion, using the light-cone analytic bootstrap~\cite{Fitzpatrick:2012yx,Komargodski:2012ek}.
The leading anomalous dimensions $\gamma_{[JJ]_m}$ for any $m$ and at finite $\ell > 1$ have been computed in~\cite[Eqs.\ 4.6--4.8]{Caron-Huot:2021kjy}, at leading order in a $1/C_J$  (i.e.\ $g^2\ped{YM}$) expansion, using the Lorentzian inversion formula~\cite{Caron-Huot:2017vep}. 
The decomposition into irreducible representations was not performed in~\cite{Caron-Huot:2021kjy}, and the results are subject to an ambiguity in the overall normalisation factor.\footnote{
    In addition to that,~\cite[Eq.\ 4.6]{Caron-Huot:2021kjy} appears to have a typo. 
    We thank Yue-Zhou Li for correspondence on this point.
} 
However, we can determine the proper factors by matching the results of~\cite{Caron-Huot:2021kjy} with the large $\ell$ ones of~\cite{Li:2015itl}, restricting both results to the special case of Yang-Mills theory in rigid AdS.
For completeness, we report here the final formulas:
\begin{equation}\label{eq:JJAnoDimFull}
    \resizebox{0.9\linewidth}{!}{\(
    \begin{aligned}
        \vec{\gamma}_{[JJ]_0} & = -\frac{g^2\ped{YM}}{4\pi^2}\vec{P}_\eta  \left(\frac{\beta^4-4\beta^3+28\beta^2-48\beta+32}{(\beta-4)(\beta-2)(\beta-1)\beta(\beta+2)}\right)  \,, \\[0.3em]
        \vec{\gamma}_{[JJ]_{2n}}^{(1)} & = \frac{g^2\ped{YM}}{2\pi^2} \vec{P}_\eta \left(\psi\left(\frac{\beta}{2}-n-2\right)-\psi\left(\frac{\beta}{2}+n\right)+\frac{4}{\beta-2n-2}-\frac{4}{(\beta-2n)(\beta-2n+2)}\right)\,, \\[0.3em]
        \vec{\gamma}_{[JJ]_{2n}}^{(2)} & = \frac{g^2\ped{YM}}{2\pi^2} \vec{P}_\eta \left(\psi\left(\frac{\beta}{2}-n\right)-\psi\left(\frac{\beta}{2}+n+2\right)+\frac{4}{\beta+2n}+\frac{4}{(\beta+2n-4)(\beta+2n-2)}\right)\,, \\[0.3em]
        \vec{\gamma}_{[JJ]_{2n-1}}^{(1)} & = \frac{g^2\ped{YM}}{2\pi^2} \vec{Q}_\eta \left(\psi\left(\frac{\beta+1}{2}-n\right)-\psi\left(\frac{\beta+1}{2}+n\right)-\frac{8}{(\beta-2n-1)(\beta-2n+1)}\right)\,, \\[0.3em]
        \vec{\gamma}_{[JJ]_{2n-1}}^{(2)} & = \frac{g^2\ped{YM}}{2\pi^2} \vec{P}_\eta \left(\psi\left(\frac{\beta+1}{2}-n\right)-\psi\left(\frac{\beta+1}{2}+n\right)+\frac{8}{(\beta+2n-1)(\beta+2n+1)}\right)\,,
    \end{aligned}
    \)}
\end{equation}
where $\psi(x)$ is the digamma function, $\beta= \Delta+\ell$, $\ell >1$, 
\begin{equation}
    \vec{P}_\eta = \vec{v}_\eta \frac{1+ c_{\eta} (-1)^\ell}{2} \, , \qquad 
    \vec{Q}_\eta = \vec{v}_\eta \frac{1- c_{\eta} (-1)^\ell}{2} \, , \qquad
    c_{s,a} = \pm 1 \, , \qquad
    \eta = s, a \, ,
\end{equation}
and the superscripts $(1)$ and $(2)$ in $\gamma_{[JJ]_m}$ refer to the two degenerate operators which appear for $n>0$.
Note that for parity-odd operators, all representations appear for any given spin, in agreement with the discussion of the previous subsection.

More recently, the anomalous dimensions for scalars with $m=2$ have been determined by Witten diagrams and by a broken conformal Ward identity~\cite{Ciccone:2024guw}:
\begin{equation}\label{eq:perturbative-scalars}
    \vec{\gamma}_{[JJ]_2}(\ell = 0) = - \frac{11 g^2\ped{YM}}{48\pi^2} \vec{v}_{s}\,.
\end{equation}
The anomalous dimensions of all the double trace operators appearing in~\eqref{eq:JJAnoDimFull} are negative, with the exception of $\gamma_{\mathbf{S\bar{S}}_{+}}$, which is positive, and of $\gamma_{\mathbf{S\bar{A}}_{c}}$, which is vanishing at leading order, both in a $1/\ell$ and in a $g^2\ped{YM}$ expansion. 
The charged operator with the largest (negative) one-loop anomalous dimension is in the adjoint representation.
We will focus our bootstrap analysis on the lightest scalar operators in the singlet, adjoint and $\mathbf{S\bar{S}}$ representations. For the $SU(3)$ case, these are perturbatively the lightest double-trace scalars, among which the singlet has the most negative $\gamma$.
For the $SU(2)$ case, we recall that perturbatively the lightest adjoint scalar is not a double-trace but rather of $[J^3]$ type with $\Delta = 7 + O(g\ped{YM}^2)$.

We also recall that the current central charge at LO in perturbation theory is~\cite{Freedman:1998tz}
\begin{equation}\label{eq:CJvsgYM}
    C_J = \frac{C_J^0}{g\ped{YM}^2} \Big(1+O(g\ped{YM}^2) \Big)\,, \qquad C_J^0 = \frac{2}{\pi^2}\,,
\end{equation}
which shows explicitly that a large central charge corresponds to weak coupling.

\section{Bootstrap setup}\label{sec:bootstrap-setup}
In this section, we review the conformal bootstrap setup of conserved currents developed in~\cite{Dymarsky:2017xzb,Reehorst:2019pzi} and extended to non-Abelian currents in~\cite{He:2023ewx}. 
In particular, we will consider a four-point function of non-Abelian conserved spin 1 currents \( J^{a}_{\mu} \) in \( 3d \), transforming in the adjoint representation of \( SU(N) \). 
The four-point function is decomposed in terms of conformal tensor structures \( \mathbb{T}_{I} \) and global symmetry tensor structures \( T_{\mathbf{r}} \):
\begin{equation}\label{eq:JJJJ-gfunctions}
    \braket{J^{a}(P_{1},Z_{1}) J^{b}(P_{2}, Z_{2}) J^{c}(P_{3},Z_{3}) J^{d}(P_{4}, Z_{3})} = \sum_{I} \sum_{\mathbf{r}} \mathbb{T}_{I}(P_{i}, Z_{i}) \, T_{\mathbf{r}}^{abcd} \, g_{I}^{\mathbf{r}}(z,\bar{z}) \, ,
\end{equation}
while the ``dynamical'' content is encoded in the \( g \)-function. 
The coordinates $(P_{i}, Z_{i})$ are those of the embedding formalism of~\cite{Costa:2011mg}. 
The sum over $I$ spans the conformal tensor structures. 
We refer to Appendix B of~\cite{He:2023ewx} for the discussion on the conformal tensor structures and the choice of linearly independent \( g \)-functions. 
The sum over $\mathbf{r}$ runs over the global symmetry singlets, whose number is $n_{4c}$, in which the correlator can be decomposed. 
Determining $n_{4c}$ is straightforward if we consider $\mathbf{adj}^{\otimes 4} = (\mathbf{adj}^{\otimes 2})\otimes(\mathbf{adj}^{\otimes 2})$ and use the decomposition~\eqref{eq:SUN-adj-decompose}, recalling that singlets can only arise in tensor products between one representation and its complex conjugate. 

In~\cite{He:2023ewx}, the factorisation of conformal and global symmetry of~\eqref{eq:YM-AdS-symmetry} was explicitly assumed, together with invariance under parity. 
Invariance under charge conjugation was not stated explicitly, but implied by~\cite[Eq.\ C.2]{He:2023ewx}, when assuming certain symmetry properties of the global symmetry tensor structures \( T_{\mathbf{r}}^{abcd} \).\footnote{
    We are grateful to Petr Kravchuk and Marten Reehorst for bringing this to our attention and for correspondence. 
    The bootstrap equations for non-Abelian conserved currents without assuming charge conjugation invariance (in $4d$) will appear in~\cite{currents4d}.
} 
At the level of $SU(N)$-invariant tensors, i.e. without imposing charge conjugation symmetry, for $N\geq 4$ we have 1 singlet each from the $\mathbf{S}, \mathbf{S\bar{S}}, \mathbf{A\bar{A}}$; 4 singlets from $(2 \, \mathbf{adj}) \otimes (2 \, \mathbf{adj})$; 2 singlets from $\mathbf{S\bar{A}}\otimes \mathbf{A\bar{S}}$ and $\mathbf{A\bar{S}}\otimes\mathbf{S\bar{A}}$, for a total of $n_{4c} = 3\times 1 + 4 + 2 = 9$.  
This number is reduced to $n_{4c}=8$ for $N=3$, and $n_{4c}=3$ for $N=2$. Independently of $N$, the two adjoint representations $\mathbf{adj}^{\otimes 2}$ have always opposite charge conjugation parity, while $\mathbf{S\bar{A}}$ and $\mathbf{\bar{S}A}$ combine into a unique irreducible representation of $PSU(N) \rtimes \Z_{2}^{\mathcal{C}}$. 
So, under charge conjugation, the 9 singlets for $N\geq 4$ decompose as $n_{4c} = n_{4c}^++n_{4c}^-$, with $n_{4c}^+ = 6$ and $n_{4c}^- = 3$.\footnote{
    For $N=3$ we have $n_{4c} = n_{4c}^++n_{4c}^-$, with $n_{4c}^+=5$ and $n_{4c}^-=3$. 
    This counting can be checked using the characters in Appendix~\ref{app:characters-charge-conj}.
} 
Imposing charge conjugation symmetry reduces the number of singlets from $n_{4c}$ to $n_{4c}^+$, and their associated tensor structures $T_{\mathbf{r}}^{abcd}$ in~\eqref{eq:JJJJ-gfunctions} do indeed satisfy~\cite[Eq.\ C.2]{He:2023ewx}.
For the $N = 2$ case, this issue does not arise since $\mathbf{adj}^{\otimes 2}$ decomposes only into self-conjugate and degeneracy-free representations.
We use symmetry tensor structures 
\begin{equation}\label{eq:Trabcd}
    T_{\mathbf{r}}^{abcd} = \frac{1}{\sqrt{\dim \mathbf{r}}} P^{abcd}_{\mathbf{r}}\,,
\end{equation}
where $\mathbf{r}$ runs over the representation labels in~\eqref{eq:SUNZ2-adj-decompose} and $P^{abcd}_{\mathbf{r}}$ are the corresponding projectors, whose explicit form is given in Appendix~\ref{sec:projectors}.

Coming back to~\eqref{eq:JJJJ-gfunctions}, each \( g \)-function can be decomposed into conformal blocks as
\begin{equation}
    g_{I}^{\mathbf{r}}(z,\bar{z}) = \sum_{\mathcal{O}_{\Delta, \ell, p, \mathbf{r}}} \lambda_{JJ\mathcal{O}}^{(a)} \lambda_{JJ\mathcal{O}}^{(b)} \, g^{(I, ab)}_{\Delta,\ell}(z,\bar{z}) \,,
\end{equation}
where \( a,b \) run over the basis of three-point tensor structures and operators exchanged are labelled by spacetime parity \( p = \pm 1 \), spin \( \ell \), conformal dimension \( \Delta \) and irreducible representation \( \mathbf{r} \) of \( PSU(N) \rtimes \Z_{2}^{\mathcal{C}} \). 
In the above decomposition, it is assumed that operators (but the current) have a unit-normalised two-point function \( \braket{\mathcal{O} \mathcal{O}} \).

The contribution of the identity and of the current itself to the block decomposition requires a separate discussion.
In embedding coordinates, we have in $d$ dimensions (see e.g.~\cite{Li:2015itl})
\begin{equation}\label{eq:CJ-definition}
    \begin{aligned}
        &\braket{J^{a}(Z_{1},P_{1}) J^{b}(Z_{2},P_{2})} = C_{J} \delta^{ab}\frac{H_{12}}{P_{12}^{d}} \, , \\[0.3em]
        &\braket{J^{a}(Z_{1},P_{1}) J^{b}(Z_{2},P_{2})J^{c}(Z_{2}, P_{3})} = \frac{C_{J}}{S_{d}} f^{abc} \frac{1}{P_{12}^{d/2} P_{23}^{d/2}P_{13}^{d/2}} \\
        &\hspace*{8em} \times \Big((d-(d+2)\alpha_{JJJ})V_{1}V_{2}V_{3} - \alpha_{JJJ}(V_{1}H_{23} + V_{2}H_{13} + V_{3} H_{23})\Big) \, ,
    \end{aligned}
\end{equation}
where $C_{J}$ is the current central charge, $S_{d} = 2\pi^{d/2}/\Gamma(d/2)$. 
The norm of the two-dimensional OPE vector $(\lambda_{JJJ}^{(1)},\lambda_{JJJ}^{(2)})$ is fixed by Ward identities to be proportional to $C_{J}/S_{d}$, while $\alpha_{JJJ}$ is a parametrization for the remaining degrees of freedom.
For reference, in \( d = 3 \), we have \( \alpha_{JJJ} = 1/2 \) for an \( SU(N) \) conserved current of \( N \) free bosons in the fundamental representation of \( SU(N) \) and \( \alpha_{JJJ} = 1 \) for free fermions~\cite{Osborn:1993cr}; for YM in $\mathrm{AdS}_{4}$ at tree level we have \( \alpha_{JJJ} = 3/4 + O(g\ped{YM}^{2}) \), see Appendix A of~\cite{Caron-Huot:2021kjy}.

The contributions of the identity and the current to the block decomposition are then
\begin{equation}\label{eq:identity-current-block-contribution}
    \begin{aligned}
        g^{\mathbf{S}_{+}}_{I}(z,\bar{z}) &= C_{J}^{2} \, g_{0,0}^{(I)}(z,\bar{z}) + \cdots \\
        g^{\mathbf{r}_{J}}_{I}(z,\bar{z}) &= C_{J} \, \theta^{(a)} \theta^{(b)} \, g_{2,1}^{(I,ab)}(z,\bar{z}) + \cdots 
    \end{aligned}
\end{equation}
where $\mathbf{r}_{J} = \mathbf{adj}$ for $N=2$, $\mathbf{r}_{J} = \mathbf{adj}_{-}$ for $N \geq 3$ odd and $\mathbf{r}_{J} = \mathbf{adj}_{+}$ for $N \geq 3$ even, and where we have defined the OPE vector
\begin{equation}
\label{lambdaJJJ}
    \theta = \begin{pmatrix}
        \theta^{(1)}\\
        \theta^{(2)}
    \end{pmatrix} =
    \frac{1}{4\pi}
    \begin{pmatrix}
        5-3\alpha_{JJJ}\\
        -\alpha_{JJJ}
    \end{pmatrix} \, .
\end{equation}

The symmetry properties of three-point functions 
\begin{equation}
    \braket{J^{a}(P_{1},Z_{1}) J^{b}(P_{2},Z_{2}) O^{A}_{\Delta, \ell, p, \mathbf{r}}(P_{3},Z_{3})} = \lambda_{JJO}^{(i)} \, \mathbb{K}^{(i)}_{\ell,p} \, C_{\mathbf{r}}^{ab,A}
\end{equation}
with respect to permutations of their operators, impose selection rules on the spin of operators exchanged in the OPE, which depend on the spacetime parity \( p \).
These are determined by the symmetry properties under operator exchange of the tensor structure \( \mathbb{K}_{\ell,p} \), and those of the tensor structure \( C_{\mathbf{r}} \) (i.e.\ under $a \leftrightarrow b$). 
The latter is related to the exchange symmetry of the four-point tensor structure $T^{abcd}_{\mathbf{r}}$, being proportional to \( C_{\mathbf{r}}^{ab,A}C_{\mathbf{r}}^{cd,B}\delta_{AB} \). 
Moreover, there exist combinations of \( (\ell,p,\mathbf{r}) \) for which there are no compatible conformal symmetry tensor structures. 
Finally, further restrictions on OPE coefficients are imposed by the conservation of the non-Abelian current. 
The selection rules were worked out in~\cite{He:2023ewx} and are summarised in Table~\ref{tab:selection-rules}.

\begin{table}[t]
    \centering
    \setlength{\tabcolsep}{1.5em}
    \begin{tabular}{ccc}
        \toprule
        parity & rep & spin \\ \midrule
        \( + \) & $\mathsf{R}_{s}$ & \( \ell \) even\\
        \( + \) & $\mathsf{R}_{a}$ & \( \ell \) odd \\
        \( - \) & $\mathsf{R}_{s}$ & \( \ell \neq 1 \) \\
        \( - \) & $\mathsf{R}_{a}$ & \( \ell \geq 1 \) \\ 
        \bottomrule
    \end{tabular}
    \caption{Spin selection rules in $\braket{JJJJ}$ depending on parity and symmetry property under exchange.\label{tab:selection-rules}}
\end{table}

The choice of crossing conditions is described in~\cite[App.\ B]{He:2023ewx} and consists of 4 bulk constraints and 1 point constraint, each of which is a system of constraints labelled by \( \mathbf{r} \). 
The crossing equations are
\begin{equation}
    \begin{aligned}
        \mathcal{M}^{\mathbf{r}}_{\mathbf{s}} \, g^{\mathbf{s}}_{[0,1,1,0]}(z,\bar{z}) &= g_{[1,1,0,0]}^{\mathbf{r}}(1-z,1-\bar{z}) \, , \\
        \mathcal{M}^{\mathbf{r}}_{\mathbf{s}} g^{\mathbf{s}}_{[0,0,0,0]}(z,\bar{z}) &= g_{[0,0,0,0]}^{\mathbf{r}}(1-z,1-\bar{z}) \, , \\
        \mathcal{M}^{\mathbf{r}}_{\mathbf{s}} \, g^{\mathbf{s}}_{[1,0,1,0]}(z,\bar{z}) &= g_{[1,0,1,0]}^{\mathbf{r}}(1-z,1-\bar{z}) \, , \\
        \mathcal{M}^{\mathbf{r}}_{\mathbf{s}} \, g^{\mathbf{s}}_{[1,1,1,1]}(z,\bar{z}) &= g_{[1,1,1,1]}^{\mathbf{r}}(1-z,1-\bar{z}) \, , \\
        \mathcal{M}^{\mathbf{r}}_{\mathbf{s}} \, g^{\mathbf{s}}_{[-1,-1,1,1]}(1/2,1/2) &= g_{[1,-1,-1,1]}^{\mathbf{r}}(1/2,1/2) \, ,
    \end{aligned}
\end{equation}
where $I = [h_1, h_2, h_3, h_4]$ are convenient labels for the conformal tensor structures (see~\cite{Kravchuk:2016qvl}), and where \( \mathcal{M} \) is the crossing matrix (see~\cite[Eq.\ C.4--C.5]{He:2023ewx} for explicit expression)\footnote{
    The crossing matrix for \( SU(2) \) is the same as the crossing matrix for \( SO(3) \).
}
\begin{equation}
    \mathcal{M}^{\mathbf{r}}_{\mathbf{s}} = \frac{\sum_{a,b,c,d} T_{\mathbf{r}}^{cbad} T_{\mathbf{s}}^{abcd}}{\sum_{a,b,c,d} T_{\mathbf{r}}^{cbad} T_{\mathbf{r}}^{cbad}} \, .
\end{equation}
Each equation is then differentiated with respect to \( (z,\bar{z}) \) at \( z=\bar{z}=1/2 \) obtaining
\begin{align}
    \mathcal{M}^{\mathbf{r}}_{\mathbf{s}} \, \partial_{z}^{m}\partial_{\bar{z}}^{n}g^{\mathbf{s}}_{[0,1,1,0]}(1/2,1/2) - (-1)^{n+m}\partial_{z}^{m}\partial_{\bar{z}}^{n} g_{[1,1,0,0]}^{\mathbf{r}}(1/2,1/2) &= 0 \, , \label{eq:crossing-constraints-derivatives-1} \\
    \left(\mathcal{M}^{\mathbf{r}}_{\mathbf{s}} - (-1)^{n+m}\delta^{\mathbf{r}}_{\mathbf{s}}\right) \partial_{z}^{m}\partial_{\bar{z}}^{n}g^{\mathbf{s}}_{[0,0,0,0]}(1/2,1/2) &= 0 \, , \label{eq:crossing-constraints-derivatives-2} \\
    \left(\mathcal{M}^{\mathbf{r}}_{\mathbf{s}} - (-1)^{n+m}\delta^{\mathbf{r}}_{\mathbf{s}}\right) \partial_{z}^{m}\partial_{\bar{z}}^{n}g^{\mathbf{s}}_{[1,0,1,0]}(1/2,1/2) &= 0 \, , \label{eq:crossing-constraints-derivatives-3} \\
    \left(\mathcal{M}^{\mathbf{r}}_{\mathbf{s}} - (-1)^{n+m}\delta^{\mathbf{r}}_{\mathbf{s}}\right) \partial_{z}^{m}\partial_{\bar{z}}^{n}g^{\mathbf{s}}_{[1,1,1,1]}(1/2,1/2) &= 0 \, , \label{eq:crossing-constraints-derivatives-4} \\
    \mathcal{M}^{\mathbf{r}}_{\mathbf{s}} \, g^{\mathbf{s}}_{[0,1,1,0]}(1/2,1/2) -g_{[1,-1,-1,1]}^{\mathbf{r}}(1/2,1/2) &= 0 \, . \label{eq:crossing-constraints-derivatives-5}
\end{align}
Note that the matrix \( \mathcal{M}^{\mathbf{r}}_{\mathbf{s}} - (-1)^{n+m}\delta^{\mathbf{r}}_{\mathbf{s}} \) does not have maximal rank. 
In particular, when \( n+m \) is even, the rank is equal to the number of anti-symmetric irreps \( r_{a} \) and when \( n+m \) is odd, the rank is equal to the number of symmetric irreps \( r_{s} \), see~\eqref{eq:SUNZ2-adj-decompose-sym-alt} and~\eqref{eq:counting-parameters} below.

All the crossing constraints can be put together into the following sum rule
\begin{equation}\label{eq:crossing-vectors}
    \sum_{\mathcal{O}_{\Delta, \ell, p, \mathbf{r}}} \lambda_{JJ\mathcal{O}}^{\mathsf{T}} \vec{\mathcal{V}}_{\Delta, \ell, p, \mathbf{r}} \lambda_{JJ\mathcal{O}} = 0 \,,
\end{equation}
where the ``crossing vector'' \( \vec{\mathcal{V}} \) is a vector of matrices built from derivatives of conformal blocks. 
Each component is a matrix (possibly \( 1 \times 1  \)) contracted with a vector of OPE coefficients \( \lambda_{JJ\mathcal{O}} = (\lambda_{JJ\mathcal{O}}^{(I)}) \). The length of these crossing vectors corresponds to the number of non-trivial equations included in~\eqref{eq:crossing-constraints-derivatives-1} through~\eqref{eq:crossing-constraints-derivatives-5}, times the number of derivatives included (\( n + m \leq \Lambda \), \( n \leq m \)). 
In particular
\begin{itemize}
    \item~\eqref{eq:crossing-constraints-derivatives-1} gives \( (r_{s} + r_{a}) \times (d\ped{even} + d\ped{odd}) \) equations
    \item~\eqref{eq:crossing-constraints-derivatives-2},~\eqref{eq:crossing-constraints-derivatives-3} and~\eqref{eq:crossing-constraints-derivatives-4} give each \( r_{s}\times d\ped{odd} + r_{a}\times d\ped{even} \) equations
    \item~\eqref{eq:crossing-constraints-derivatives-5} gives \( r_{s} + r_{a} \) equations
\end{itemize}
where
\begin{equation}\label{eq:counting-parameters}
    \begin{gathered}
        r_{s} =
        \begin{cases}
            2 & N=2 \\
            3 & N \geq 3
        \end{cases}
        \quad\,,\qquad\qquad\qquad
        r_{a} =
        \begin{cases}
            1 & N=2 \\
            2 & N=3 \\
            3 & N \geq 4
        \end{cases}\quad\,,
        \\[0.5em]
        d\ped{even} = \frac{1}{2}\left(\floor{\frac{\Lambda}{2}}+1\right) \left(\floor{\frac{\Lambda}{2}}+2\right) \,, \qquad
        d\ped{odd} = \frac{1}{2} \floor{\frac{\Lambda+1}{2}}\left(\floor{\frac{\Lambda+1}{2}}+1\right) \, .
    \end{gathered}
\end{equation}
Thus, we find a total of \( (d\ped{even}+4d\ped{odd}+1)r_{s} + (4d\ped{even}+d\ped{odd}+1)r_{a} \) components.

In~\eqref{eq:crossing-vectors}, two particular contributions will play a role in our bootstrap bounds: the identity \( \id \) and the current itself \( J \). 
The contribution of these two particular operators to the block decomposition is given in~\eqref{eq:identity-current-block-contribution}. 
We single out these terms in the sum rule~\eqref{eq:crossing-vectors} and divide by the identity OPE coefficient to obtain
\begin{equation}\label{eq:crossing-vectors-id-J}
    \vec{\mathcal{V}}_{\id} + \frac{1}{C_{J}} \theta^{\mathsf{T}} \vec{\mathcal{V}}_{J} \theta + \sum_{\mathcal{O}_{\Delta, \ell, p, \mathbf{r}}} \frac{1}{C_{J}^{2}}  \lambda_{JJ\mathcal{O}}^{\mathsf{T}} \vec{\mathcal{V}}_{\Delta, \ell, p, \mathbf{r}} \lambda_{JJ\mathcal{O}} = 0.
\end{equation}

As usual in the numerical conformal bootstrap, each component of the crossing vector can be approximated arbitrarily well by rational functions in the scaling dimension \( \Delta \). 
More precisely,
\begin{equation}
    \vec{\mathcal{V}}_{\Delta, \ell, p, \mathbf{r}} \simeq \chi_{\ell}(\Delta) \vec{M}_{\ell, p, \mathbf{r}}(\Delta) \, ,
\end{equation}
where \( \chi_{\ell}(\Delta) =(4 r_{*})^{\Delta}/\prod_{i}(\Delta - p_{i}) \geq 0 \) for \( \Delta \) above the unitarity bound, with $r_{*} = 3 - 2\sqrt{2} $ is the modulus of the radial coordinate~\cite{Hogervorst:2013sma} at the crossing symmetric point, and the \( \vec{M}_{\ell, p, \mathbf{r}}(\Delta) \) are vectors of polynomial matrices in \( \Delta \).

In the Semi-Definite Programs (SDPs) that we will formulate, the action of functionals on \( \vec{\mathcal{V}} \) translates into acting on the vector of polynomial matrices \( \vec{M} \). 
Since \( \chi_{\ell} \geq 0 \) in unitary theories, the conditions of positivity on \( \vec{\mathcal{V}} \) are equivalent to positivity conditions on \( \vec{M} \).

\paragraph{Notation} 
We denote as \( s \), \( a \) and \( s\bar{s} \) the lowest dimensional parity-even scalars \( (\ell,p) = (0,+) \) that appear in the \( \mathbf{S}_{+} \), \( \mathbf{adj}_{+} \) and \( \mathbf{S\bar{S}}_{+} \) representations respectively. 
We use primes to indicate sub-leading operators in a given selection sector, determined by spin, spacetime parity and global symmetry representation. As an exception, we denote by \( T_{\mu\nu}' \) the first parity-even, singlet, \( \ell = 2 \) operator (recall that the stress tensor is not present in the boundary CFT of a QFT in rigid AdS). 
The unitarity bound is denoted by
\begin{equation}
    \Delta\ped{UB}(\ell) =
    \begin{dcases}
        \frac{d-2}{2} &\qquad \ell = 0\,, \\
        \ell + d - 2 & \qquad\ell \geq 1\,,
    \end{dcases}
    \qquad \qquad (d=3) \, .
\end{equation}

\section{Bounds on \texorpdfstring{\( C_{J} \)}{CJ}}\label{sec:bootstrap-CJ}

In order to address the decoupling scenario discussed earlier, it will be useful to set up an SDP suited for OPE coefficient extremisation. 
This is because if we can bound $C_{J}$ to be strictly above zero, the two-point function of $J$ will be non-zero for any choice of the AdS radius $L$, ruling out the decoupling scenario. 
In particular, we look for the optimal functional $\alpha$ satisfying:
\begin{equation}\label{eq:SDP-CJbound-SU3}
    \begin{aligned}
        \text{objective} \colon \quad
        & \alpha(\vec{\mathcal{V}}_{\id}) \\
        \text{normalisation} \colon  \quad
        & \alpha(\theta^{\mathsf{T}}\vec{\mathcal{V}}_{J} \theta) \\
        \text{conditions} \colon \quad
        & \alpha(\vec{\mathcal{V}}_{\Delta, \ell, +, \mathbf{r}}) \succeq 0 \qquad && \Delta \geq  \Delta\ped{UB}(\ell) && \mathbf{r} \in\mathsf{R}_{s} && \ell = 0, 2, 4,  \ldots  \\
        & \alpha(\vec{\mathcal{V}}_{\Delta, \ell, +, \mathbf{r}}) \succeq 0 \qquad && \Delta \geq  \Delta\ped{UB}(\ell) && \mathbf{r} \in \mathsf{R}_{a} && \ell = 1,3,5, \ldots  \\
        & \alpha(\vec{\mathcal{V}}_{\Delta, \ell, -, \mathbf{r}}) \succeq 0 \qquad && \Delta \geq  \Delta\ped{UB}(\ell) && \mathbf{r} \in\mathsf{R}_{s} && \ell = 0,2,3 \ldots  \\
        & \alpha(\vec{\mathcal{V}}_{\Delta, \ell, -, \mathbf{r}}) \succeq 0 \qquad && \Delta \geq  \Delta\ped{UB}(\ell) && \mathbf{r} \in \mathsf{R}_{a} && \ell = 1,2,3 \ldots
    \end{aligned}
\end{equation}

This SDP is parametric in \( \theta \), namely in \( \alpha_{JJJ} \), see~\eqref{lambdaJJJ}. 
The optimal solution minimises the objective. 
Let \( P = \max \alpha(\vec{\mathcal{V}}_{\id})\) be the optimal objective for a given value of \( \alpha_{JJJ} \). 
Acting with a functional satisfying the SDP constraints on the crossing equation~\eqref{eq:crossing-vectors-id-J}, we have
\begin{equation}
    \frac{1}{C_{J}} = - \alpha(\mathcal{V}_{\id}) - \sum_{\mathcal{O}_{\Delta, \ell, p, \mathbf{r}}} \frac{1}{C_{J}^{2}}  \lambda_{JJ\mathcal{O}}^{\mathsf{T}} \alpha(\vec{\mathcal{V}}_{\Delta, \ell, p, \mathbf{r}}) \lambda_{JJ\mathcal{O}} \leq  - \alpha(\mathcal{V}_{\id}) \,,
\end{equation}
and thus
\begin{equation}
    C_{J} \geq \max\left(-\frac{1}{\alpha(\vec{\mathcal{V}}_{\id})}\right) = - \frac{1}{P} \, .
\end{equation}
We recall that \( \alpha(\vec{\mathcal{V}}_{\id}) < 0 \) is required for the SDP~\eqref{eq:SDP-CJbound-SU3} to be feasible.

Hence, for each value of \( \alpha_{JJJ} \) we obtain a lower bound on \( C_{J} \). 
A priori, the problem could be unbounded and there could exist values of \( \alpha_{JJJ} \) for which \( P \to -\infty \).
This happens in our setup for all \( \alpha_{JJJ} \) at low derivative order, e.g.\ at \( \Lambda = 4 \), and is known to generically happen when bounding OPE coefficients from the bootstrap of scalar four-point functions for large values of the external operator dimension.
Fortunately, the problem is bounded in this case, and we establish a non-trivial rigorous lower bound on the current central charge \( C_{J} \).
This is analogous to the lower bounds on the stress tensor central charge $C_{T}$ that have been obtained from the bootstrap of four stress tensors~\cite{Dymarsky:2017yzx} or four abelian conserved currents~\cite{Dymarsky:2017xzb}.

We normalise our numerical bounds by the current central charge in a theory of a Free Boson (FB) in the fundamental of $SU(N)$, denoted by $C_{J}\ap{FB}$ (see Appendix~\ref{app:CJ-normalisation} for further details).
In the plots of the lower bounds as a function of $\alpha_{JJJ}$, we add for reference blue dots corresponding to this theory, $(\alpha_{JJJ}, C_{J}/C_{J}\ap{FB}) = (1/2, 1)$, and to the theory of a free fermion in the fundamental of $SU(N)$, $(C_{J}\ap{FF}/C_{J}\ap{FB}, \alpha_{JJJ}) = (1, 2\sqrt{2})$~\cite{Osborn:1993cr}.

\begin{figure}[p!]
    \centering
    \includegraphics[width=0.7\linewidth]{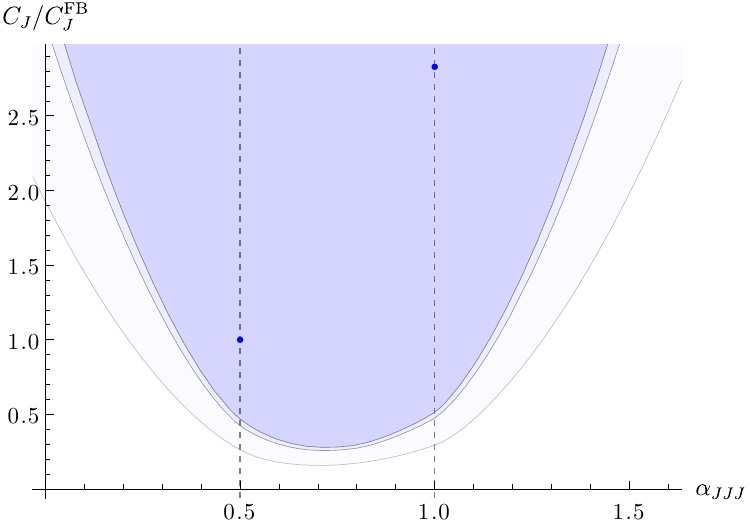}
    \caption{Lower bound on \( C_{J} \) in the case of \( SU(2) \) as a function of \( \alpha_{JJJ} \). The shaded area is allowed, darker colour corresponding to higher $\Lambda$, we have \( \Lambda = 11,19\) and $23$ respectively. The dashed vertical lines and the blue dots at \( \alpha_{JJJ} = 1/2, 1 \) denote the free boson and free fermion theory, respectively. We see that $C_J=0$ is excluded for any choice of $\alpha_{JJJ}$.
    \label{fig:SU2_CJ-bound_pd-11-19-23}}
\end{figure}

\begin{figure}[p!]
    \centering
    \includegraphics[width=0.7\linewidth]{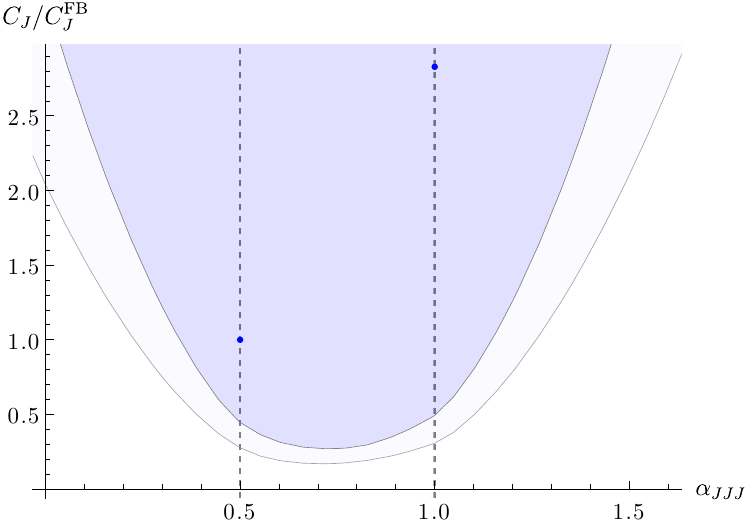}
    \caption{Lower bound on \( C_{J} \) in the case of \( SU(3) \) as a function of \( \alpha_{JJJ} \). The shaded area is allowed, the darker area corresponding to $\Lambda=19$ (and the rest to $\Lambda=11$). The dashed vertical lines and the blue dots at \( \alpha_{JJJ} = 1/2, 1 \) denote the free boson and free fermion theory, respectively. As in the $SU(2)$ case, $C_J=0$ is excluded for any choice of $\alpha_{JJJ}$.\label{fig:SU3_CJ-bound_pd-11-19}}
\end{figure}

\begin{figure}[t!]
    \centering
    \includegraphics[width=0.6\linewidth]{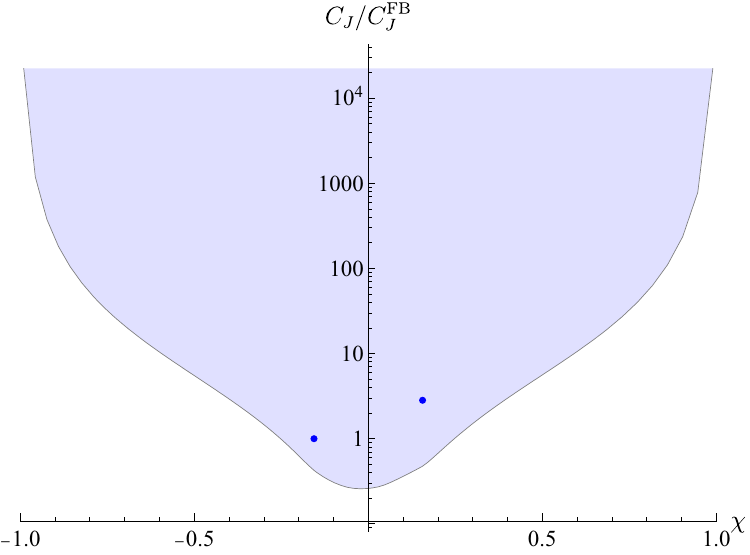}
    \caption{Lower bound on \( C_{J} \) in the case of \( SU(2) \) as a function of a compact parametrization of the OPE ratio $\chi\in[-1,1]$. The shaded area is allowed and corresponds to the $\Lambda=19$ computation of Figure~\ref{fig:SU2_CJ-bound_pd-11-19-23}. The vertical axis is on a logarithmic scale.\label{fig:SU2_CJ-bound_compact}}
\end{figure}

The  \( SU(2) \)  bound is reported in Figure~\ref{fig:SU2_CJ-bound_pd-11-19-23} with \( \Lambda = 11, 19, 23 \), and for \( SU(3) \) in Figure~\ref{fig:SU3_CJ-bound_pd-11-19}, with \( \Lambda = 11, 19 \). 
As no rigorous analytic bound is known on the variable $\alpha_{JJJ}$,\footnote{
    This is to contrast with the parameters of the OPE ratios entering $\braket{TTT}$ and $\braket{JJT}$, which are known to satisfy the ``conformal collider bounds''~\cite{Hofman:2008ar,Hofman:2016awc}. 
    It is known that $1/2 \leq \alpha_{JJJ} \leq 1$ if and only if the binding energy from non-Abelian gauge interactions for well-separated neutral 2-particle states is attractive, see~\cite[Eq.\ 4.23]{Li:2015itl}, where in our conventions $\lambda_{JJJ} = \frac{C_{J}}{4\pi} \alpha_{JJJ}$.
}
we also present in Figure~\ref{fig:SU2_CJ-bound_compact} the same $\Lambda = 23$ bound in terms of the compact variable $\chi = \frac{2}{\pi} \arctan\left(\alpha_{JJJ} - \frac{3}{4}\right)$, to show that the bound monotonically increases as $\alpha_{JJJ} \to \pm \infty$, i.e.\ $\chi \to \pm 1$.

Note that for \( SU(2) \) the SDP is numerically cheaper than the one in~\eqref{eq:SDP-CJbound-SU3}, since fewer representations are exchanged in the tensor product of two adjoints~\eqref{eq:SU2-adj-decompose-sym-alt}, resulting in fewer sum rules obtained from a given crossing equation.

\begin{figure}[t!]
    \centering
    \begin{subfigure}[c]{0.7\linewidth}
        \includegraphics[width=\linewidth]{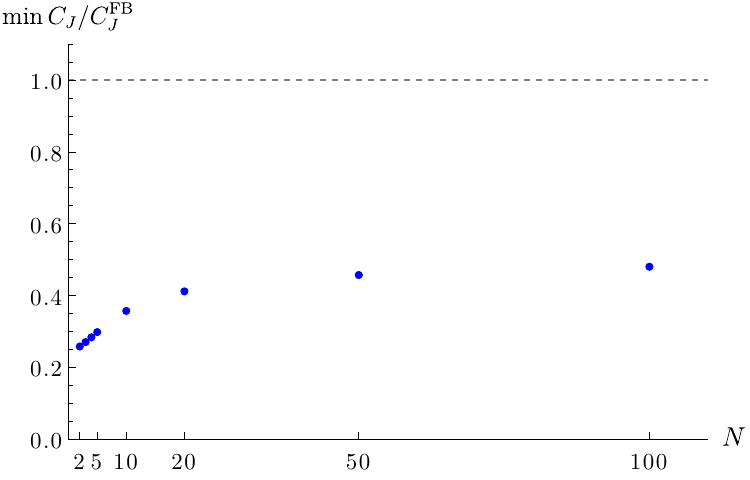}
    \end{subfigure}
    \hspace*{1em}
    \begin{subfigure}[c]{0.2\linewidth}
        \begin{tabular}{cl}
            \toprule
            $N$ & $\alpha_{JJJ}^{*}$ \\ \midrule
            \( 2 \) & \( 0.717 \) \\
            \( 3 \) & \( 0.727 \) \\
            \( 4 \) & \( 0.736 \) \\
            \( 5 \) & \( 0.743 \) \\
            \( 10 \) & \( 0.744 \) \\
            \( 20 \) & \( 0.731 \) \\
            \( 50 \) & \( 0.697 \) \\
            \( 100 \) & \( 0.673 \) \\
            \bottomrule
    \end{tabular}
    \end{subfigure}
    \caption{Absolute lower bound of $C_J$ as a function of $N$ for $SU(N)$ non-Abelian conserved currents. The bound is obtained by scanning over $\alpha_{JJJ}$. The minimum is reached for $\alpha_{JJJ}^{*}$, whose value is reported on the table to the right of the plot. The bound was obtained at $\Lambda = 19$.\label{fig:CJmin}}
\end{figure}

Finally, we have obtained the absolute lower bound on $C_{J}$ for higher rank $SU(N)$ groups at $\Lambda = 19$. 
In this case, we have not mapped the entire lower bound as a function of $\alpha_{JJJ}$, but used instead the Navigator method~\cite{Reehorst:2021ykw} to minimize the objective $\max \alpha(\mathcal{V}_{\id})$ (inversely proportional to the $C_{J}$ lower bound) as a function of $\alpha_{JJJ}$. 
A plot of $\min C_{J}$ as a function of $N$ is reported in Figure~\ref{fig:CJmin}, together with the value of $\alpha_{JJJ}^{*}$ for which the minimum is attained.

As shown in these figures, the values of the minima and the corresponding $\alpha_{JJJ}$ seem to converge to a fixed value in the large $N$ limit. 
This mirrors the behaviour of $C_J$ in theories such as the $O(N)$ vector model and Gross–Neveu–Yukawa, as observed in~\cite{Diab:2016spb}. 
By contrast, it differs from the expected behaviour of the central charge $C_T$~\cite{Osborn:1993cr}, which instead scales linearly with $N$.

The assumptions listed in~\eqref{eq:SDP-CJbound-SU3} hold in any unitary theory, making our bounds applicable to any CFT with global $SU(N)$ symmetry. 
In this work, however, we focus on the dual description of YM in AdS, which is a non-local CFT without a stress tensor. 
In principle, the bounds presented in this section could be strengthened by removing $T_{\mu\nu}$ from the spectrum and imposing a gap in the spin-2, parity-even singlet sector. 
The challenge is identifying the appropriate gap: when $g\ped{YM}\rightarrow 0$, the first operator is (schematically) $J_\mu J_\nu$ with dimension $\Delta=4$, but as the coupling grows, this operator acquires an anomalous dimension. 
Since the minimum of $C_J$ should correspond to a strong-coupling regime,
this anomalous dimension could be $O(1)$, making the analysis highly sensitive to the assumed gap. We leave this investigation to future work, since it would not alter the main conclusion: the decoupling scenario is ruled out.

\section{Bounds on scaling dimensions of scalars}\label{sec:bootstrap-dimensions}

Having rigorously excluded the decoupling scenario in the previous subsection, we now turn to the merging–annihilation and Higgsing scenarios.
Our strategy is to impose constraints on operator dimensions. 
In both cases, the key question is which operators can cross marginality, and at what point this occurs.
We start in Section~\ref{generalbounds} by obtaining general exclusion bounds assuming nothing but the unitarity bound, and mild twist gaps used for the stability of the numerical algorithm.
These bounds are clearly generic, as we have not yet supplied any input that would single out specific theories. 
Thus, in order to get more constraining results, we must direct the bootstrap to focus on theories that arise as boundary limits of YM in AdS, rather than on generic $3d$ CFTs with a conserved current. 
We implement this refinement in two steps. 
First, in Sections~\ref{sec:bootstrap-dimesions-feasibility} and~\ref{sec:bootstrap-dimensions-feasibility} we inject information about the theory in the form of gaps on a handful of sectors and on sub-leading operators after those we are trying to bound. 
Second, in Section~\ref{sec:bootstrap-dimensions-GFV}, we probe the ``flow'' itself by first demanding that leading operators sit at the values predicted by the GFV, and then relaxing this assumption progressively in terms of a parameter $\delta$. 
This parameter provides an indirect probe for $L$.

Since the numerics involved in this section are quite expensive, we consider only $SU(2)$ and $SU(3)$ theories.
The precise Semi-Definite Programs (SDP) used to obtain the bounds of this section are deferred to Appendix~\ref{app:numerics}.

\subsection{Upper bounds}\label{generalbounds}

\begin{figure}[t!]
    \centering
    \includegraphics[width=0.6\linewidth]{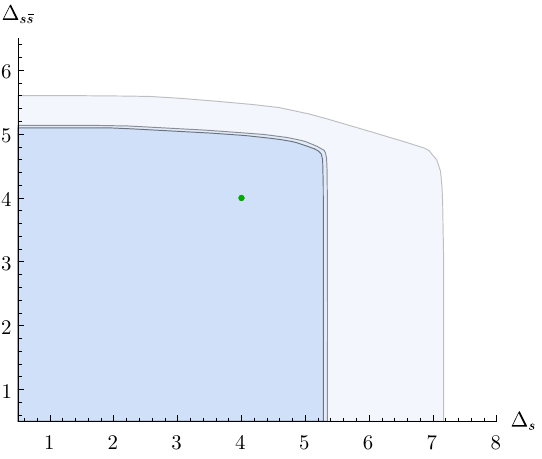}
    \caption{Gap maximisation bound for \( (\Delta_{s},\Delta_{s\bar{s}}) \) in the case of \( SU(2) \). The shaded area is allowed, darker colour means higher \( \Lambda = 11, 19, 23 \). The green dot at \( (\Delta_{s},\Delta_{s\bar{s}}) = (4,4) \) denotes the  generalised free vector current theory. The precise spectrum assumptions are reported in~\eqref{eq:SDP-gap-maximisation-SU2}.}
    \label{fig:JJJJ_SU2_Pe-S_Pe-SSb_bound_pd11-19-23}
\end{figure}

\begin{figure}[p]
    \centering
    \includegraphics[scale=0.9]{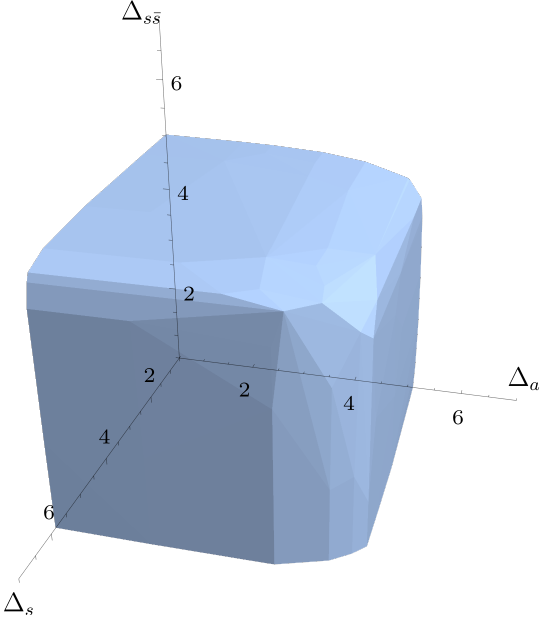}
    \caption{Gap maximisation bound for \( (\Delta_{s},\Delta_{a},\Delta_{s\bar{s}}) \) in the case of \( SU(3) \) for $\Lambda = 23$. The shaded coloured region is allowed. The precise spectrum assumptions are reported in~\eqref{eq:SDP-gap-maximisation-SU3}.\label{fig:JJJJ_SU3_Pe-S_Pe-adjS_Pe-SSb_bound_pd-23}}
\end{figure}
\begin{figure}[p]
    \centering
    \includegraphics[width=\textwidth]{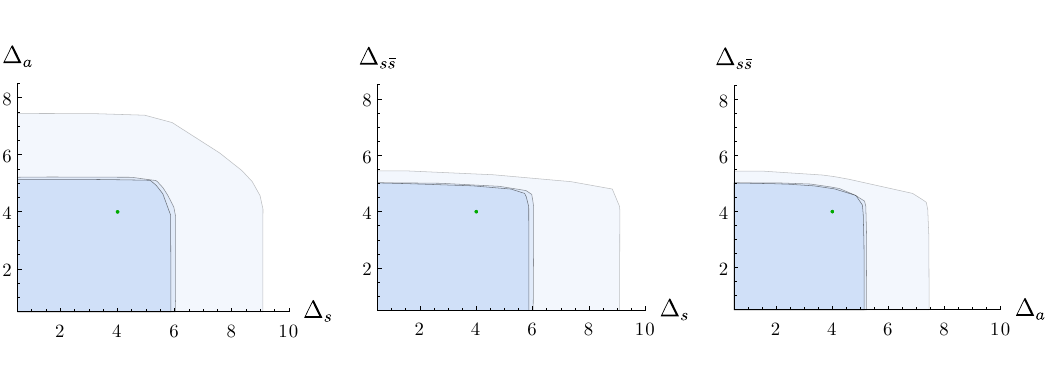}
    \vspace*{-2em}
    \caption{Projections of the gap maximisation bound in Figure~\ref{fig:JJJJ_SU3_Pe-S_Pe-adjS_Pe-SSb_bound_pd-23} for \( (\Delta_{s},\Delta_{a},\Delta_{s\bar{s}}) \) in the case of \( SU(3) \), to the three planes $(\Delta_{s},\Delta_{a})$, $(\Delta_{s},\Delta_{s\bar{s}})$ and $(\Delta_{a},\Delta_{s\bar{s}})$ respectively. The shaded area is allowed; darker colours indicate higher values of \( \Lambda = 11, 19, 23 \). The green dot at \( (\Delta_{s},\Delta_{a},\Delta_{s\bar{s}}) = (4,4,4) \) denotes the generalised free vector current theory. The precise spectrum assumptions are reported in~\eqref{eq:SDP-gap-maximisation-SU3}.\label{fig:JJJJ_SU3_Pe-S_Pe-adjS_Pe-SSb_bound_projection_pd-11-19-23}}
\end{figure}

We start by providing upper bounds on the scaling dimensions of the leading parity-even scalar operators.
These are $s$, $a$, and $s\bar s$ for $SU(3)$, while for $SU(2)$ we have only $s$ and $s\bar s$.\footnote{
    As discussed in Section~\ref{sec:spectrum-GFV} and shown in Table~\ref{tab:spectrum-SU2}, the first adjoint scalar in the $SU(2)$ GFV has dimension $\Delta = 7$, which makes this charged scalar sector less relevant for the Higgsing scenario.
}
In Figure~\ref{fig:JJJJ_SU2_Pe-S_Pe-SSb_bound_pd11-19-23} we report the allowed region in the ($\Delta_{s}$, $\Delta_{s\bar{s}}$) plane for $SU(2)$, and in Figure~\ref{fig:JJJJ_SU3_Pe-S_Pe-adjS_Pe-SSb_bound_pd-23} in ($\Delta_{s}$, $\Delta_a$, $\Delta_{s\bar{s}}$) space for $SU(3)$.
In Figure~\ref{fig:JJJJ_SU3_Pe-S_Pe-adjS_Pe-SSb_bound_projection_pd-11-19-23} we provide two-dimensional projections of the bounds in Figure~\ref{fig:JJJJ_SU3_Pe-S_Pe-adjS_Pe-SSb_bound_pd-23}.

These bounds assume only unitarity and apply to any CFT with conserved $SU(2)$ or $SU(3)$ conserved currents. 
Note that since a generalised free current is a consistent solution to crossing, the points \( (\Delta_{s},\Delta_{s\bar{s}}) = (4,4) \) for $SU(2)$ and \( (\Delta_{s},\Delta_{a},\Delta_{s\bar{s}}) = (4,4,4) \) for $SU(3)$ are always allowed. 
Moreover, since these are gap maximisation bounds, any point in the square \( (\Delta_{s},\Delta_{s\bar{s}}) \in [1/2,4] \times [1/2,4]  \) or in the box \( (\Delta_{s},\Delta_{a},\Delta_{s\bar{s}}) \in [1/2,4] \times [1/2,4] \times [1/2,4] \) is also allowed. 
As expected, we cannot infer much about YM from these bounds, given that all the scalars we considered are allowed to be both relevant and irrelevant.

\begin{figure}[t!]
    \centering
    \includegraphics[width=0.6\linewidth]{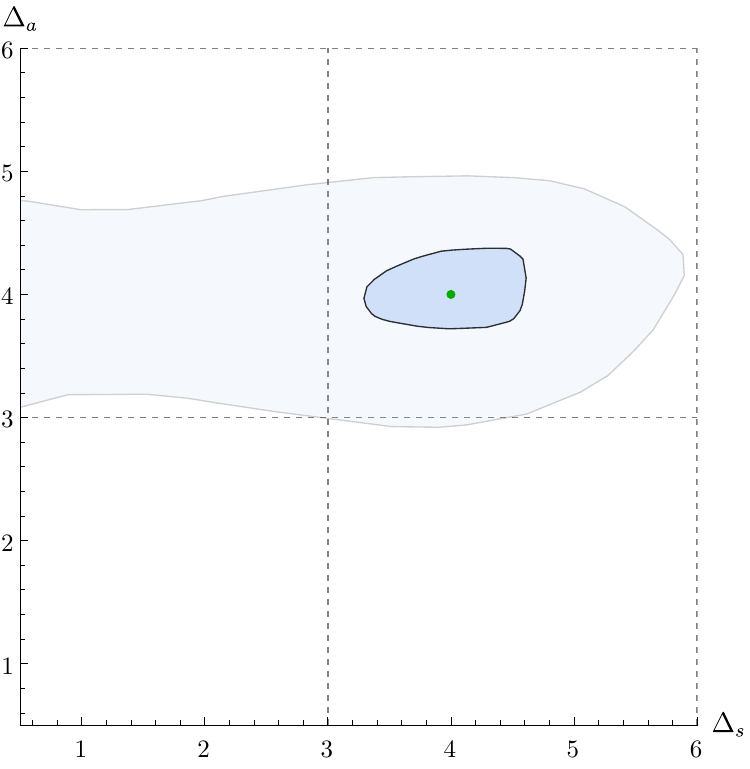}
    \caption{Allowed region for \( (\Delta_{s},\Delta_{a}) \) in the case of \( SU(3) \) assuming $\Delta_{s'} \geq 6$, $\Delta_{a'} \geq 6 $ and $ \Delta_{T_{\mu\nu}'} \geq 4$ (see~\eqref{eq:SDP-strong-gaps}). The shaded areas correspond to $\Lambda=11$ (lighter) and $\Lambda = 19$ (darker). The green dot \( (\Delta_{s},\Delta_{a}) = (4,4) \) corresponds to the GFV. The dashed lines denote either marginality or the value of the gap imposed on the sub-leading scalar operators.\label{fig:SU3_Pe-S_Pe-adjS_allowed-region_strong-gaps_pd-11-19}}
\end{figure}

\subsection{Allowed region with strong assumptions}\label{sec:bootstrap-dimesions-feasibility}

\begin{figure}[t!]
    \centering
    \includegraphics[width=0.6\linewidth]{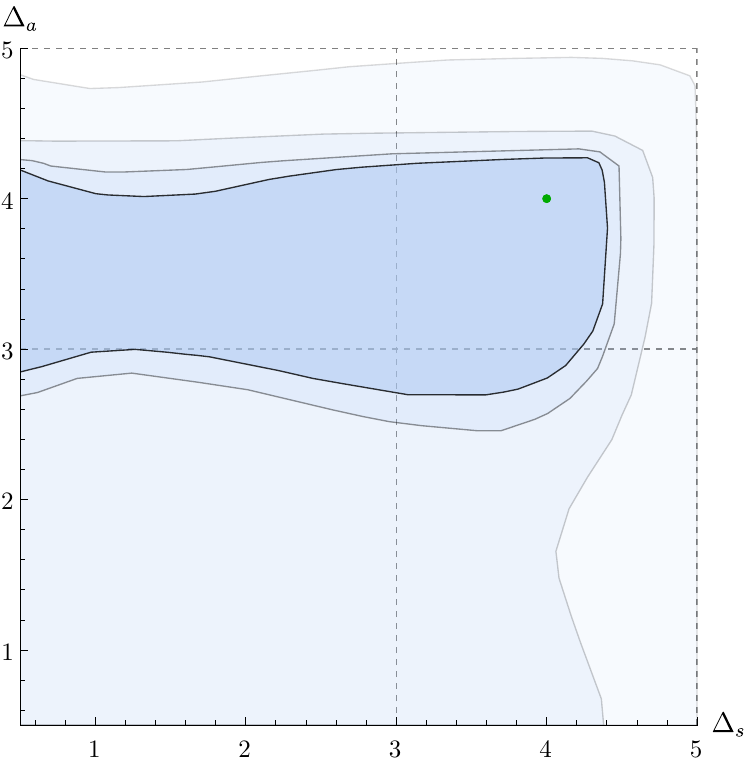}
    \caption{Allowed region for \( (\Delta_{s},\Delta_{a}) \) in the case of \( SU(3) \) with the gaps from~\eqref{eq:sdp-gfv-medium-gaps}. The shaded areas (from lighter to darker) correspond to \( \Lambda = 11,15,19, 23 \). The green dot \( (\Delta_{s},\Delta_{a}) = (4,4) \) denotes the GFV. The dashed lines denote either marginality or the value of the gap imposed on the sub-leading operators. 
    \label{fig:SU3_Pe-S_Pe-adjS_allowed-region_GFV-like-gaps_pd-11-15-19}}
\end{figure}

Before considering more physically motivated gaps, we study what kind of constraints arise for $\Delta_s$ and $\Delta_a$ when imposing very aggressive gaps, which essentially correspond to the spectral gaps of the UV GFV. 
In particular, we take $s$ and $a$ to be isolated in the spectrum and we impose \( \Delta_{s'} \geq 6 \), \( \Delta_{a'} \geq 6 \) and \( \Delta_{T_{\mu\nu}'} \geq 4 \).\footnote{
    Recall that the boundary CFT does not have a stress tensor, and $T_{\mu\nu}'$ denotes the lightest spin 2 parity-even operator.
} 
The result, reported in Figure~\ref{fig:SU3_Pe-S_Pe-adjS_allowed-region_strong-gaps_pd-11-19} for $SU(3)$, is an isolated island of allowed parameter space around the generalised free vector (GFV) theory.
The observation that the entire allowed region can collapse into an island centred around the UV of the RG flow, provided sufficiently strong assumptions are made, is encouraging and motivated the methodology presented in Section~\ref{sec:bootstrap-dimensions-GFV}. 
As one might expect, the resulting bound already rules out both the Higgsing and merging scenarios. 
It is noteworthy that this occurs already for $\Lambda=19$. 
If we believe that at least one of these scenarios must be realised, then at least one of our assumptions cannot hold across the full flow. 
In more precise terms, at least one among the operators $s'$, $a'$, and $T_{\mu\nu}'$ must acquire a negative anomalous dimension.

\subsection{Allowed region with moderate assumptions}\label{sec:bootstrap-dimensions-feasibility}

We now take a middle route between Sections~\ref{generalbounds} and~\ref{sec:bootstrap-dimesions-feasibility}, assuming gaps inspired by the GFV, but at the same time allowing operators to obtain anomalous dimensions.
In particular, we assume a handful of gaps \( \Delta \geq \Delta_{\ell, p, \mathbf{r}}\ap{gap} \) in sectors $(\ell, p, \mathbf{r})$ with small spin $\ell$, which lie between the unitarity bound and spectral gaps of the GFV. 
The precise gap assumptions are reported in Table~\ref{tab:gfv-medium-gaps}. 
We impose that $s$ and $a$, the lightest singlet and adjoint scalars, are isolated in the spectrum, and \( \Delta_{s'}, \Delta_{a'} \geq 5 \) as gaps on the sub-leading operators. 
The allowed region in $(\Delta_s, \Delta_a)$ space, which has been calculated for \( \Lambda = 11, 15, 19 \) and $23$, is given in Figure~\ref{fig:SU3_Pe-S_Pe-adjS_allowed-region_GFV-like-gaps_pd-11-15-19}.
The SDP used to obtain this allowed region is reported in~\eqref{eq:sdp-gfv-medium-gaps}.
While in this case we do not observe the creation of an island, there seems to be a trend with increasing constraining power, measured by $\Lambda$, towards pushing $\Delta_a$ above the value of $\Delta_a=3$. 
Figure~\ref{fig:SU3_Pe-S_Pe-adjS_allowed-region_GFV-like-gaps_pd-11-15-19} provides the first faint evidence in favour of the merging scenario, which will be strengthened by the results of next section.

\subsection{Tracing the flow}\label{sec:bootstrap-dimensions-GFV}

We now implement the idea of ``following'' the RG flow of YM in AdS, alluded to in the previous subsections, starting from the GFV and flowing to what is (potentially) the transition point.\footnote{
    In addition to the analysis presented in the preceding sections, the strategy and gap assumptions adopted in this section were partially inspired by~\cite{Antunes:2024hrt} and the hybrid bootstrap study of~\cite{Su:2022xnj}.
}
Recall that as long as the Dirichlet bc exists in the form of a unitary and stable boundary CFT, the boundary limit of YM in AdS is described by this one-parameter family of CFTs. 
The parameter of the flow is \( g\ped{YM} \) (or \( L \Lambda\ped{YM} \)) and continuously connects to the GFV at \( g\ped{YM} = 0 \). 
The above idea is implemented by assuming a spectrum for operators exchanged in the OPE \( J \times J \) which is parametrically close to that of the GFV. 
This parametric proximity is quantified in terms of a parameter that we call $\delta$. 
At $\delta=0$ the spectrum of the first few operators appearing in \( J \times J \) is forced to be that of the GFV (see precise details below). 
As we relax $\delta$ to some non-zero value, the scaling dimensions of operators are allowed to live in an interval centred around the GFV value, whose precise size depends on the chosen value of $\delta$. 
This, then, allows the operators to attain an anomalous dimension, the permissible size of which depends on $\delta$. 
It is in this sense that $\delta$ behaves as a probe for the RG flow.
At a technical level, this type of spectrum is implemented in the bootstrap using what is now referred to as ``interval positivity'', introduced in~\cite[Sec.\ 5.4]{Go:2020ahx}.

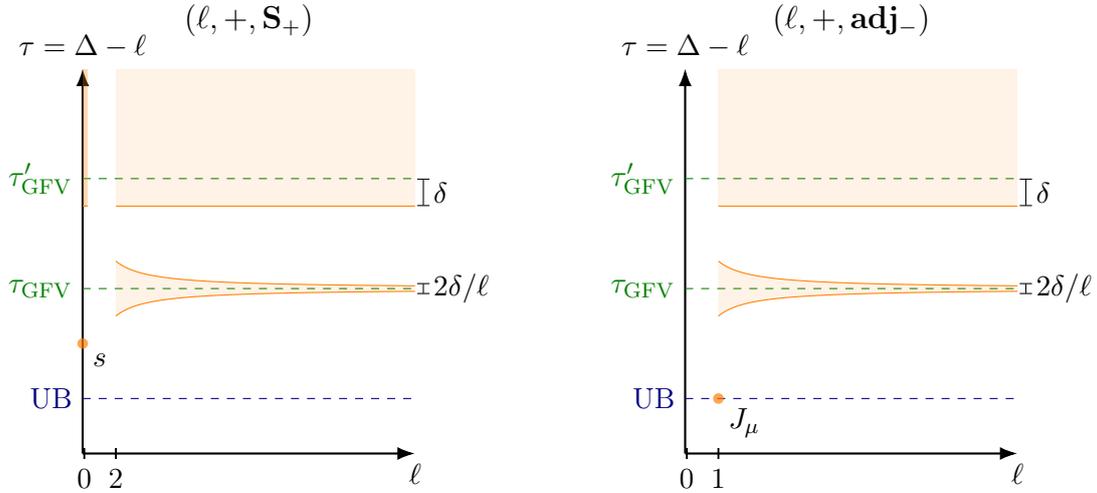
\begin{figure}[t!]
    \centering
    \begin{subfigure}[c]{0.45\linewidth}
    \centering
        \begin{tikzpicture}[x={0.03\linewidth},y={0.2\linewidth}]
        \pgfmathsetmacro{\ub}{1};
        \pgfmathsetmacro{\GFVlead}{2};
        \pgfmathsetmacro{\GFVsub}{3};
        \pgfmathsetmacro{\lmin}{0};
        \pgfmathsetmacro{\lmax}{20};
        \pgfmathsetmacro{\taumin}{0.5};
        \pgfmathsetmacro{\taumax}{4};
        \pgfmathsetmacro{\del}{0.25};

        \node[above, font=\large] at ({(\lmin+\lmax)/2}, \taumax+0.2) {$(\ell,+,\bf{S_+})$};

        \draw[-{Latex},line width=0.8pt] (\lmin,\taumin) -- (\lmin,\taumax) node[above] {\( \tau = \Delta - \ell \)};
        \draw[-{Latex},line width=0.8pt] (\lmin,\taumin) -- (\lmax,\taumin) node[below] {\( \ell \)};
        \draw[line width=0.7pt] (2, \taumin+0.05) -- (2, \taumin - 0.05)     node[below] {2};
        \draw[line width=0.7pt] (0.1, \taumin+0.05) -- (0.1, \taumin - 0.05)     node[below] {0};

        \draw[dashed,color=blue!50!black] (\lmin,\ub) node[left] {UB} -- (\lmax,\ub);

        \draw[dashed,color=green!50!black] (\lmin,\GFVlead)  node[left] {$\tau\ped{GFV}$} -- (\lmax,\GFVlead);
        \draw[dashed,color=green!50!black] (\lmin,\GFVsub)  node[left] {$\tau\ped{GFV}'$} -- (\lmax,\GFVsub);

        \draw[domain={\lmin+2:\lmax}, samples=500, color=orange, name path=upper] plot (\x, {\GFVlead + (2*\del)/\x});
        \draw[domain={\lmin+2:\lmax}, samples=500, color=orange, name path=lower] plot (\x, {\GFVlead - (2*\del)/\x});
        \tikzfillbetween[of=upper and lower]{orange, opacity=0.1};

        \draw[color=orange, name path=subleading] (\lmin+2,{\GFVsub-\del}) -- (\lmax,{\GFVsub-\del});
        \draw[transparent, name path=top] (\lmin+2,\taumax) -- (\lmax,\taumax);
        \tikzfillbetween[of=subleading and top]{orange, opacity=0.1};

        \draw[color=orange, name path=subleading] (\lmin,{\GFVsub-\del}) -- (\lmin+0.3,{\GFVsub-\del});
        \draw[transparent, name path=top] (\lmin,\taumax) -- (\lmin+0.3,\taumax);
        \tikzfillbetween[of=subleading and top]{orange, opacity=0.3};

        \fill[orange,opacity=0.7] (\lmin, 1.5) circle (2pt);
        \node[below right] at (\lmin, 1.5) {$s$};

        \draw[|-|] (\lmax+0.5, {\GFVlead-\del/5}) to node[right]{\( 2 \delta/\ell \)} (\lmax+0.5, {\GFVlead+\del/5});
        \draw[|-|] (\lmax+0.5, {\GFVsub-\del}) to node[right]{\( \delta \)} (\lmax+0.5, \GFVsub);
\end{tikzpicture}
    \end{subfigure}
    \hspace*{1em}
    \begin{subfigure}[c]{0.45\linewidth}
    \centering
        \begin{tikzpicture}[x={0.03\linewidth},y={0.2\linewidth}]
        \pgfmathsetmacro{\ub}{1};
        \pgfmathsetmacro{\GFVlead}{2};
        \pgfmathsetmacro{\GFVsub}{3};
        \pgfmathsetmacro{\lmin}{0};
        \pgfmathsetmacro{\lmax}{20};
        \pgfmathsetmacro{\taumin}{0.5};
        \pgfmathsetmacro{\taumax}{4};
        \pgfmathsetmacro{\del}{0.25};

        \node[above, font=\large] at ({(\lmin+\lmax)/2}, \taumax+0.2) {$(\ell,+,\bf{adj_-})$};

        \draw[-{Latex},line width=0.8pt] (\lmin,\taumin) -- (\lmin,\taumax) node[above] {\( \tau = \Delta - \ell \)};
        \draw[-{Latex},line width=0.8pt] (\lmin,\taumin) -- (\lmax,\taumin) node[below] {\( \ell \)};
        \draw[line width=0.7pt] (2, \taumin+0.05) -- (2, \taumin - 0.05)     node[below] {1};
        \draw[line width=0.7pt] (0.1, \taumin+0.05) -- (0.1, \taumin - 0.05)     node[below] {0};

        \draw[dashed,color=blue!50!black] (\lmin,\ub) node[left] {UB} -- (\lmax,\ub);

        \draw[dashed,color=green!50!black] (\lmin,\GFVlead)  node[left] {$\tau\ped{GFV}$} -- (\lmax,\GFVlead);
        \draw[dashed,color=green!50!black] (\lmin,\GFVsub)  node[left] {$\tau\ped{GFV}'$} -- (\lmax,\GFVsub);

        \draw[domain={\lmin+2:\lmax}, samples=500, color=orange, name path=upper] plot (\x, {\GFVlead + (2*\del)/\x});
        \draw[domain={\lmin+2:\lmax}, samples=500, color=orange, name path=lower] plot (\x, {\GFVlead - (2*\del)/\x});
        \tikzfillbetween[of=upper and lower]{orange, opacity=0.1};

        \draw[color=orange, name path=subleading] (\lmin+2,{\GFVsub-\del}) -- (\lmax,{\GFVsub-\del});
        \draw[transparent, name path=top] (\lmin+2,\taumax) -- (\lmax,\taumax);
        \tikzfillbetween[of=subleading and top]{orange, opacity=0.1};

        \fill[orange,opacity=0.7] (\lmin+2, 1) circle (2pt);
        \node[below right] at (\lmin+2, 1) {$J_\mu$};

        \draw[|-|] (\lmax+0.5, {\GFVlead-\del/5}) to node[right]{\( 2 \delta/\ell \)} (\lmax+0.5, {\GFVlead+\del/5});
        \draw[|-|] (\lmax+0.5, {\GFVsub-\del}) to node[right]{\( \delta \)} (\lmax+0.5, \GFVsub);
    \end{tikzpicture}
    \end{subfigure}
    \caption{Schematic representation of the $\delta$-gaps for two examples, the $(\ell,+,\bf{S}_+)$ sector (on the left) and the $(\ell,+,\bf{adj}_-)$ sector (on the right). The dots correspond to isolated operators, while bands correspond to intervals or gaps. We represented the spin $\ell$ as a continuum (except for $\ell=0$ in the left panel, which has to be treated separately) even though the two sectors only contain even and odd integer spins, respectively. Note that the twist of the leading and sub-leading operators in the GFV (the dashed green lines) is not constant at small spin, so the vertical position in the picture is just schematic in this case.
    }
\end{figure}

The dependence on the parameter $\delta$ of our gaps is inspired by the universal light-cone bootstrap result stating that anomalous dimensions of spin $\ell$ \( [J J] \) operators in $3d$ scale as \( 1/\ell \) at large spin~\cite{Komargodski:2012ek,Fitzpatrick:2012yx}. 
As discussed in Section~\ref{sec:spectrum-GFV}, the first operator in each \( (\ell, p, \mathbf{r}) \) sector is of the type \( [JJ] \), with the exception of the identity \( \id \) and the current itself \( J \), which have fixed dimensions. 
Moreover, thanks to the result reviewed in Section~\ref{sec:spectrum-perturbative}, we know that the \( (\ell, p, \mathbf{S\bar{A}}_{c}) \) sector anomalous dimensions scale as \( 1/\ell^{a} \) with $a \geq 2$. 
The precise set of gap assumptions for the case of $SU(3)$ is as follows:
\begin{itemize}
    \item In each \( (\ell, p, \mathbf{r}) \) sector exchanged in \( \braket{JJJJ} \), except for \( \ell = 0 \) or \( \mathbf{S\bar{A}}_{c} \), we assume that
    \begin{itemize}
        \item the first operator can be in an interval \( [\Delta\ped{GFV} - \delta/\ell, \Delta\ped{GFV} + \delta/\ell] \) of width \( 2\delta/\ell \) centered around the first operator in the GFV spectrum \( \Delta\ped{GFV} \) in the \( (\ell, p, \mathbf{r}) \) sector;
        \item the second operator can have dimension \( \Delta \geq \Delta\ped{GFV}' - \delta \), where \( \Delta\ped{GFV}' \) is the dimension of the sub-leading operator in the \( (\ell, p, \mathbf{r}) \) of the GFV spectrum.\footnote{
            As reviewed in Section~\ref{sec:spectrum-perturbative}, the sub-leading operator is typically a ``\( J^{3} \)'' operator for which the scaling of anomalous dimensions in $1/\ell$ is not known---but see~\cite{Harris:2024nmr,Fardelli:2024heb,Kravchuk:2024wmv} for recent work on this topic.
        }
    \end{itemize}
    \item In each \( (\ell, p, \mathbf{S\bar{A}}_{c}) \) sector we assume the first operator can be in an interval \( [\Delta\ped{GFV} - \delta/\ell^{2}, \Delta\ped{GFV} + \delta/\ell^{2}] \), while the second operator can have dimension \( \Delta \geq \Delta\ped{GFV}' - \delta \)
    \item In the \( (0, -, \mathbf{r}) \) sector we assume that the first operator can be in an interval \( [\Delta\ped{GFV}-\delta,\Delta\ped{GFV}+\delta] \), while the second operator can have dimensions \( \Delta \geq \Delta\ped{GFV}' - \delta \)
    \item In the \( (0, +, \mathbf{S}_{+}) \), \( (0, +, \mathbf{adj}_{+}) \) and \( (0, +, \mathbf{S\bar{S}}_{+}) \) sectors we assume first an isolated operator of dimension \( \Delta_{s} \), \( \Delta_{a} \) and \( \Delta_{s\bar{s}} \) respectively, and a second operator that can have dimension \( \Delta \geq \Delta\ped{GFV}' - \delta = 6 - \delta \)
    \item The current itself is exchanged in the \( (1,+,\mathbf{adj}_{-}) \) sector at \( \Delta = 2 \) with OPE ratio parametrised by \( \alpha_{JJJ} \) as described in~\eqref{lambdaJJJ}.
\end{itemize}
We will refer to this set of gap assumptions as \emph{\( \delta \)-gaps}.

The gap assumptions depend on 5 parameters \( (\delta, \Delta_{s}, \Delta_{a}, \Delta_{s\bar{s}}, \alpha_{JJJ}) \): we will choose and array of values for $\delta$ and scan over the remaining ones.
For this, we will use the Navigator method~\cite{Reehorst:2021ykw} to explore the space of allowed scaling dimensions \( (\Delta_{s},\Delta_{a},\Delta_{s\bar{s}}) \) as well as the OPE ratio \( \alpha_{JJJ} \). 
For a fixed value of \( \delta \), the allowed region in  \( (\Delta_{s},\Delta_{s},\Delta_{s\bar{s}}) \) has the interpretation of containing the one-parameter family of conformal data of the Dirichlet boundary CFT for all values of the coupling \( 0 \leq g\ped{YM} \leq g\ped{YM}(\delta) \),\footnote{
    If all anomalous dimensions are monotonic for $g\ped{YM} \leq g\ped{YM}(\delta)$.
}
where we have defined $g\ped{YM}(\delta)$ to be the maximum value of $g\ped{YM}$ that can be probed with our gap assumptions at a given $\delta$. 
By definition, the allowed region with \( \delta \)-gaps \( \delta = \delta_{1} \) contains the region with \( \delta = \delta_{2} < \delta_{1} \).
We will not map the allowed region in \( (\Delta_{s}, \Delta_{a}, \Delta_{s\bar{s}}, \alpha_{JJJ}) \) space to its full extent, as this would be numerically rather expensive. 
Instead, we will maximise and minimise scaling dimensions, which will provide a ``bounding box'' of the actual allowed region.

In the case of $SU(2)$ we adopt analogous gap assumptions, with the exception that the $\mathbf{S\bar{A}}_{c}$ representation is not present in this case, and the search space is just $(\Delta_{s}, \Delta_{s\bar{s}}, \alpha_{JJJ})$.

The strategy, then, is to look at how the allowed region changes as we increase \( \delta \). 
We will consider as evidence in favour of the ``merger and annihilation'' scenario the fact that \( \Delta_{s} \leq 3 \) is allowed in this CFT space for a smaller value of \( \delta \) with respect to the value of $\delta$ for which either \( \Delta_{a} \) or \( \Delta_{s\bar{s}} \) are allowed to be below $3$. 
This is non-rigorous evidence, since we are not guaranteed that the boundary CFT of YM in AdS at \( g\ped{YM} = g\ped{YM}(\delta) \) extremizes (or is close to extremizing) \( \Delta_{s} \) in this CFT space.

Recall that, in the case of $SU(2)$, the leading order anomalous dimension of the lightest $s$ and $s\bar{s}$ operators are respectively negative and positive, as we reviewed in Section~\ref{sec:spectrum-perturbative}, while the lightest adjoint scalar has dimension $\Delta = 7 + O(g\ped{YM}^2)$, as we reviewed in Section~\ref{sec:spectrum-GFV}. 
Thus, the Higgsing scenario is ``perturbatively'' disfavoured in this case. 
In the case of $SU(3)$, instead, the $a$ operator is much lighter in the GFV ($\Delta = 4$) and has a negative anomalous dimension at leading order, which makes Higgsing a more likely possibility.
We have thus decided to invest more computational resources in the case of $SU(3)$, i.e.\ performing the present computation for a denser array of $\delta$ values.

\begin{table}[t!]
    \centering
    \begin{tabular}{l|ll}
        \toprule
        \( \delta \) & \( (\min \Delta_{s}, \Delta_{s\bar{s}}, \alpha_{JJJ}) \) & \( (\Delta_{s}, \min \Delta_{s\bar{s}}, \alpha_{JJJ}) \) \\ \midrule
        0 & \( (3.970, 3.999, 0.867) \) & \( (3.989, 3.987, -) \) \\
        0.1 & \( (3.780, 4.101, 0.708) \) & \( (3.971, 3.931, -) \) \\
        0.3 & \( (3.521, 4.180, 0.713) \) & \( (3.883, 3.851, -) \) \\
        0.5 & \( (3.227, 4.229, 0.724) \) & \( (3.746, 3.754, -) \) \\
        0.7 & \( (2.865, 4.135, 0.730) \) & \( (3.504, 3.611, -) \) \\
        1 & \( (1.956, 3.421, 0.771) \) & \( (2.698, 3.172, -) \) \\
        \bottomrule
    \end{tabular}
    \caption{Results from the minimization of either \( \Delta_{s} \) or \( \Delta_{s\bar{s}} \) for different \( \delta \)-gaps for \( SU(2) \) at \( \Lambda = 19 \). In the $\Delta_{s\bar{s}}$ minimisation, the $\alpha_{JJJ}$ direction has a flat gradient for most of the minimization (i.e.\ smaller than the duality gap); the value of $\alpha_{JJJ}$ is thus not meaningful and is not reported.\label{tab:delta-gaps-minimization-SU2}}
\end{table}

\begin{figure}[t!]
    \centering
    \includegraphics[width=0.8\linewidth]{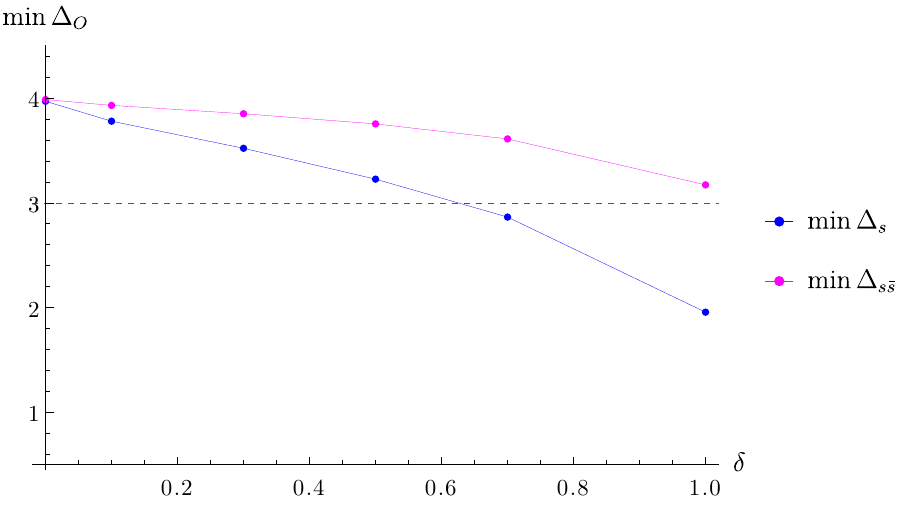}
    \caption{Comparison of $\min \Delta_{s}$ and $\min \Delta_{s\overline{s}}$ for different $\delta$-gaps in the case of $SU(2)$ at $\Lambda=19$.\label{fig:JJJJ_SU2_Pe-S_Pe-SSb_delta-gaps_min-compare_pd19}}
\end{figure}

\begin{figure}[t!]
    \centering
    \includegraphics[width=0.85\linewidth]{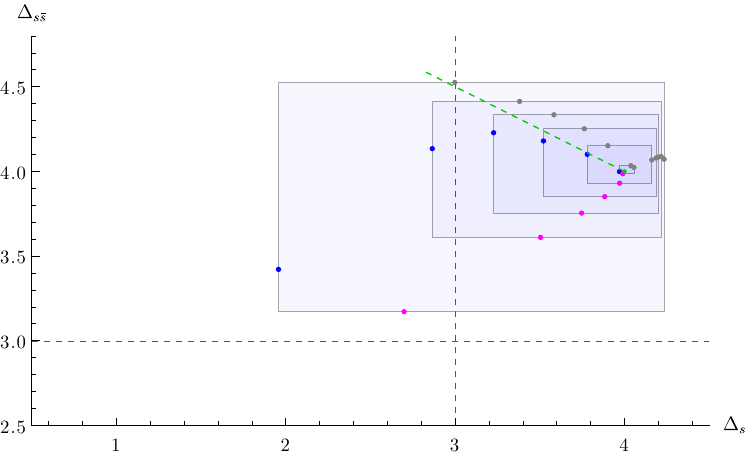}
    \caption{Projection in the $(\Delta_s, \Delta_{s\bar{s}})$ plane of the allowed region in $(\Delta_s, \Delta_{s\bar{s}}, \alpha_{JJJ})$ space for different $\delta$-gaps for the case of \( SU(2) \) at \( \Lambda=19 \). Shaded areas, from darker to lighter, correspond to $\delta=0,\, 0.1,\, 0.03,\, 0.5, \,0.7, \,1$. Note that this overestimates the size of the actual allowed region since it's obtained by only minimising/maximising $\Delta_s$ and $\Delta_{s\bar{s}}$. The blue and magenta dots denote the endpoints of the minimisation of $\Delta_{s}$ and \ $\Delta_{s\bar{s}}$, respectively. The grey dots denote the endpoints of the maximisation procedure. The dashed green line corresponds to the anomalous dimension~\eqref{eq:perturbative-scalars}. The dashed grey lines denote the marginality condition.\label{fig:JJJJ_SU2_Pe-S_Pe-SSb_delta-gaps_pd19}}
\end{figure}

\subsubsection{\texorpdfstring{$SU(2)$}{SU(2)} case}

Let us start with the $SU(2)$ case. In Table~\ref{tab:delta-gaps-minimization-SU2} we report the results from the minimization of $\Delta_{s}$ and $\Delta_{s\bar{s}}$ in  $(\Delta_{s}, \Delta_{s\bar{s}}, \alpha_{JJJ})$ space for different values of $\delta$ that we have explored.
The bootstrap computation is done at $\Lambda = 19$.
A plot of $\min \Delta_{s}$ and $\min \Delta_{s\bar{s}}$ as a function of $\delta$ is reported in Figure~\ref{fig:JJJJ_SU2_Pe-S_Pe-SSb_delta-gaps_min-compare_pd19}, where we observe that the singlet is allowed to cross the marginality condition for $\delta_{*} \simeq 0.6$, while the $s\bar{s}$ operator cannot be relevant until at least $\delta = 1$, an thus we expect $\delta_{**} \gtrsim 1$ in the notation used in Section~\ref{sec:intro-results}.

In Figure~\ref{fig:JJJJ_SU2_Pe-S_Pe-SSb_delta-gaps_pd19} we draw bounding boxes for the allowed region in $(\Delta_{s}, \Delta_{s\bar{s}})$ space for different values of $\delta$, as reported in Table~\ref{tab:delta-gaps-minimization-SU2}. 
For a given $\delta$, the actual allowed region for the scaling dimensions is a two-dimensional island-like shape that is tangent to the boxes at the corresponding dots. 
For $\delta = 0$, the allowed region is essentially a small island centred at the GFV values $(\Delta_{s}, \Delta_{s\bar{s}}) = (4,4)$.
As $\delta$ increases, the island enlarges asymmetrically, allowing the singlet to cross marginality while the charged $s\bar{s}$ scalar is still bounded to be irrelevant.

The SDP used to obtain Figure~\ref{fig:JJJJ_SU2_Pe-S_Pe-SSb_delta-gaps_min-compare_pd19} and~\ref{fig:JJJJ_SU2_Pe-S_Pe-SSb_delta-gaps_pd19} is reported in~\eqref{eq:sdp-gfv-delta-gaps-SU2}. 
The SDP is parametric in \( (\Delta_{s}, \Delta_{s\bar{s}}, \alpha_{JJJ}, \delta) \).  
For a fixed value of \( \delta \), we used the Navigator method to minimise/maximise each direction in scaling dimensions, keeping the remaining parameters free. 
Namely, we find the minimal and maximal values of either \( \Delta_{s} \) or \( \Delta_{s\bar{s}} \) for which the navigator function \( \mathcal{N}(\Delta_{s},\Delta_{s\bar{s}},\alpha_{JJJ}) = \max \alpha(\vec{\mathcal{V}}_{\id}) \) is negative.

\subsubsection{\texorpdfstring{$SU(3)$}{SU(3)} case}

\begin{table}[p]
    \centering
    \begin{tabular}{l|lll}
        \toprule
        \( \delta \) & \( (\min \Delta_{s}, \Delta_{a}, \Delta_{s\bar{s}}, \alpha_{JJJ}) \) & \( (\Delta_{s}, \min \Delta_{a}, \Delta_{s\bar{s}}, \alpha_{JJJ}) \)& \( (\Delta_{s}, \Delta_{a}, \min \Delta_{s\bar{s}}, \alpha_{JJJ}) \) \\ \midrule
        0 & \( (3.943, 3.984, 3.997, 0.913) \) & \( (3.963, 3.975, 3.998, 0.912) \) & \( (3.969, 3.994, 3.986, -) \) \\
        0.0125 & \( (3.881, 3.972, 4.014, 0.840) \) & \( (3.969, 3.944, 4.021, 0.820) \) & \( (3.957, 4.014, 3.974, -) \) \\
        0.05 & \( (3.775, 3.945, 4.034, 0.790) \) & \( (3.969, 3.899, 4.064, 0.784) \) & \( (3.910, 4.029, 3.956, -) \) \\
        0.1 & \( (3.664, 3.925, 4.065, 0.778) \) & \( (3.918, 3.848, 4.093, 0.758) \) & \( (3.876, 4.033, 3.937, -) \) \\
        0.2 & \( (3.491, 3.892, 4.112, 0.756) \) & \( (3.807, 3.765, 4.127, 0.731) \) & \( (3.820, 4.016, 3.903, -) \) \\
        0.3 & \( (3.320, 3.847, 4.135, 0.751) \) & \( (3.696, 3.687, 4.156, 0.726) \) & \( (3.742, 3.992, 3.865, -) \) \\
        0.4 & \( (3.139, 3.796, 4.149, 0.752) \) & \( (3.586, 3.604, 4.183, 0.729) \) & \( (3.670, 3.958, 3.825, -) \) \\
        0.5 & \( (2.944, 3.676, 4.152, 0.757) \) & \( (3.448, 3.513, 4.206, 0.731) \) & \( (3.518, 3.938, 3.779, -) \) \\
        0.6 & \( (2.728, 3.600, 4.116, 0.760) \) & \( (3.308, 3.413, 4.227, 0.733) \) & \( (3.318, 3.912, 3.724, -) \) \\
        0.7 & \( (2.471, 3.536, 4.028, 0.765) \) & \( (3.172, 3.300, 4.245, 0.734) \) & \( (3.116, 3.885, 3.658, -) \) \\
        0.8 & \( (2.121, 3.467, 3.891, 0.775) \) & \( (3.055, 3.172, 4.259, 0.736) \) & \( (2.852, 3.856, 3.575, -) \) \\
        0.9 & \( (1.177, 3.507, 3.655, 0.831) \) & \( (2.947, 3.023, 4.268, 0.738) \) & \( (2.502, 3.808, 3.471, -) \) \\
        1 & \( (0.5000, -, -, -) \) & \( (2.844, 2.846, 4.265, 0.742) \) & \( (2.018, 3.731, 3.337, -) \) \\
        \bottomrule
    \end{tabular}
    \caption{Results from the minimization of either \( \Delta_{s} \), \( \Delta_{a} \) or \( \Delta_{s\bar{s}} \) for different \( \delta \)-gaps for \( SU(3) \) at \( \Lambda = 19 \). In some cases, the gradient in one or more directions is flat for most of the minimisation (i.e.\ below the duality gap) and thus the value in that direction is not reported.\label{tab:delta-gaps-minimization-SU3}}
\end{table}

\begin{figure}[p]
    \centering
    \includegraphics[width=0.8\linewidth]{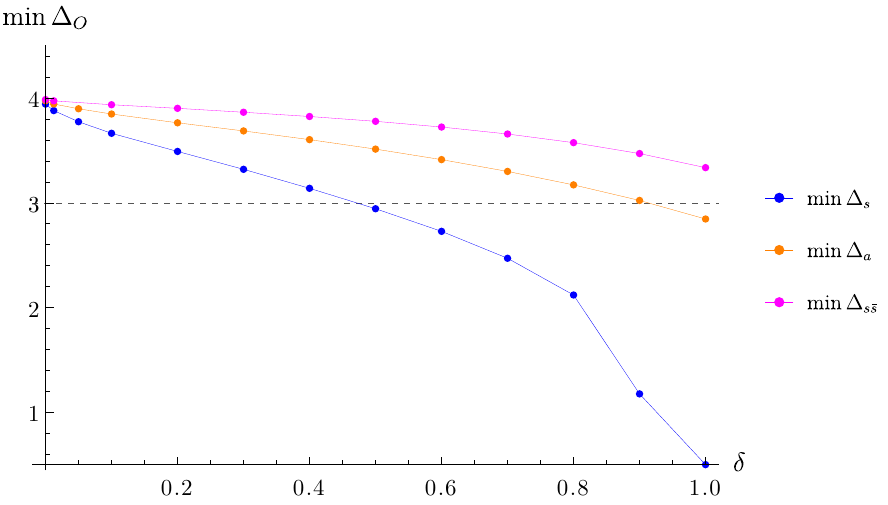}
    \caption{Comparison of $\min \Delta_{s}$, $\min \Delta_{a}$ and $\min \Delta_{s\overline{s}}$ for different values of the $\delta$ parameter in the case of $SU(3)$ at $\Lambda = 19$.\label{fig:JJJJ_SU3_Pe-S_Pe-adjS_delta-gaps_min-compare_pd19}}
\end{figure}

\begin{figure}[p!]
    \centering
    \begin{subfigure}[c]{\linewidth}
        \centering
        \includegraphics[width=0.85\linewidth]{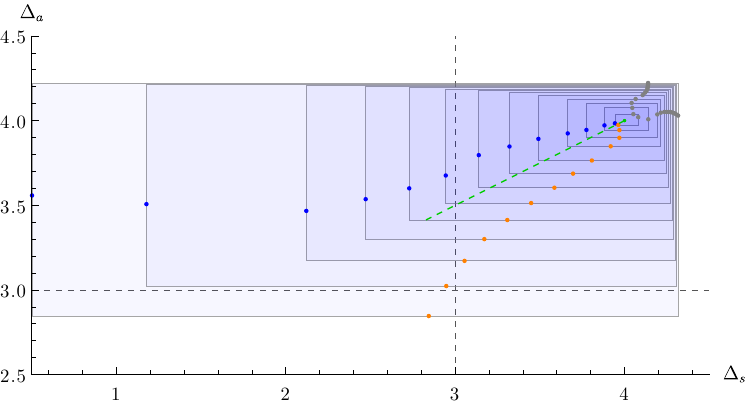}
        \caption{}
    \end{subfigure}
    \begin{subfigure}[c]{\linewidth}
        \centering
        \includegraphics[width=0.85\linewidth]{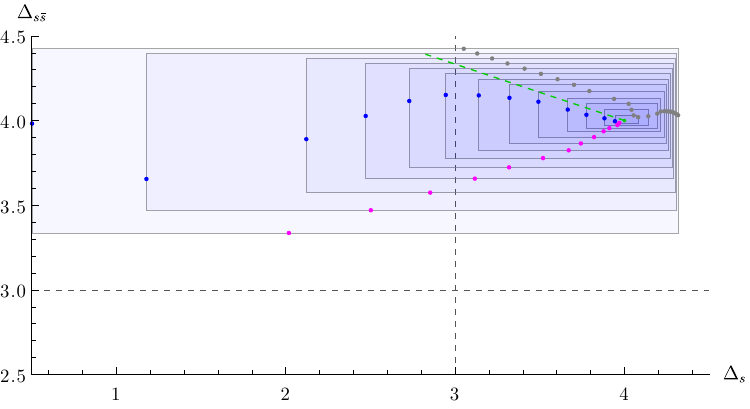}
        \caption{}
    \end{subfigure}
    \caption{Projection to the $(\Delta_s, \Delta_{a})$ and $(\Delta_s, \Delta_{s\bar{s}})$ planes in panels (a) and (b), respectively, of the allowed region in $(\Delta_s, \Delta_a, \Delta_{s\bar{s}}, \alpha_{JJJ})$ space for different $\delta$-gaps in the case of $SU(3)$ at $\Lambda=19$. Shaded areas, from darker to lighter, correspond to $\delta=0.0125,\, 0.05,\, 0.1, \,0.2,\ldots,\,1$. Note that this overestimates the actual size of the allowed region since it is obtained by only extremising $\Delta_s, \Delta_a$ and $\Delta_{s\bar{s}}$. The blue, orange and magenta dots denote the endpoints of the minimisation of $\Delta_{s}$,  $\Delta_{a}$ and \ $\Delta_{s\bar{s}}$, respectively. The grey dots denote the endpoints of the maximisation procedure. The dashed green line corresponds to the anomalous dimension~\eqref{eq:perturbative-scalars}. The dashed grey lines denote the marginality condition.\label{fig:JJJJ_SU3_Pe-S_Pe-adjS_Pe-SSb_delta-gaps_pd19}
    }
\end{figure}

We now proceed to $SU(3)$. 
The results from the minimization of $\Delta_{s}$, $\Delta_{a}$ and $\Delta_{s\bar{s}}$ in $(\Delta_{s},\Delta_{a},$ $ \Delta_{s\bar{s}}, \alpha_{JJJ})$ space are reported in Table~\ref{tab:delta-gaps-minimization-SU3} for different values of $\delta$ that we have explored. 
The bootstrap computation is done at $\Lambda = 19$.
A plot of $\min \Delta_{s}$, $\min \Delta_{a}$ and $\min \Delta_{s\bar{s}}$ as a function of $\delta$ is reported in Figure~\ref{fig:JJJJ_SU3_Pe-S_Pe-adjS_delta-gaps_min-compare_pd19}, where we observe that the singlet is allowed to cross marginality for $\delta_{*} \simeq 0.5$, the adjoint for $\delta_{**} \simeq 0.9$, while the $s\bar{s}$ operator cannot be relevant until at least $\delta = 1$.

In panel (a) of Figure~\ref{fig:JJJJ_SU3_Pe-S_Pe-adjS_Pe-SSb_delta-gaps_pd19} we draw bounding boxes for the allowed region in the $(\Delta_{s}, \Delta_{a})$ space for different values of $\delta$, as reported in Table~\ref{tab:delta-gaps-minimization-SU3}.
For a given $\delta$, the actual allowed region for the scaling dimensions is a three-dimensional island-like shape, whose two-dimensional projection is tangent to the boxes at the corresponding dots. 
For $\delta = 0$, the allowed region is essentially a small island centred at the GFV values $(\Delta_{s}, \Delta_{a}) = (4,4)$. 
As $\delta$ increases, the island enlarges asymmetrically, allowing the singlet to cross marginality while the charged $s\bar{s}$ scalar is still bounded to be irrelevant. 
When $\delta=1$, $\Delta_{s}$ is permitted to lie at the unitarity bound, and the allowed region is roughly comparable to the peninsula in Figure~\ref{fig:SU3_Pe-S_Pe-adjS_allowed-region_GFV-like-gaps_pd-11-15-19}.
In panel (b) of Figure~\ref{fig:JJJJ_SU3_Pe-S_Pe-adjS_Pe-SSb_delta-gaps_pd19} we consider instead the projection to the $(\Delta_{s}, \Delta_{s\bar{s}})$ plane. 

The SDP used to obtain Figures~\ref{fig:JJJJ_SU3_Pe-S_Pe-adjS_delta-gaps_min-compare_pd19} and~\ref{fig:JJJJ_SU3_Pe-S_Pe-adjS_Pe-SSb_delta-gaps_pd19} is reported in~\eqref{eq:sdp-gfv-delta-gaps}. 
This SDP is parametric in \( (\Delta_{s}, \Delta_{a}, \Delta_{s\bar{s}}, \alpha_{JJJ}, \delta) \). 
For a fixed value of \( \delta \), we used the Navigator method to minimise/maximise each scaling dimension, keeping the remaining parameters free. 
Namely, find the minimal and maximal values of either \( \Delta_{s} \), \( \Delta_{a} \) or \( \Delta_{s\bar{s}} \) for which the navigator function \( \mathcal{N}(\Delta_{s},\Delta_{a},\Delta_{s\bar{s}},\alpha_{JJJ}) = \max \alpha(\vec{\mathcal{V}}_{\id}) \) is negative.

To summarise, for both $SU(2)$ and $SU(3)$, the singlet operator is permitted to cross marginality at somewhat small values of $\delta$, roughly $\delta_{*} \sim 1/2$. 
As noted in the introduction, this does not guarantee that it will actually do so. 
By contrast, charged scalars require values of $\delta$ that are nearly twice as large. 
This indicates that there exist solutions to crossing in which a scalar singlet becomes marginal while the spectrum remains GFV (or YM)-like. 
In contrast, theories with a marginal charged operator lie much farther from such a regime. 
Our analysis at least suggests that a merger–annihilation scenario remains feasible even with a spectrum resembling the UV one, while Higgsing would necessitate more substantial corrections across a larger portion of the spectrum. 

We envision that pushing the numerics to higher $\Lambda$ would make the distinction between the two scenarios even more pronounced,\footnote{
    Since these are lower bounds, when increasing $\Lambda$, the value of $\delta$ where a scalar is allowed to cross marginality can only increase (or remain constant).}
and possibly start showing hints of a special point in CFT space in the form of a kink or a feature of some form.

\section{Conclusions and Outlook}

In this work, we have studied the constraints imposed by unitarity and crossing on the correlator of four non-Abelian conserved currents in three dimensions. 
Our spectral assumptions are compatible with the conformal theory induced on the boundary of AdS$_4$ by pure Yang-Mills theory with $SU(N)$ gauge group and Dirichlet boundary condition. Since the boundary CFT has a global symmetry, it is incompatible with colour confinement and should cease to exist in the flat space limit. 
Conversely, rigorously showing that no boundary condition with a conserved $SU(N)$ current survives in this limit would amount to solving the confinement problem.  
To this end, we tested the three currently available scenarios for confinement in AdS: (a) decoupling, (b) merging and annihilation, (c) Higgsing. 
We were able to exclude scenario (a) by showing that the two-point function of currents $\braket{JJ}$ is bounded to be strictly positive, hence $J$ cannot become null and decouple from the theory. 
We have not been able to sharply rule out scenarios (b) or (c), but our constraints strongly favour (b), since a scalar singlet operator is allowed to become marginal long before any charged operator does. 

Given that our results depend sensitively on the gaps imposed in the operator spectrum, in this paper, we erred on the side of caution, and our assumptions simply parametrise any continuous deformation of free gluons in AdS. 
While this makes our results fully rigorous and of interest beyond the context of QFT in AdS, better isolating Yang-Mills would be necessary to discriminate between scenarios (b) and (c), or at least to prove that the Dirichlet boundary condition does cease to exist beyond a finite radius.

The bootstrap excels at isolating theories with large gaps in the spectrum. 
While the requirement may be ambiguous, the prototypical example is the $3d$ Ising model.
The interacting equations of motion of the Wilson-Fisher fixed point~\cite{Rychkov:2015naa} justify the presence of a single relevant scalar in the $\mathbb{Z}_2$ odd sector, in contrast to the generalised free theory of a field with the same dimension. 
The role of the equations of motion is, in general, to render some operator redundant. These redundancy channels~\cite{Kousvos:2025ext} are sometimes able to isolate special solutions to crossing from a discrete or continuous family with otherwise similar spectra.
As remarked in Section~\ref{sec:scenarios}, a similar mechanism is at work in the Higgsing scenario, where the current multiplet recombines with a charged operator. 
It would be interesting to explore whether the related redundancy channels can help isolate (or rule out) the corresponding boundary CFT.

As for the Dirichlet boundary condition, its spectrum is a continuous deformation of a Generalised Free Vector, with no operator recombining perturbatively close to the free point. 
Therefore, redundancy channels are absent and cannot distinguish the interacting theory from free gluons or from a different deformation thereof. 
Our spectral assumptions do exclude generic $SU(N)$ gauge theories with matter, and in particular, crucially, QCD in the conformal window, whose boundary spectrum includes a protected displacement operator~\cite{Billo:2016cpy,Ciccone:2025dqx}. 
To do better, one option might be to inject the presence of AdS directly by including bulk operators in the bootstrap. 
A system of correlators, amenable to the bootstrap, involving bulk and boundary operators was proposed in~\cite{Meineri:2023mps} for $\mathrm{AdS}_{2}$ -- see also~\cite{Levine:2023ywq,Ghosh:2025sic} for related work. 
One of its advantages is a sum rule for the bulk UV central charge: a four-dimensional analogue would allow us to exclude any additional UV degree of freedom beyond the gluons. 

Other avenues to isolate Yang-Mills theory include the study of Wilson lines ending on the boundary, or, moving away from Dirichlet and onto Neumann boundary conditions, Wilson lines fully embedded in the three-dimensional boundary, as recently studied in~\cite{Gabai:2025hwf}. 
This would lead us into the territory of the bootstrap in defect CFT.

A related weakness of our setup is that the value of $g\ped{YM}$ (or equivalently $L \Lambda\ped{YM}$) was not a direct input. 
While $C_{J}$ acts naturally as a physical parametrisation of the bulk coupling, and its value can be injected into the bootstrap, a way to parametrise gap assumptions in terms of $C_{J}$ would still be needed. 
Of course, one can put an upper bound on the leading scaling dimension in each symmetry sector as a function of $C_{J}$, and simply assume that $C_{J}$ is monotonically decreasing with $L \Lambda\ped{YM}$.
While we have explored this strategy, it unfortunately does not discriminate between merging and Higgsing in a conclusive way, since either the singlet or a charged operator crosses marginality first, depending on the value of $\alpha_{JJJ}$. 

In Section~\ref{sec:bootstrap-dimensions-GFV}, we instead used parametric gap assumptions as an indirect proxy for the RG scale. 
However, this method can only bound $L \Lambda\ped{YM}$ from above, and for instance does not distinguish the interacting theory from the free UV limit, as is visible in the plots of that section. 
Tracking the value of an exactly marginal coupling through the bootstrap bound is not without precedent: in $\mathcal{N}=4$ SYM~\cite{Chester:2021aun,Chester:2023ehi} and $\mathcal{N}=2$ SQCD~\cite{Chester:2022sqb}, this was done with the help of constraints coming from localisation.
A different idea, tailored for QFT in AdS, but so far only implemented in $\mathrm{AdS}_{2}$, is to combine perturbation theory with the rigid structure of the OPE to derive a set of coupled differential equations which determines the dependence of the CFT data on the AdS radius~\cite{LoParco:2025Bootstrap}.  
An approach closer to the one we adopted in this paper would be, in the spirit of~\cite{Su:2022xnj}, to allow the leading operator in each sector to lie in an interval centred around the dimensions predicted by the Lorentzian Inversion Formula~\cite{Caron-Huot:2017vep}, which were computed to first order in the bulk coupling in~\cite{Caron-Huot:2021kjy}. 
Any method that permits injecting information about the coupling, or (part of) the spectrum of the interacting theory, would also alleviate the need to scan over $\delta$, thus making the problem much more numerically tractable.

Beyond the effort to isolate the boundary CFT corresponding to the Dirichlet boundary condition, it would be interesting to formulate bootstrap assumptions which place the putative $\mathrm{D}^\star$ boundary condition in the allowed region. 
An optimistic possibility is that the corresponding bound would reveal the presence of two kinks, associated with Dirichlet and $\mathrm{D}^\star$, which approach each other, merge and disappear. 
Hints of such a phenomenon have been observed in~\cite{Chester:2022hzt,Reehorst:2024vyq}.
Such a finding would go a long way towards a conclusive answer for the nature of the deconfinement transition. 
However, at the moment, too little is known about the nature of $\mathrm{D}^\star$ to turn this hope into a concrete strategy. 
For instance, it could be that $\mathrm{D}^\star$ does not exist as a unitary boundary condition at weak coupling, but only appears at $g\ped{YM}\sim 1$, before disappearing again together with D. 
Analytical studies of the possible boundary conditions at weak coupling, or explicit constructions of candidates for $\mathrm{D}^\star$, can help clarify what to hope for in future bootstrap analyses.

\section*{Acknowledgments}

We thank A.~Antunes, R.~Ciccone, V.~Gorbenko, P.~Kravchuk, J.~Penedones, M.~Reehorst and  N.~Su  for discussions. 
We are especially grateful to N.~Su for many valuable discussions on the numerical setup and on the code used in~\cite{He:2023ewx}.
SRK, MM, AP and MS thank the organisers of the conference Bootstrap 2024 at Complutense University of Madrid, where the idea of this project took form, for the hospitality. 
SRK, MM and AP thank the organisers of the conference Bootstrap 2025 at Instituto Principia in S\~{a}o Paulo for the hospitality and for providing a stimulating environment where part of this work was carried out. 
LDP, AP and  AV thank the organisers of the long-term workshop Progress of Theoretical Bootstrap at Yukawa Institute for Theoretical Physics (YITP-T-25-01) in Kyoto for the hospitality during the final stages of this work.
LDP, SRK, MM, AP and MS are partially supported by the INFN ``Iniziativa Specifica'' ST\&FI. MM is supported by the Italian Ministry of University and Research (MUR) under the FIS grant BootBeyond (CUP: D53C24005470001).
SRK has received support from the Marie Sk\l{}odowska-Curie Action (MSCA) High energy Intelligence (HORIZON-MSCA-2023-SE-01-101182937-HeI). 
SRK thanks LPENS and Perimeter Institute for their hospitality during extended stays.
The computations presented here were conducted in the SISSA HPC cluster Ulysses and the INFN Pisa HPC cluster Theocluster Zefiro. 

\appendix

\section{Higgsing transition in the Abelian Higgs model}\label{app:AHM}

We consider first the theory of a complex scalar field $\Phi$ in AdS, and later couple it to a dynamical gauge field.

We take the following action 
\begin{equation}
    S = \int\dif{x} \, \left(|\nabla_M \Phi|^2 + m^2 |\Phi|^2 +\frac{\lambda}{2} |\Phi|^4\right) \, ,
\end{equation}
where the integration is over AdS as in~\eqref{eq:AdS-integral}, and we use $M$ to denote an index in the tangent space. 
This action has two real parameters $m^2$ and $\lambda$, the latter assumed to be strictly positive to ensure global stability. 
The action is invariant under a $U(1)$ symmetry under which $\Phi$ has charge 1, namely the transformation is $\Phi\to e^{i \alpha} \Phi$ with parameter $\alpha \sim \alpha+2\pi$. 

As long as we can treat $\lambda$ as a small interaction, meaning that the classical analysis of the potential is reliable, we have the following vacuum structure for the theory as a function of $m^2$:
\begin{itemize}
    \item For $L^2 m^2 \geq L^2 m^2\ped{BF} = - \frac{d^2}{4}$~\cite{Breitenlohner:1982jf} there is a stable vacuum in the origin $\Phi=0$. 
    Note that this vacuum is a global minimum of the potential for $m^2 \geq 0$ but becomes a local maximum for $m^2\ped{BF} \leq m^2 <0$. 
    The fluctuations around this vacuum are simply given by the weakly interacting scalar field $\Phi$ of mass $m^2$, which admits a $U(1)$ invariant boundary condition: $\Phi(z,\vec{x})\underset{z\to 0}{\sim} z^{\Delta} \hat{\Phi}(\vec{x})$, with $\Delta(\Delta -d) = L^2 m^2$. 

    In the range $L^2 m^2\ped{BF} \leq L^2 m^2 \leq L^2 m^2\ped{BF}+1$ there are actually two choices of $U(1)$ invariant boundary conditions, because both solutions $\Delta_-$ and $\Delta_+ = d-\Delta_-\geq \Delta_-$ to the quadratic equation for $\Delta$ are above the unitarity bound.

    The boundary conformal theory is understood as a Generalised Free Scalar theory of the operator $\hat{\Phi}(\vec{x})$, which has scaling dimension $\Delta$ and charge 1 under the $U(1)$ symmetry, perturbed by the small interaction $\lambda$.
    Note that from the point of view of the boundary theory, the $U(1)$ symmetry is non-locally realised. 
    It is inherited from the bulk symmetry, but it does not have an associated local conserved current operator.

    \item For $m^2 < 0$, there is a circle of vacua outside of the origin at the local minimum of the potential, namely $|\Phi| = v \equiv \sqrt{-\lambda/m^2}$. 
    Expanding around one of the possible expectation values on the circle, say $v e^{i\theta}$, with $\theta \sim \theta + 2\pi$, we write 
    \begin{equation}\label{eq:expPhi}
        \Phi(z,\vec{x}) = \left(v + \frac{\rho(z,\vec{x})}{\sqrt{2}}\right)e^{i\left(\theta+ \frac{\pi(z,\vec{x})}{\sqrt{2}\,v}\right)} \, .
    \end{equation}
    The fluctuations are the radial mode $\rho$ of mass $m^2_\rho = 2(-m^2) \geq 0$ and the massless Goldstone boson $\pi(z,\vec{x})\sim \pi(z,\vec{x}) + 2 \pi\,\sqrt{2}\,v $. 
    The boundary conditions of these modes are $\rho(z,\vec{x}) \underset{z\to 0}{\sim} z^{\Delta_\rho}\hat{\rho}(\vec{x})$, with $\Delta_\rho(\Delta_\rho -d)= L^2 m^2_\rho$, and $\pi(z,\vec{x}) \underset{z\to 0}{\sim} z^d\hat{\pi}(\vec{x})$. 
    The resulting boundary condition for $\Phi$ is not invariant under a $U(1)$ transformation, because it approaches the constant expectation value $v e^{i\theta}$ at the boundary. 

    A boundary condition that breaks a bulk symmetry is the AdS version of spontaneous symmetry breaking. 
    The boundary theory contains the operator $\hat{\rho}$ of dimension $\Delta_\rho$ and the operator $\hat{\pi}$ of exactly marginal dimension $d$. 
    The latter dimension is a consequence of the bulk symmetry breaking~\cite{Copetti:2023sya} and therefore does not receive corrections in $\lambda$. 
    The operator $\hat{\pi}$ is referred to as ``tilt operator'', borrowing the terminology used in the context of defect and boundary CFT~\cite{Padayasi:2021sik}. 
    Deforming the action by a boundary term proportional to the tilt induces a shift in the constant $\theta$ in the boundary condition of $\Phi$, i.e.\ it moves the theory to a nearby point in the circle of vacua. 
    This means that the space of bulk vacua parametrised by $\theta$ corresponds to a continuous family of boundary conformal theories, parametrised by the exactly marginal coupling associated to the tilt~\cite{Carmi:2018qzm}.
\end{itemize}

It is useful to repackage the above information by listing the various options for the boundary conditions as a function of $m^2$, keeping $\lambda > 0$ fixed and small as before. 
\begin{itemize}
    \item For $L^2 m^2\geq0$, the only available boundary condition is the $U(1)$ preserving boundary condition with scaling dimension $\Delta_+$.
    The choice $\Delta_-$ is AdS invariant but leads to a ``real non-unitary'' theory.
    \item For $L^2 m^2\ped{BF}+1\leq L^2 m^2 < 0$, in addition to the option above that keeps existing, a continuous family (circle) of boundary conditions that breaks the $U(1)$ symmetry becomes available. 
    Note that this additional continuous set of boundary theories starts existing precisely when the lightest $U(1)$ charged operator $\hat{\Phi}$ in the boundary spectrum of the $U(1)$ preserving boundary condition crosses marginality, i.e.\ when we cross $m^2 = 0$. 
    Indeed, the new $U(1)$ breaking boundary condition can also be defined as the endpoint of the boundary RG flow obtained by perturbing the $U(1)$ preserving one with the relevant operator $\hat{\Phi}$. 
    The phase of the complex coupling associated with this deformation determines which vacuum is selected.
    \item For $L^2 m^2\ped{BF}\leq L^2 m^2 < L^2 m^2\ped{BF}+1$, in addition to the two options above, there is one more $U(1)$ preserving boundary condition, the one with dimension $\Delta_-$, which becomes unitary in this range of $m^2$. 
    Both the $\Delta_+$ and the $\Delta_-$ can be connected to the symmetry-breaking boundary condition by an RG flow triggered by the relevant charged operator $\hat{\Phi}$. 
    Moreover, they can be connected by an RG flow triggered by the singlet operator $|\hat{\Phi}|^2$, starting from the $\Delta_-$ boundary condition where this operator is relevant, and ending at the $\Delta_+$ boundary condition where it is irrelevant.
    \item For $L^2 m^2 < L^2 m^2\ped{BF}$ only the continuous family of symmetry-breaking boundary conditions is available. 
    Indeed, as we approach the limit $m^2  \to m^2\ped{BF}$ from above, the two available symmetry-preserving boundary conditions, with scaling dimensions $\Delta_-$ and $\Delta_+$, merge and annihilate with each other~\cite{Kaplan:2009kr, Hogervorst:2021spa}. 
    The singlet operator $|\hat{\Phi}|^2$ is marginal at the merging point. 
    Moving to more negative values of $m^2$, these boundary conditions can only exist as a pair of conjugate complex conformal theories, or as real boundary conditions that break AdS isometries.
\end{itemize}

\begin{figure}[t!]
    \centering
    \begin{subfigure}[b]{0.48\linewidth}
        \centering
        \begin{tikzpicture}[x={0.45\linewidth},y={0.45\linewidth}]
            \coordinate (BF) at (-0.4,0.85);
            \coordinate (BFp1) at (0.1,1.1);
            \coordinate (zero) at (0.5,0.6);
            \draw[-{Latex},line width=0.8pt] (-1,0) -- (1,0) node[below,yshift=-2] {\( L^{2} m^{2} \)};

            \draw[dashed, line width=0.5pt] (BF) -- (-0.4,0) node[below,yshift=-2] {\( L^{2} m^{2}\ped{BF} \)};
            \draw[dashed, line width=0.5pt] (BFp1) -- (0.1,0) node[below,yshift=-2] {\( L^{2} m^{2}\ped{BF}+1 \)};
            \draw[dashed, line width=0.5pt] (zero) -- (0.5,0) node[below,yshift=-2] {\( 0 \)};

            \draw[color=red!80!black,line width=1.2pt] (BF) to [out=90,in=-180] (BFp1) node[above left, xshift=-2ex] {\( \Delta_{-} \)};
            \draw[color=red!80!black,line width=1.2pt, dashed] (BFp1) -- (0.8,1.1);

            \draw[color=green!50!black,line width=1.2pt] (BF) to [out=-90,in=-180] (0.1,0.6) -- (zero) -- (0.9,0.6) node[above left] {\( \Delta_{+} \)} ;

            \draw[color=orange, line width=1.2pt] (zero) to [out=-120,in=0] (-0.9,0.1);
            \draw[color=orange, line width=1.2pt] (zero) to [out=-140,in=0] (-0.9,0.3) node[above, xshift=2ex] {\( U(1) \to 1 \)};

            \draw[color=orange, line width=1.2pt, dotted] (-0.7, 0.2) ellipse (0.05 and 0.1);
            \draw[color=orange, line width=1.2pt, dotted] (-0.3,0.2125) ellipse (0.045 and 0.09);
            \draw[color=orange, line width=1.2pt, dotted] (0,0.265) ellipse (0.04 and 0.075);
            \draw[color=orange, line width=1.2pt, dotted] (0.25,0.375) ellipse (0.03 and 0.055);

            \filldraw (BF) circle (0.015);
            \filldraw (BFp1) circle (0.015);
            \filldraw (zero) circle (0.015);

             \node[left,text width=3cm,align=right]  at (1,1.5) {flat space \\ unbroken phase};
            \node[right,text width=3cm,align=left]  at (-1,1.5) {flat space \\ broken phase};
        \end{tikzpicture}
        \caption{}
    \end{subfigure}
    \hfill
    \begin{subfigure}[b]{0.48\linewidth}
        \centering
        \begin{tikzpicture}[x={0.45\linewidth},y={0.45\linewidth}]
            \coordinate (BF) at (-0.4,0.85);
            \coordinate (BFp1) at (0.1,1.1);
            \coordinate (zero) at (0.5,0.6);
            \draw[-{Latex},line width=0.8pt] (-1,0) -- (1,0) node[below,yshift=-2] {\( L^{2} m^{2} \)};

            \draw[dashed, line width=0.5pt] (BF) -- (-0.4,0) node[below,yshift=-2] {\( L^{2} m^{2}\ped{BF} \)};
            \draw[dashed, line width=0.5pt] (BFp1) -- (0.1,0) node[below,yshift=-2] {\( L^{2} m^{2}\ped{BF}+1 \)};
            \draw[dashed, line width=0.5pt] (zero) -- (0.5,0) node[below,yshift=-2] {\( 0 \)};

            \draw[color=red!80!black,line width=1.2pt] (BF) to [out=90,in=-180] (BFp1) node[above left, xshift=-2ex] {\( \Delta_{-} \)};
            \draw[color=red!80!black,line width=1.2pt, dashed] (BFp1) -- (0.8,1.1);

            \draw[color=green!50!black,line width=1.2pt] (BF) to [out=-90,in=-180] (0.1,0.6) -- (zero) -- (0.9,0.6) node[above left] {\( \Delta_{+} \)} ;

            \draw[color=orange, line width=1.2pt] (zero) to [out=-130,in=0] (-0.9,0.2) node[above, xshift=2ex] {\( U(1) \to 1 \)};

            \filldraw (BF) circle (0.015);
            \filldraw (BFp1) circle (0.015);
            \filldraw (zero) circle (0.015);

            \node[left,text width=3cm,align=right]  at (1,1.5) {flat space \\ un-Higgsed phase};
            \node[right,text width=3cm,align=left]  at (-1,1.5) {flat space \\ Higgsed phase};
        \end{tikzpicture}
        \caption{}
    \end{subfigure}
    \caption{Illustration of the possible boundary conditions as a function of $L^{2} m^2$. Panel (a) refers to the theory of the complex scalar: at $m^2 = 0$, the boundary charged operator $\hat{\Phi}$ of the $\Delta_{+}$ boundary condition crosses marginality, and the theory develops a circle of symmetry-breaking boundary conditions. The symmetry-preserving boundary condition keeps existing past that point, until it annihilates with the $\Delta_{-}$ boundary condition. Upon gauging, as illustrated in panel (b), the circle of symmetry-breaking boundary conditions gets replaced by a single Higgsed boundary condition, in which the conservation of the boundary $U(1)$ current is violated. Note that the vertical axis does not represent a definite quantity; it is merely used to juxtapose the different options.\label{fig:abhig}
    }
\end{figure}
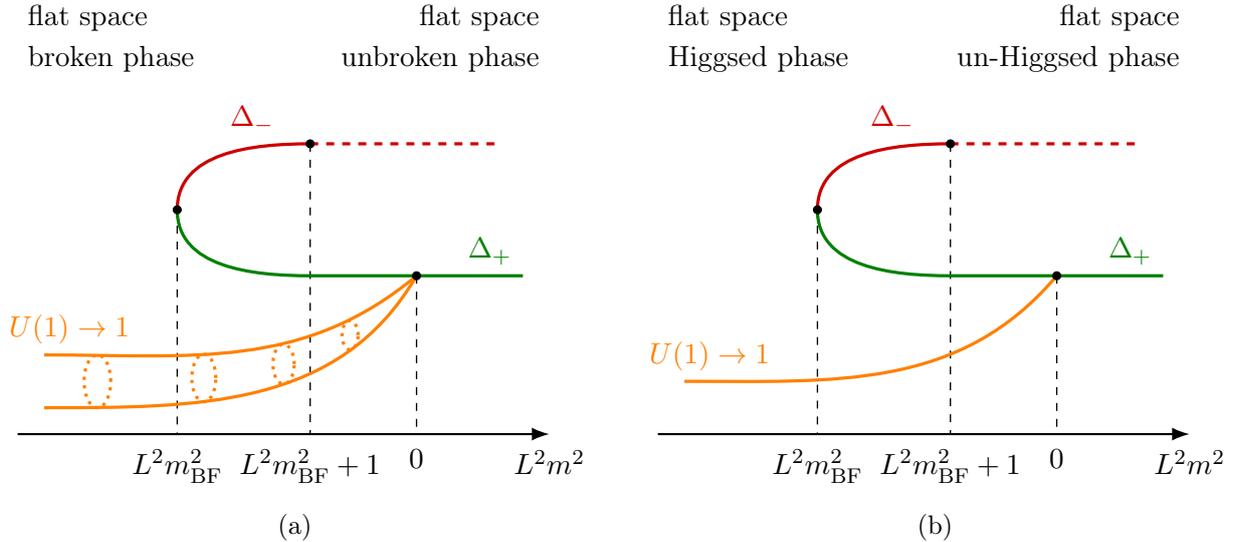

These boundary conditions are illustrated in panel (a) of Figure~\ref{fig:abhig}.

We now couple $\Phi$ to a dynamical $U(1)$ gauge field $A_M$, i.e.\ a Maxwell field, by promoting the derivative to a gauge covariant one $\nabla_M \Phi \to D_M \Phi\equiv (\nabla_M + i A_M)\Phi$ and adding to the action the kinetic term for the gauge field
\begin{equation}
    S \to S\big\vert_{\nabla \Phi\to D\Phi}+\int \dif{x} \, \frac{1}{4e^2} F^{MN} F_{MN} \, ,
\end{equation}
where $e^2$ is the gauge coupling and $F_{MN} = \nabla_M A_N - \nabla_N A_M$. 
In the interest of illustrating the Higgsing transition in AdS in this simpler example, let us focus on the Dirichlet boundary condition for the gauge field.
Neglecting for a moment the interaction with matter, this boundary condition gives a $U(1)$ conserved current operator $j^\mu$ of protected dimension $d-1$. 

In the range $m^2\geq m^2\ped{BF}$, by picking a $U(1)$ preserving boundary condition for the matter fields, the interaction with matter preserves the current conservation and forces us to identify the $U(1)$ symmetry associated to the boundary current with the $U(1)$ symmetry acting on the operators from the matter sector.
Therefore, the $U(1)$ symmetry under which $\hat{\Phi}$ is charged has become locally realised on the boundary as a consequence of the bulk gauging.

Let us now ask what happens in the range $m^2 < 0$ when we pick a symmetry-breaking boundary condition for the matter field. 
In the bulk, one can repeat the treatment familiar from flat space, by expanding the complex scalar field around its expectation value as in~\eqref{eq:expPhi}, and reach the same conclusion that the gauge field gains a positive mass-squared $m^2_A = 2 e^2 v^2$. This means that the current operator stops being conserved and it gains some positive anomalous dimension. 

Let us explain this phenomenon from the boundary point of view. 
As we already observed above in the discussion of the theory before gauging, the symmetry-breaking boundary condition becomes available precisely as we cross $m^2 = 0$, i.e.\ when the lightest charged operator $\hat{\Phi}$ of the symmetry-preserving boundary condition crosses marginality. 
We also remarked above that the symmetry-breaking boundary condition can be reached, starting from the symmetry-preserving one, by perturbing with the relevant operator $\hat{\Phi}$. This is analogous to the deformation~\eqref{explicitsymmetrybreaking} considered in the case of Yang-Mills.
As we turn on this deformation, since it is charged, the boundary current stops being conserved 
\begin{equation}
\partial_\mu j^\mu \propto e^{-i \theta}\hat{\Phi} + e^{i \theta} \hat{\Phi}^{*}~,
\end{equation}
where $\theta$ is the phase of the complex boundary coupling. The analogous equation for Yang-Mills is~\eqref{brokencurrent}.
This means that, as we flow to the IR and reach the new AdS invariant boundary condition, the operator $j^\mu$ is no longer conserved and can have a generic dimension. 
This boundary RG flow can be followed reliably if $m^2$ is negative but small. 

A complementary way to reach the Higgsed boundary condition is to start from the matter sector with a symmetry-breaking boundary condition and turn on a small gauge coupling $e$, treating it perturbatively.
Doing so, one finds at leading order $\partial_\mu j^\mu \propto e v L \,\hat{\pi}$, expressing that there is a recombination between the conserved current operator and the tilt operator. This shows again that the current stops being conserved and gets a non-zero anomalous dimension. 
This relation is also useful to clarify that, since the tilt operator $\hat{\pi}$  is replaced in the spectrum by a descendant of the current,
it cannot be used any more to give rise to a continuous family of boundary conditions: the circle of boundary theories collapses to a single point after gauging, see panel (b) of Figure~\ref{fig:abhig}.

Putting things together, we see that this example is a simple realisation of the Higgsing transition as we described for YM in AdS in Section~\ref{sec:scenarios}. 
The starting point is a family of symmetry-preserving boundary conditions, which has only irrelevant charged operators ($\Delta_+$ boundary condition for $m^2 > 0$ here, Dirichlet boundary condition at small coupling in Section~\ref{sec:scenarios}). 
At some point, as we move along this family (decreasing $m^2$ here, increasing $g\ped{YM}$ in Section~\ref{sec:scenarios}) a charged boundary operator crosses marginality (in the current example at $m^2 = 0$, in Section~\ref{sec:scenarios} possibly at $(L\Lambda\ped{YM})\ped{H}$). 
When this happens, a new line of Higgsed boundary conditions emerges, in which the current stops being conserved. 
We see clearly in the example discussed here that the symmetry-preserving boundary condition keeps existing past the point at which the Higgsed boundary condition emerges. 
When it stops existing, this is actually due to a merger and annihilation with a boundary condition with the same symmetry ($\Delta_+$ boundary condition here, $\mathrm{D}^{\star}$ in Section~\ref{sec:scenarios}).

\section{Spectrum of \texorpdfstring{\( SU(N) \)}{SU(N)} GFV}\label{app:characters-GFV}

In this appendix, we compute the low-lying spectrum of primary operators in the Mean Field Theory generated by a non-Abelian \( SU(N) \) conserved current in \( 3d \). 
We do this by using character techniques introduced in~\cite{Sundborg:1999ue,Aharony:2003sx}. 
This section provides an extension of the results presented in Appendix A of~\cite{Ciccone:2024guw}. 
In particular, we distinguish between parity-even and parity-odd sectors and show how to identify the intrinsic charge under charge conjugation for self-conjugate representations. 
The latter identification requires a somewhat more sophisticated formalism, which will be discussed in some detail. 
In Section~\ref{app:characters-parity} we discuss characters of the $3d$ conformal group in the presence of parity. 
In Section~\ref{app:characters-charge-conj} we discuss $SU(N)$ characters in the presence of charge conjugation. 
In Section~\ref{app:spectrum-general}, we finally review the general technique for obtaining the spectrum.
At the end of the appendix, in Table~\ref{tab:spectrum-SU2} and~\ref{tab:spectrum-SU3} we explicitly report the spectrum of the first primary operators for \( SU(2) \) and \( SU(3) \).

\subsection{Characters of \texorpdfstring{\( SO(4,1)_{0} \rtimes \Z_{2}^{\mathcal{P}} \)}{SO(4,1) rtimes Z2P}}\label{app:characters-parity}

The orthogonal group $O(d)$ has two disconnected components $O(d)_{+}$ and $O(d)_{-}$, the one connected to the identity being $SO(d)$. 
The other component can be realised as $O(d)_{-} = \{R P : R \in SO(d)\}$ where $P$ is an orthogonal matrix with $\det{P} = -1$. 
As a group, we have $O(d) \cong SO(d) \times \Z_{2}$ for $d$ odd and $O(d) \cong SO(d)\rtimes\Z_{2}$ for $d$ even, where $\Z_{2} = \{\id, P\}$. 
The choice of $P$ is arbitrary and gives rise to different realisations of $O(d)$, which are all isomorphic.\footnote{
    This is not the case for $Pin(d)$, the universal covering group of $O(n)$, which is known to have two inequivalent choices, $Pin_\pm(d)$. 
    Primary operators of the boundary theory considered here are, however, all bosonic and only in tensor representations; therefore, the two actions are equivalent.
}
A commonly used convention is to pick $P = P_{i}$ to be a reflection of the $i$-th coordinate in $\R^{d}$, since this works for any $O(d)$.\footnote{
    This convention is also useful when making contact with parity and time-reversal symmetry in Lorentzian QFTs.
} 
This is, however, not the most convenient choice for the present discussion, since, even in odd $d$, such $P_{i}$-s would not commute with all the $SO(d)$ generators.
In $d = 3$ (or in general for odd $d$), it is instead more convenient to choose a reflection of all the coordinates as a representative
\begin{equation}
    \mathcal{P} = P_{1} P_{2} P_{3} =
    \begin{pmatrix}
        -1 & 0 & 0 \\
        0 & -1 & 0 \\
        0 & 0 & -1
    \end{pmatrix}
    = -\id \, .
\end{equation}
This is related to $P_{i}$ by $ P_{i} = e^{i \pi M_{i}} \mathcal{P}$, where $M_{i}$ is the generator of rotations in the $i$-th direction of $\R^{3}$.

We define primary operators of the $SO(4,1)_{0}$ conformal group to transform under parity as in~\eqref{eq:parityDef}.
Irreducible representations of \( SO(4,1)_{0} \rtimes \Z_{2}^{\mathcal{P}} \) are labelled by \( (\Delta, \ell, p) \), where \( \Delta \) is the scaling dimension, \( \ell \) is the \( SO(3) \) spin, and \( p = \pm \) is the intrinsic parity.
Let us denote by $g\in \{\id, \mathcal{P}\}$ the two elements of $\Z_{2}^{\mathcal{P}}$. 
We define the character of ``long'' representations of \( SO(4,1)_{0} \rtimes \Z_{2}^{\mathcal{P}} \) as
\begin{equation}
    \chi_{\Delta,\ell,p}(q,x,g) = \tr\left[q^{D} x^{L} g\right] \, ,
\end{equation}
where the trace is over the \( (\Delta,\ell,p) \) representation, \( D \) is the generator of dilations and \( L \) is the Cartan generator of \( SO(3) \), \( q \) and \( x \) being the corresponding fugacities. 
These characters are given explicitly by
\begin{equation}\label{eq:character-conformal-parity}
    \begin{aligned}
        \chi_{\Delta,\ell,p}(q, x, \id) &= \tr\left[q^{D} x^{L} \id\right] = \frac{q^{\Delta} \chi_{\ell}(x)}{(1-q)(1-qx)(1-q/x)} \, , \\[1em]
        \chi_{\Delta,\ell,p}(q, x, \mathcal{P}) &= \tr\left[q^{D} x^{L} \mathcal{P}\right] = \frac{p (-1)^{\ell} q^{\Delta} \chi_{\ell}(x)}{(1+q)(1+qx)(1+q/x)} \,,
    \end{aligned}
\end{equation}
where \( \chi_{\ell}(x) = \sum_{m=-\ell}^{\ell} x^{m} \) is the \( SO(3) \) character for a spin $\ell$ representation. 
When \( g = \id \), the character in~\eqref{eq:character-conformal-parity} is the ordinary conformal character for a primary operator of spin $\ell$ in a $3d$ CFT.
When \( g = \mathcal{P} \), we obtain a ``twisted'' character. 
Its form is easily derived from the untwisted character by recalling that \( D \) and the $SO(3)$ subgroup commute with \( \mathcal{P} \), so the contribution to the trace of the $SO(3)$ highest-weight sub-module \( \ket{\Delta, \ell, p} \) is \( p (-1)^{\ell} q^{\Delta}\chi_{\ell}(x) \).
The contributions of the descendants \( P_{\mu_{1}}\cdots P_{\mu_{k}}\ket{\Delta, \ell, p} \) picks a \( (-1)^{k} \) factor compared to the ``untwisted'' case since \( \mathcal{P} P_{\mu} = -P_{\mu}\mathcal{P} \). 
This explains the form of $\chi_{\Delta,\ell,p}(q, x, \mathcal{P})$ in~\eqref{eq:character-conformal-parity}.

Characters of ``short'' representations for which \( \Delta \) saturates the unitarity bound, i.e.\ the identity, the free scalar, and conserved spin currents, are given by
\begin{equation}\label{eq:character-conformal-short-parity}
    \begin{aligned}
        \chi_{0,0,p}\ap{(short)}(q,x, g) &= 1 \, , \\
        \chi_{1/2,0,p}\ap{(short)}(q,x, g) &= \chi_{1/2,0,p}(q,x,g)-\chi_{5/2,0,p}(q,x,g) \, , \\
        \chi_{\ell+1,\ell,p}\ap{(short)}(q,x,g) &= \chi_{\ell+1,\ell,p}(q,x,g)-\chi_{\ell+2,\ell-1,p}(q,x,g) \, .
    \end{aligned}
\end{equation}

\subsection{Characters of \texorpdfstring{$PSU(N) \rtimes \Z_{2}^{\mathcal{C}}$}{PSU(N) rtimes Z2C}}\label{app:characters-charge-conj}

The determination of the characters of $\tilde{PSU}(N) =PSU(N) \rtimes \Z_{2}^{\mathcal{C}}$, where $\mathbb{Z}_{2}^{\mathcal{C}} = \{1, \mathcal{C}\}$, is more involved than that of \( SO(4,1)_{0} \rtimes \Z_{2}^{\mathcal{P}} \) because the action of ${\mathcal{C}}$ on $SU(N)$ cannot be ``trivialized'' as in the case for parity. 
The action of $\Z_{2}^{\mathcal{C}}$ on $PSU(N)$ is given by a map from $\mathbb{Z}_{2}$ to the automorphism group of $PSU(N)$, $\mathrm{Out}(PSU(N))$, isomorphic to $\mathbb{Z}_{2}$. 
The map $\varphi \colon \mathbb{Z}_{2} \to \mathrm{Out}(PSU(N))$ is given by $\varphi_{\id}(g) = g$, $\varphi_{\mathcal{C}}(g) = g^{*}$, with $g \in PSU(N)$ and ${}^{*}$ denotes complex conjugation. 
We consider here $N \geq 3$, since $\mathrm{Out}(PSU(2))$ is trivial. 

The action of charge conjugation on primary fields is given by~\eqref{eq:CCDef}. 
When the primary operator $\mathcal{O}$ is in a non-self-conjugate representation $\lambda\neq \lambda^*$, the intrinsic charge conjugation phase $\mathfrak{c}$ in~\eqref{eq:CCDef} can be set to one by a unitary transformation. 
In contrast, if $\lambda^{*} = \lambda$, we can induce two inequivalent representations $\lambda_{\pm}$ which are isomorphic to $\lambda$ as $PSU(N)$ representations and where $\mathcal{C}$ acts as $\pm C_{\lambda}$, where $C_{\lambda}$ is the intertwiner between the representation $R_{\lambda}(g)$, $g \in PSU(N)$, and the twisted representation $\tilde{R}_{\lambda}(g) = R_{\lambda}(\varphi(g))$  (i.e.\ if $\lambda^{*} = \lambda$, then there exists $C_{\lambda}$ such that $\tilde{R}_{\lambda}(g) = C_{\lambda} R_{\lambda}(g) {C_{\lambda}}^{-1}$). 
For example, for the adjoint representation charge conjugation may be represented by $ {(C_{\lambda})}^{ab} = (-1)^{N+1} C^{ab}$, with $C^{ab}$ as given in~\eqref{eq:Cab}, in the $(1,0,\ldots,0,1)_{+}$ representation.\footnote{
    The conventional choice for the $(-1)^{N+1}$ factor reflects the conventions used for the outer automorphism used in~\cite{Fuchs:1995zr}.
}

Complex conjugation acts on anti-hermitian Cartan generators as $H_{i} \to - H_{i}$, thus mapping weights to their opposite: $\mu\to -\mu$. The highest-weight state is then not mapped to itself but to the lowest-weight state.
It also does not preserve the canonical decomposition $\mathfrak{g} = \mathfrak{g}_{+}\oplus \mathfrak{h} \oplus \mathfrak{g}_{-}$, where $\mathfrak{h}$ is the Cartan subalgebra of $\mathfrak{g}$, but swaps positive and  negative roots: $\mathfrak{g}_{+} \leftrightarrow \mathfrak{g}_{-}$.
For these reasons, as for parity, it is convenient for the construction of characters to use another definition of charge conjugation. 
This is the natural definition at the level of the Lie algebra, and corresponds to the symmetry of the Dynkin diagram of $A_{N-1}$, reflecting the simple roots $\alpha_{i} \to \alpha_{N-i}$, with $i=1, \ldots, N-1$. 
We then have $\varphi(H_{i}) = H_{N-i}$ and on the step operators of the simple root $\varphi(E_{\pm \alpha_{i}}) = E_{\pm \alpha_{N-i}}$. 
The action on any other step operator is determined by $\varphi([x,y]) = [\varphi(x),\varphi(y)]$. The action on any weight of the Lie algebra follows, and is reported in~\eqref{eq:DynkinCharge} in the main text.

Note that the group extension of $\mathrm{Out}(SU(N))$ by $SU(N)$ is unique for odd $N$, while for even $N$ there are two inequivalent extensions, depending on the representative chosen for $\mathrm{Out}(SU(N))$, see the discussion e.g.\ in~\cite{Arias-Tamargo:2019jyh,Henning:2021ctv}. 
In contrast, for $PSU(N)$ there is a unique extension for any $N$, which is isomorphic to the one induced by complex conjugation or, equivalently, by the Lie algebra automorphism $\varphi$ that swaps the simple roots.
 
The group $\tilde{PSU}(N)$ has two disconnected components. 
We can define the character of an irreducible representation $\lambda_{(c, \pm)}$ by taking the trace of $(e^{h}, \id)\in \tilde{PSU}(N)$ or $(e^{h}, \mathcal{C})\in \tilde{PSU}(N)$ over the representation, with $h$ in the Cartan subalgebra of $\mathfrak{su}(N)$.

For $\lambda_{c}$ representations, $(e^{h}, \id)$ elements are represented as the direct sum $R_{\lambda}(e^{h}) \oplus R_{\lambda^{*}}(e^{h})$, while the elements $(e^{h},\mathcal{C})$ are off-diagonal matrices. We thus have
\begin{equation}
    \begin{aligned}
        \chi_{\lambda_{c}}(e^{h}, \id) &= \chi_{\lambda}(e^{h}) + \chi_{\lambda^{*}}(e^{h}) \, , \\
        \chi_{\lambda_{c}}(e^{h}, \mathcal{C}) &= 0 \, ,
    \end{aligned}
\end{equation}
where $\chi_{\lambda}(e^{h})= {\rm tr}\, R_\lambda(e^h)$ is the usual character of $SU(N)$ representations, see later~\eqref{eq:character-determinant} and~\eqref{eq:character-matrix-ABC}.

For $\lambda_{\pm}$ representations, $(e^{h},\id)$ elements are represented as $R_{\lambda}(e^{h})$, while $(e^{h},\mathcal{C})$ are represented by $\pm C_{\lambda} R_{\lambda}(e^{h})$. Hence
\begin{equation}
    \begin{aligned}
        \chi_{\lambda_{\pm}}(e^{h}, \id) &= \chi_{\lambda}(e^{h}) \, , \\
        \chi_{\lambda_{\pm}}(e^{h}, \mathcal{C}) &= \pm \tr\!\Big[C_{\lambda} R_{\lambda}(e^{h})\Big] = \pm \chi_{\lambda}^{\varphi}(e^{h})\, .
    \end{aligned}
\end{equation}

The insertion of $C_{\lambda}$ in the trace defines a so-called ``twisted'' or ``twining'' character $\chi_{\lambda}^{\varphi}$.
The computation of this character requires an in-depth analysis of how the outer automorphisms of the $\mathfrak{su}(N)$ algebra act on the irreducible representations, and was worked out in~\cite{Fuchs:1995zr} for general outer automorphisms of Kac-Moody Lie algebras.\footnote{
    See also~\cite{Henning:2017fpj,Graf:2020yxt} for some recent literature that uses these characters.
}
The key result of~\cite{Fuchs:1995zr} that we need here is that the twining character $\chi^{\varphi}$ for a Lie algebra $\mathfrak{g}$ is identical to the ordinary character $\breve{\chi}$ of the ``orbit Lie algebra'' $\breve{\mathfrak{g}}$:
\begin{equation}\label{eq:twiningChara}
    \chi_{\lambda}^{\varphi}(e^{h}) = \breve{\chi}_{\breve{\lambda}}(e^{\breve{h}}) \, .
\end{equation}
Note that this relation holds for characters of infinite-dimensional Verma modules as well as for finite-dimensional representations (see the Example later, where the identity can be easily worked out for $\mathfrak{su}(3)$ Verma modules).

In the case of $\mathfrak{g} = A_{N-1} = \mathfrak{su}(N)$ Lie algebra, the orbit Lie algebra is
\begin{equation}
    \breve{\mathfrak{g}} = 
    \begin{cases}
        C_{n} = \mathfrak{sp}(2n) \,,& N = 2n+1\,, \\
        B_{n+1} = \mathfrak{so}(2n+3)\,,  & N = 2n+2\,.
    \end{cases}
\end{equation}
The difference between even and odd $N$ arises from the fact that for odd $N$ the action of $\varphi$ is free, while it is not for even $N$, as one simple root -- the one in the middle of the Dynkin diagram -- is left invariant.

The Dynkin labels $\lambda$ and $\breve{\lambda}$, and the Cartans $h$ and $\breve{h}$ appearing in~\eqref{eq:twiningChara} are related as follows:
\begin{equation}
    \begin{aligned}
        N = 2n+1 \ \colon \ & \qquad
        \begin{aligned}
        \lambda &= (\lambda_{1},\ldots,\lambda_{n},\lambda_{n},\ldots,\lambda_{1}) \, , \\
        h &= x_{1} H_{1} + \cdots + x_{2n} H_{2n} \, , \\[0.3em]
        \breve{\lambda} &= (\lambda_{1},\ldots,\lambda_{n}) \, ,\\
        \breve{h} &= (x_{1}+x_{2n}) \breve{H}_{1} + \cdots + (x_{n}+x_{n+1}) \breve{H}_{n} \, ;
        \end{aligned}
        \\[0.5em]
        N = 2n+2 \ \colon \ & \qquad
        \begin{aligned}
        \lambda &= (\lambda_{1},\ldots,\lambda_{n},\lambda_{n+1},\lambda_{n},\ldots,\lambda_{1}) \, , \\
        h &= x_{1} H_{1} + \cdots + x_{2n+1} H_{2n+1}  \, , \\[0.3em]
        \breve{\lambda} &= (\lambda_{1},\ldots,\lambda_{n},\lambda_{n+1}) \, ,  \\
        \breve{h} &= (x_{1}+x_{2n}) \breve{H}_{1} + \cdots + (x_{n}+x_{n+2}) \breve{H}_{n} + x_{n+1} \breve{H}_{n+1} \, ;
        \end{aligned}
    \end{aligned}
\end{equation}
where $H_{i}$ and $\breve{H}_{i}$ are the Cartan generators of $\mathfrak{g}$ and $\breve{\mathfrak{g}}$ in the Cartan-Weyl basis, respectively.

The characters for any representation of the classical Lie algebras $A_n$, $B_n$, $C_n$ are well-known and can be written by a ratio of determinants as follows:
\begin{equation}\label{eq:character-determinant}
    \chi_{\lambda}(y) = \frac{\det M(p, y)}{\det M(0, y)} \, ,
\end{equation}
where (see e.g.~\cite{Fulton:2004uyc})
\begin{equation}\label{eq:character-matrix-ABC}
   \begin{aligned}
        A_{n} \ \colon &&
         M(p, y)_{ij} &= y_{j}^{p_{i}+n+1-i} \, , \qquad i,j = 1,\ldots, n+1 \, , \\
        && p &= (\lambda_{1}+\cdots+\lambda_{n}, \lambda_{2}+\cdots+\lambda_{n},\ldots,\lambda_{n}, 0) \, , \\
        && y_{n+1} &= 1/(y_{1}\cdots y_{n}) \, ;
        \\[1em]
         B_{n} \ \colon  &&
         M(p, y)_{ij} &= y_{j}^{p_{i} + n + \frac{1}{2}-i} - y_{j}^{-\big(p_{i} + n+\frac{1}{2}-i\big)} \, , \qquad i,j = 1,\ldots,n  \, , \\
        && p &= (\lambda_{1}+\cdots+\lambda_{n-1}+\lambda_{n}/2,\lambda_{2}+\cdots+\lambda_{n-1}+\lambda_{n}/2, \lambda_{n}/2) \, ;
        \\[1em]
               C_{n} \ \colon  &&
        M(p, y)_{ij} &= y_{j}^{p_{i} + n+1-i} - y_{j}^{-(p_{i} + n+1-i)} \, , \qquad i,j = 1,\ldots,n \, , \\
        && p &= (\lambda_{1}+\cdots+\lambda_{n},\lambda_{2}+\cdots+\lambda_{n}, \ldots, \lambda_{n}) \, .
   \end{aligned}
\end{equation}
Here $e^{h} = y_{1}^{L_{1}} \cdots y_{n}^{L_{n}}$ are the fugacities in a orthonormal basis, and $p_i$ are the weights of the highest-weight state in this basis: $L_i |\lambda \rangle = p_i |\lambda\rangle$. 
These are related to the fugacities in the Cartan-Weyl basis $e^{h} = z_{1}^{H_{1}} \cdots z_{n}^{H_{n}}$, $z_i = e^{x_i}$, where $H_i |\lambda \rangle = \lambda_i |\lambda\rangle$, by
\begin{equation}\label{eq:coroot-to-dual-orthonormal}
   \begin{aligned}
        A_{n} \ \colon \ & y_{1} = z_{1} \, , \quad y_{i} = \frac{z_{i}}{z_{i-1}}\, , \ (i = 2,\ldots,n) \\
        B_{n} \ \colon \ & y_{1} = z_{1} \, , \quad y_{i} = \frac{z_i}{z_{i-1}} \, , \ (i = 2,\ldots,n-1) \, , \quad y_{n} = \frac{z_{n}^2}{z_{n-1}} \, , \\
        C_{n} \ \colon \ & y_{1} = z_{1} \, , \quad y_{i} = \frac{z_{i}}{z_{i-1}}\, , \ (i = 2,\ldots,n) \, , 
   \end{aligned}
\end{equation}

Summarising, we have the following character formulas
\begin{equation}\label{eq:SUNZ2C-characters}
    \begin{aligned}
        \chi_{\lambda_{c}}(z, \id) &= \chi_{\lambda}(z) + \chi_{\lambda^*}(z) \,, \\
        \chi_{\lambda_{c}}(z, \mathcal{C}) &= 0\,,
    \end{aligned}
\end{equation}
and
\begin{equation}
    \begin{aligned}
        \chi_{\lambda_{\pm}}(z, \id) &= \chi_{\lambda}(z)\,, \\[0.3em]
        \chi_{\lambda_{\pm}}(z, \mathcal{C}) &= 
        \begin{cases}
            \pm\chi_{\breve{\lambda}}^{(C_{n})}(\breve{z})\,, & N = 2n+1 \,, \\[0.5em]
            \pm\chi_{\breve{\lambda}}^{(B_{n+1})}(\breve{z})\,,  & N = 2n+2 \,,
        \end{cases}
    \end{aligned}
\end{equation}
where
\begin{equation}
    \begin{aligned}
        N = 2n+1 \ \colon \ & \breve{z} = (z_{1} z_{2n}, z_{2} z_{2n-1},\ldots, z_{n} z_{n+1})\,, \\
        N = 2n+2 \ \colon \ & \breve{z} = (z_{1} z_{2n+1}, z_{2} z_{2n},\ldots, z_{n} z_{n+2}, z_{n+1})\,.
    \end{aligned}
\end{equation}

The intertwiner $C_{\lambda}$ can be explicitly constructed by the action on the universal enveloping algebra $U(\mathfrak{g})$, and thus on any representation, by the Poincaré-Brikhoff-Witt (PBW) theorem. 
Namely, we recall that given a basis $\{F_{1},\ldots, F_{r}\}$ for $\mathfrak{g}_{-} $, the Verma module $V_{\lambda}$ of highest-weight state $v_{\lambda}$ is spanned by the ordered monomials in $U(\mathfrak{g}_{-})$
\begin{equation}
    V_{\lambda}\cong \mathrm{Span}\{F_{1}^{n_{1}}\cdots F_{r}^{n_{r}} v_{\lambda} \ \colon  H_{i} v_{\lambda} = \lambda_{i} v_{\lambda} , \ U(\mathfrak{g}_{+}) v_{\lambda} = 0 \} \, .
\end{equation}
The intertwiner in the PBW basis is then
\begin{equation}\label{eq:Clambda-varphi-pbw}
    C_{\lambda} F_{1}^{n_{1}}\cdots F_{r}^{n_{r}} v_{\lambda} = \varphi(F_{1})^{n_{1}}\cdots \varphi(F_{r})^{n_{r}} v_{\lambda} \, ,
\end{equation}
Since $\varphi$ is an algebra automorphism, this can always be rewritten as a linear combination of elements of a basis of $U(\mathfrak{g}_{-})$.
A similar construction applies to finite-dimensional irreducible representations, which are quotients of Verma modules.

\paragraph{Example} 
We consider the case of infinite-dimensional Verma modules of $SU(3)$. The Cartan matrix is
\begin{equation}
    A = 
    \begin{pmatrix}
        2 & -1 \\
        -1 & 2
    \end{pmatrix} \, .
\end{equation}
The simple roots are $\{\alpha_{1}, \alpha_{2}\}$, the positive roots are $\Delta_{+} = \{\alpha_{1},\alpha_{2},\alpha_{1}+\alpha_{2}\}$. 
In the basis of fundamental weights $\{\omega_1,\omega_2\}$ we have $\alpha_1 = 2 \omega_1 - \omega_2$, $\alpha_2 = - \omega_1 + 2 \omega_2$ and $\alpha_1 + \alpha_2 = \omega_1 + \omega_2$. 
The Cartan-Weyl basis of $\mathfrak{su}(3)$ is
\begin{equation}
    \begin{gathered}
        \{H_{1}, H_{2}\}, \, \qquad
        \{E_{1} = E_{\alpha_{1}}, E_{2} = E_{\alpha_{2}}, E_{3} = E_{\alpha_{1}+\alpha_{2}} = [E_{\alpha_{1}}, E_{\alpha_{2}}]\}  \, , \\
        \{F_{1} = E_{-\alpha_{1}}, F_{2} = E_{-\alpha_{2}}, F_{3} = E_{-\alpha_{1}-\alpha_{2}} =- [E_{-\alpha_{2}}, E_{-\alpha_{1}}]\} \, ,
    \end{gathered}
\end{equation}
which span $\mathfrak{h}$, $\mathfrak{g}_{+}$ and $\mathfrak{g}_{-}$, respectively.
The outer automorphism acts by swapping the two simple roots $\alpha_{1} \leftrightarrow \alpha_{2}$. 
On the Cartan-Weyl basis
\begin{equation}
    \begin{gathered}
        \varphi(H_{1}) = H_{2} \, , \qquad \varphi(H_{2}) = H_{1} \, , \\
        \varphi(E_{1}) = E_{2} \, , \qquad \varphi(E_{2}) = E_{1}  \, , \qquad \varphi(E_{3}) = - E_{3} \, , \\
        \varphi(F_{1}) = F_{2} \, , \qquad \varphi(F_{2}) = F_{1} \, , \qquad \varphi(F_{3}) = - F_{3} \, .
    \end{gathered}
\end{equation}
Let $v_{\lambda}$ be a highest-weight state with Dynkin labels \ $(\lambda_1,\lambda_2)$. 
The PBW basis of the Verma module is given by $F_{1}^{n_{1}}F_{2}^{n_{2}} F_{3}^{n_{3}}v_{\lambda}$ with $n_{i}$ integers.
The group element $e^{h}$, with $h$ in the Cartan sub-algebra $h \in \mathfrak{h}$, acts diagonally as
\begin{equation}
    e^{h} F_{1}^{n_{1}}F_{2}^{n_{2}} F_{3}^{n_{3}}v_{\lambda} = e^{\mu(h)} F_{1}^{n_{1}}F_{2}^{n_{2}} F_{3}^{n_{3}}v_{\lambda} \qquad
    \mu = \lambda - n_1 \alpha_1 - n_2 \alpha_2 - n_3 (\alpha_1+\alpha_2) \, ,
\end{equation}
Recall that if $h = x_{1} H_{1} + x_{2} H_{2}$ and $\mu = \mu_1 \omega_1 + \mu_2 \omega_2$, then $\mu(h) = x_1 \mu_1 + x_2 \mu_2$. 
From this, we recover the well-known expression for Verma module characters
\begin{equation}
    \begin{aligned}
        \mathcal{V}_{\lambda}(e^{h})
        = \sum_{n_1, n_2, n_3=0}^{+\infty} e^{\lambda(h)} e^{-n_1\alpha_1(h)}e^{-n_2\alpha_2(h)}e^{-n_3(\alpha_1+\alpha_2)(h)} 
        = e^{\lambda(h)} \prod_{\alpha \in \Delta_{+}} \frac{1}{1-e^{-\alpha(h)}} \, .
    \end{aligned}
\end{equation}
For a symmetric weight $\lambda = \lambda_1\omega_1+\lambda_1\omega_2$, the action of $C_\lambda$ on the Verma module is given in~\eqref{eq:Clambda-varphi-pbw} and in this case reads
\begin{equation}
    C_\lambda F_{1}^{n_{1}}F_{2}^{n_{2}} F_{3}^{n_{3}}v_{\lambda} 
    = (-1)^{n_{3}} F_{1}^{n_{2}}F_{2}^{n_{1}} F_{3}^{n_{3}}v_{\lambda} 
    + \sum_{k=0}^{\min\{n_1,n_2\}} c_{k} F_{1}^{n_{2}-k}F_{2}^{n_{1}-k} F_{3}^{n_{3}+k}v_{\lambda} \, ,
\end{equation}
for some coefficients $c_{k}$ that are determined by the commutation relations of the $\mathfrak{su}(3)$ algebra, but whose precise form is not important here. 
Thus, only the states with $n_{1} = n_{2}$ can contribute to the twisted character, and those contribute with $(-1)^{n_3} e^{\mu(h)}$. 
We then have
\begin{equation}
    \mathcal{V}_{\lambda}^{\varphi}(e^{h}) 
    = \tr\!\Big[C_\lambda R_\lambda(e^{h})\Big] 
    = \!\!  \sum_{n_1, n_3=0}^{+\infty} (-1)^{n_3}e^{\lambda(h)} e^{-n_1(\alpha_1 + \alpha_2)(h)}e^{-n_3(\alpha_1+\alpha_2)(h)} 
    = \frac{e^{\lambda(h)}}{1-e^{-(2\alpha_1+2\alpha_2)(h)}} \, .
\end{equation}
Writing $h = x_1 H_1 + x_2 H_2$ and the weights in the fundamental basis, this is
\begin{equation}
    \mathcal{V}_{\lambda}^{\varphi}(e^{h}) = \frac{e^{(x_1+x_2)\lambda_1}}{1-e^{-2(x_1+x_2)}} \,,
\end{equation}
which coincides with the character of the $C_{1} = \mathfrak{sp}(2) = \mathfrak{su}(2)$ Verma module
\begin{equation}
    \breve{\mathcal{V}}_{\breve{\lambda}}(e^{\breve{h}}) = \tr\!\Big[R_{\breve{\lambda}}(e^{\breve{h}})\Big] = \frac{e^{\breve{\lambda}(\breve{h})}}{1-e^{-\breve{\alpha}(\breve{h})}} = \frac{e^{\breve{x}_1 \breve{\lambda}_1}}{1-e^{-2\breve{x}_1}}\,, \qquad \breve{\lambda} = \breve{\lambda}_1\breve{\omega}_1 \,, \qquad \breve{h} = \breve{x}_1 \breve{H}_1 \,,
\end{equation}
upon identifying $\breve{\lambda}_{1} = \lambda_{1}$ and $\breve{x}_{1} = x_1 + x_2$. 

\subsection{Technique}\label{app:spectrum-general}

In Section~\ref{app:characters-parity} and~\ref{app:characters-charge-conj} we presented the characters of the $3d$ conformal group in the presence of parity \( SO(4,1)_{0} \rtimes \Z_{2}^{\mathcal{P}} \) and of $PSU(N)\rtimes \Z_{2}^{\mathcal{C}}$, respectively. 
Characters for representations of~\eqref{eq:YM-AdS-symmetry} can be easily written down with the knowledge of these two ingredients. 
For a representation $(\Delta, \ell, p, \lambda_{(c,\pm)})$ these are
\begin{equation} \label{eq:character-conformal-parity-SUn}
    \chi_{\Delta,\ell,p,\lambda_{(c,\pm)}}(q,x,y,g_{\mathcal{P}},g_{\mathcal{C}}) =
    \begin{cases}
        \chi_{\Delta,\ell,p}\ap{(short)}(q,x,g_{\mathcal{P}}) \chi_{\lambda_{(c,\pm)}}(y,g_{\mathcal{C}}) & \text{if } (\Delta,\ell) = (0,0), (1/2,0), (\ell+1,\ell) \, , \\
        \chi_{\Delta,\ell,p}(q,x,g_{\mathcal{P}}) \chi_{\lambda_{(c,\pm)}}(y,g_{\mathcal{C}}) & \text{otherwise} \, ,
    \end{cases}
\end{equation}
where $g_{\mathcal{P}} \in \{\id, \mathcal{P}\}$, $g_{\mathcal{C}} \in \{\id, \mathcal{C}\}$, \( \chi_{\Delta,\ell,p} \) and \( \chi_{\Delta,\ell,p}^{(\text{short})} \) are given by~\eqref{eq:character-conformal-parity} and~\eqref{eq:character-conformal-short-parity}, and $\chi_{\lambda_{(c,\pm)}}$ is given by~\eqref{eq:SUNZ2C-characters}. 
We denote by $(q, x)$ the fugacities for the $SO(4,1)_{0}$ Cartan generators, while $y = (y_1,\ldots,y_{N-1})$ are the fugacities for the $\mathfrak{su}(N)$ Cartan subalgebra in the orthonormal basis~\eqref{eq:coroot-to-dual-orthonormal}.

The generating function for the spectrum of a free bosonic field in a representation $R$ of a group $G$ is given by the generating function of symmetric products
\begin{equation}\label{eq:generating-function}
    Z_{R}(\eta, g) = \sum_{n=0}^{+\infty} \eta^{n} \chi_{\mathrm{Sym}^{n}R}(g) = \frac{1}{\det(1-\eta R(g))} = \exp\left(\sum_{n=1}^{+\infty} \frac{\eta^{n}}{n} \chi_{R}(g^{n})\right) \, , \qquad g \in G \, ,
\end{equation}
where $\eta$ is a fugacity for the particle number.
In the GFV case we use $R = (2,1,+, \mathbf{adj}_{\pm}) \equiv J$ for $N$ even/odd, and
\begin{equation}\label{eq:gDEF}
    g = \Big((q^{D}x^{L}, g_{\mathcal{P}}), (z^H, g_{\mathcal{C}})\Big) \in G = \Big(SO(4,1)_{0}\rtimes \Z_{2}^{\mathcal{P}}\Big) \times \Big(PSU(N) \rtimes \Z_{2}^{\mathcal{C}}\Big) \, ,
\end{equation}
where $z^H = \prod_{i=1}^{N-1} z_i^{H_i}$. 
Taking powers of $g$ requires some care because of the semi-direct product. 
$\mathcal{P}$ commutes with $D$ and $L$, so $(q^{D}x^{L}, g_{\mathcal{P}})^{n} = (q^{n D}x^{n L}, g_{\mathcal{P}}^n)$, while for $(z^{H}, g_{\mathcal{C}})^{n}$ we have to distinguish between $n$ even and $n$ odd when $g_{\mathcal{C}} = \mathcal{C}$. 
The key identity is
\begin{equation}
    (z^{H}, \mathcal{C})^{2} = (z^{H} z^{\varphi(H)}, \id) \equiv (\tilde z^{H}, \id) \, ,
\end{equation}
where
\begin{equation}
    \begin{aligned}
        N = 2n+1 \ \colon \ & \tilde{z} = (z_{1} z_{2n}, z_{2} z_{2n-1},\ldots, z_{2n} z_{1}) \, , \\
        N = 2n+2 \ \colon \ & \tilde{z} = (z_{1} z_{2n}, z_{2} z_{2n-1},\ldots,z_{n+1}^{2}, \ldots ,z_{2n} z_{1}) \, .
    \end{aligned}
\end{equation}
We then have
\begin{equation}
    \begin{aligned}
        (z^{H}, \mathcal{C})^{2m} &= (\tilde{z}^{\, m H}, \id) \, , \\
        (z^{H}, \mathcal{C})^{2m+1} &= (z^{H}\, \tilde{z}^{\, mH} , \mathcal{C}) \,,
    \end{aligned}
\end{equation}
which can be mapped back to the $y$-fugacities using~\eqref{eq:coroot-to-dual-orthonormal}. 
We can now explicitly determine the form of the characters $\chi_J(g^n)$ appearing in~\eqref{eq:generating-function} in terms of the character of a single current $J$, which will be denoted by $z_{J}(q, x, y, g_{\mathcal{P}},g_{\mathcal{C}})$.
To avoid cluttered notation, let us split the element $g$ in~\eqref{eq:gDEF} into four pieces, and define
\begin{equation}
    \begin{aligned}
        g_{++} &= ((q^D x^L,\id), (z^{H}, \id)) \, , \quad 
        && g_{+-} &= ((q^D x^L,\id), (z^{H}, \mathcal{C})) \, , \\
        g_{-+} &= ((q^D x^L,\mathcal{P}), (z^{H}, \id)) \, , \quad 
        &&  g_{--} &= ((q^D x^L,\mathcal{P}), (z^{H}, \mathcal{C})) \, .
    \end{aligned} 
\end{equation}
We then have
\begin{equation}\label{eq:character-gn}
    \begin{aligned}
        \chi_{J}(g_{++}^n) &= z_{J}(q^{n}, x^{n}, y^{n}, \id, \id) \, , \\
        \chi_{J}(g_{-+}^{2m}) &= z_{J}(q^{2m}, x^{2m}, y^{2m}, \id, \id) \, , \qquad 
        \chi_{J}(g_{-+}^{2m+1} )  = z_{J}(q^{2m+1}, x^{2m+1}, y^{2m+1}, \mathcal{P}, \id) \, , \\
        \chi_{J}(g_{+-}^{2m} ) &= z_{J}(q^{2m}, x^{2m}, \tilde{y}^{\,m}, \id, \id) \, , \qquad
        \chi_{J}(g_{+-}^{2m+1}  ) = z_{J}(q^{2m+1}, x^{2m+1}, y \tilde{y}^{\,m}, \id, \mathcal{C}) \, , \\
        \chi_{J}(g_{--}^{2m}) &= z_{J}(q^{2m}, x^{2m}, \tilde{y}^{\, m}, \id, \id) \, , \qquad
        \chi_{J}(g_{--}^{2m+1} ) = z_{J}(q^{2m+1}, x^{2m+1}, y {\tilde{y}}^{\, m}, \mathcal{P}, \mathcal{C}) \, ,
    \end{aligned}
\end{equation}
where $y^{n} = (y_1^{n},\ldots, y_{N-1}^{n})$, and similarly for $\tilde{y}^{\, m}$ and $y \tilde{y}^{\, m}$.

We then define a vector for generating functions on each connected component of the group 
\begin{equation}\label{eq:ZJdef}
    \vec{Z}_{J}(\eta, q, x, y) = 
    \begin{pmatrix}
        Z_{J}(\eta, q, x, y, \id, \id) \\
        Z_{J}(\eta, q, x, y, \mathcal{P}, \id) \\
        Z_{J}(\eta, q, x, y, \id, \mathcal{C}) \\
        Z_{J}(\eta, q, x, y, \mathcal{P}, \mathcal{C})
    \end{pmatrix} \, .
\end{equation}
Each entry is computed by combining~\eqref{eq:generating-function} with~\eqref{eq:character-gn}:
\begin{equation}
    \resizebox{0.9\linewidth}{!}{\(
    \begin{aligned}
        Z_{J}(\eta, q, x, y, \id, \id) 
        &= \exp\left(\sum_{n=1}^{+\infty} \frac{\eta^{n}}{n} z_{J}(q^{n}, x^{n}, y^{n}, \id, \id)\right) \, , \\[0.5em]
        Z_{J}(\eta, q, x, y, \mathcal{P}, \id) 
        &= \exp\left(\sum_{m=0}^{+\infty} \frac{\eta^{2m+1}}{2m+1} z_{J}(q^{2m+1}, x^{2m+1}, y^{2m+1}, \mathcal{P}, \id) + \sum_{m=2}^{+\infty} \frac{\eta^{2m}}{2m} z_{J}(q^{2m}, x^{2m}, y^{2m}, \id, \id)\right) \, , \\[0.5em]
        Z_{J}(\eta, q, x, y,\id, \mathcal{C}) 
        &= \exp\left(\sum_{m=0}^{+\infty} \frac{\eta^{2m+1}}{2m+1} z_{J}(q^{2m+1}, x^{2m+1}, y \tilde{y}^{\, m}, \id, \mathcal{C}) + \sum_{m=2}^{+\infty} \frac{\eta^{2m}}{2m} z_{J}(q^{2m}, x^{2m}, \tilde{y}^{\, m}, \id, \id)\right) \, , \\[0.5em]
        Z_{J}(\eta, q, x, y,\mathcal{P}, \mathcal{C}) 
        &= \exp\left(\sum_{m=0}^{+\infty} \frac{\eta^{2m+1}}{2m+1} z_{J}(q^{2m+1}, x^{2m+1}, y \tilde{y}^{\, m}, \mathcal{P}, \mathcal{C}) + \sum_{m=2}^{+\infty} \frac{\eta^{2m}}{2m} z_{J}(q^{2m}, x^{2m}, \tilde{y}^{\, m}, \id, \id)\right) \, . \\[0.5em]
    \end{aligned}
    \)}
\end{equation}
We decompose~\eqref{eq:ZJdef} as a sum of characters of irreducible representations, 
\begin{equation}
    \vec{Z}_{J}(\eta, q, x, y) = \sum_{R,n} d_{R,n} \, \eta^{n} \, \vec{\chi}_{R}(q, x, y) \, , \qquad
    \vec{\chi}_{R}(q, x, y) =
    \begin{pmatrix}
        \chi_{R}(q,x,y, \id, \id) \\
        \chi_{R}(q,x,y, \mathcal{P}, \id) \\
        \chi_{R}(q,x,y, \id, \mathcal{C}) \\
        \chi_{R}(q,x,y, \mathcal{P}, \mathcal{C})
    \end{pmatrix}
    \, ,
\end{equation}
where $R = (\Delta, \ell, p, \lambda_{(c,\pm)})$ and $d_{R,n}$ denotes the degeneracy of a primary operator in the representation $R$ built out of $n$ currents $J$ (i.e.\ the degree of the symmetric product).

To perform this decomposition, we work recursively in a formal power series in \( q \) up to a given cut-off in scaling dimension. Given the spectrum \( S_{K} \) up to \( \Delta = K \) we consider the difference
\begin{equation}
    \vec{Z}_{J}(\eta,q,x,y) - \sum_{(R, n) \in S_{K}} d_{R,n} \, \eta^{n} \, \vec{\chi}_{R}(q,x,y) = \vec{f}_{K+1}(\eta,x,y) q^{K+1} + O(q^{N+2}) \, .
\end{equation}
The function \( \vec{f}_{K+1}(\eta,x,y) q^{K+1} \) is decomposed in characters by matching the leading term in the \( \eta, q, x, y_{1},\ldots,y_{N-1}  \to 0 \) limit with the behaviour of the character~\eqref{eq:character-conformal-parity-SUn} in the same limit
\begin{equation}
    \begin{aligned}
        \vec{\chi}_{\Delta,\ell,p,\lambda_{\pm}}(q,x,y)
        &\sim
        \frac{q^{\Delta}}{x^{\ell} y_{1}^{\lambda_{1}+\cdots+\lambda_{N}} y_{2}^{\lambda_{1}+\cdots+\lambda_{N-2}}\cdots y_{N-1}^{\lambda_{1}}}
        \begin{pmatrix}
      1 \\
      p (-1)^{\ell} \\
      \pm 1 \\
      \pm p (-1)^{\ell}
    \end{pmatrix} \, , \\[0.3em]
    \vec{\chi}_{\Delta,\ell,p,\lambda_{c}}(q,x,y)
        &\sim
        \frac{q^{\Delta}}{x^{\ell} y_{1}^{\lambda_{1}+\cdots+\lambda_{N}} y_{2}^{\lambda_{1}+\cdots+\lambda_{N-2}}\cdots y_{N-1}^{\lambda_{1}}}
        \begin{pmatrix}
      1 \\
      p (-1)^{\ell} \\
      0 \\
      0
    \end{pmatrix} \, ,
    \end{aligned}
\end{equation}
and recursively adding this contribution to $S_{K}$. 
In practice, there can be degeneracies between operators with the same $(\Delta = K+1, \ell, \lambda, n)$ but different behaviour under parity and charge conjugation.
To disentangle the contributions, we use the leading behaviour of the first component of $\vec{f}_{K+1}(\eta,x,y)q^{K+1}$ to fix the $(\Delta, \ell, \lambda, n)$ quantum numbers. 
We then consider the coefficient associated with these quantum numbers in the full vector, i.e.
\begin{equation}
    \vec{f}_{K+1}(\eta,x,y) q^{K+1} \sim \frac{\eta^{n} q^{K+1}}{x^{\ell} y_{1}^{\lambda_{1}+\cdots+\lambda_{N}}\cdots y_{N-1}^{\lambda_{1}}} \begin{pmatrix}
        a_{1} \\
        a_{2} \\
        a_{3} \\
        a_{4}
    \end{pmatrix}
    \, .
\end{equation}
\begin{table}[t!]\small{
    \centering
    \begin{NiceTabular}[t]{ccc|ccc}
        \CodeBefore
        \rowcolor{orange!50}{3,5,6}
        \Body
        \toprule
        \( p \) & \( \ell \) & \( \lambda_{1} \) & \( \Delta \)  \\ \midrule
        \(+\) & 0 & 0 & 0 & 4 & 6\\
        \(+\) & 0 & 2 & 7 & 9 & 10\\
        \(+\) & 0 & 4 & 4 & 6 & 7\\
        \(+\) & 0 & 6 & 9 & 10\\
        \(+\) & 0 & 8 & 8 & 10\\
        \(+\) & 1 & 2 & 2 & 5 & 6\\
        \(+\) & 2 & 0 & 4 & 6 & 7\\
        \(+\) & 2 & 4 & 4 & 6 & 7\\
        \(+\) & 3 & 2 & 5 & 6 & 7\\
        \(+\) & 4 & 0 & 6 & 8 & 9\\
        \(+\) & 4 & 4 & 6 & 7 & 8\\
        \(+\) & 5 & 2 & 7 & 8 & 9\\
        \(+\) & 6 & 0 & 8 & 9 & 10\\
        \(+\) & 6 & 4 & 8 & 9 & 10\\
        \(+\) & 7 & 2 & 9 & 10\\
        \(+\) & 8 & 0 & 10\\
        \(+\) & 8 & 4 & 10\\
        \bottomrule
        \CodeAfter
    \end{NiceTabular}
    \hspace{2cm}
    \begin{NiceTabular}[t]{ccc|ccc}
        \toprule
        \( p \) & \( \ell \) & \( \lambda_{1} \) & \( \Delta \)  \\ \midrule
        \(-\) & 0 & 0 & 5 & 6 & 7\\
        \(-\) & 0 & 4 & 5 & 7 & 8\\
        \(-\) & 1 & 2 & 4 & 6 & 7\\
        \(-\) & 2 & 0 & 5 & 7 & 8\\
        \(-\) & 2 & 2 & 5 & 6 & 7\\
        \(-\) & 2 & 4 & 5 & 6 & 7\\
        \(-\) & 3 & 0 & 6 & 7 & 8\\
        \(-\) & 3 & 2 & 6 & 7 & 8\\
        \(-\) & 3 & 4 & 6 & 7 & 8\\
        \(-\) & 4 & 0 & 7 & 8 & 9\\
        \(-\) & 4 & 2 & 7 & 8 & 9\\
        \(-\) & 4 & 4 & 7 & 8 & 9\\
        \(-\) & 5 & 0 & 8 & 9 & 10\\
        \(-\) & 5 & 2 & 8 & 9 & 10\\
        \(-\) & 5 & 4 & 8 & 9 & 10\\
        \(-\) & 6 & 0 & 9 & 10\\
        \(-\) & 6 & 2 & 9 & 10\\
        \(-\) & 6 & 4 & 9 & 10\\
        \(-\) & 7 & 0 & 10\\
        \(-\) & 7 & 2 & 10\\
        \(-\) & 7 & 4 & 10\\
        \bottomrule
    \end{NiceTabular}
    \caption{\( SU(2) \) GFV  spectrum up to \( \Delta = 10 \), showing only sectors exchanged in $\braket{JJJJ}$. In the parity-even scalar sector, non-exchanged sectors are highlighted in orange.\label{tab:spectrum-SU2}}}
\end{table}
If $\lambda\neq \lambda^*$, we have $a_{3}=a_{4} = 0$ and we solve the system
\begin{equation}
    \begin{cases}
        a_{1} = d_{+} + d_{-} \,,\\
        a_{2} = (-1)^{\ell}\left(d_{+} - d_{-}\right) \,,
    \end{cases}
\end{equation}
where $d_{\pm} = d_{K+1,\ell,\pm, \lambda_{c},n}$. If $\lambda= \lambda^*$ we solve
\begin{equation}
    \begin{cases}
        a_{1} = d_{+,+} + d_{+,-} + d_{-,+} + d_{-,-}\,, \\
        a_{2} = (-1)^{\ell} \left(d_{+,+} + d_{+,-} - d_{-,+} - d_{-,-}\right) \,,\\
        a_{3} = d_{+,+} - d_{+,-} + d_{-,+} - d_{-,-}\,, \\
        a_{4} = (-1)^{\ell} \left(d_{+,+} - d_{+,-} - d_{-,+} + d_{-,-}\right)\,,
    \end{cases}
\end{equation}
where $d_{\pm,\pm} = d_{K+1,\ell,\pm, \lambda_{\pm},n}$. A consistent solution requires all the degeneracies to be non-negative integers.
When $\vec{f}_{K+1}$ vanishes, the new set $S_{K}'$ will be $S_{K+1}$, namely we will have determined the spectrum at level \( \Delta = K+1 \).
This strategy can be implemented algorithmically, e.g.\ in \texttt{Mathematica}, for a fixed value of \( N \).
\begin{table}[t!]\small{
    \centering
    \begin{NiceTabular}[t]{ccl|ccc}
        \CodeBefore
        \rowcolor{orange!50}{3,5,6,8,9,10,11,12,13,14}
        \Body
        \toprule
        \( p \) & \( \ell \) & \( (\lambda_{1}, \lambda_{2})_{(c,\pm)} \) & \( \Delta \)  \\ \midrule
        \(+\) & 0 & \( (0,0)_{+} \) & 0 & 4 & 6\\
        \(+\) & 0 & \( (0,0)_{-} \) & 9 & 10\\
        \(+\) & 0 & \( (1,1)_{+} \) & 4 & 6 & 7\\
        \(+\) & 0 & \( (1,1)_{-} \) & 7 & 9 & 10\\
        \(+\) & 0 & \( (3,0)_{c} \) & 7 & 8 & 9\\
        \(+\) & 0 & \( (2,2)_{+} \) & 4 & 6 & 7\\
        \(+\) & 0 & \( (2,2)_{-} \) & 7 & 8 & 9\\
        \(+\) & 0 & \( (4,1)_{c} \) & 7 & 8 & 9\\
        \(+\) & 0 & \( (3,3)_{+} \) & 8 & 10\\
        \(+\) & 0 & \( (3,3)_{-} \) & 9 & 10\\
        \(+\) & 0 & \( (6,0)_{c} \) & 8 & 10\\
        \(+\) & 0 & \( (5,2)_{c} \) & 10\\
        \(+\) & 0 & \( (4,4)_{+} \) & 8 & 10\\
        \(+\) & 1 & \( (1,1)_{-} \) & 2 & 5 & 6\\
        \(+\) & 1 & \( (3,0)_{c} \) & 5 & 6 & 7\\
        \(+\) & 2 & \( (0,0)_{+} \) & 4 & 6 & 7\\
        \(+\) & 2 & \( (1,1)_{+} \) & 4 & 6 & 7\\
        \(+\) & 2 & \( (2,2)_{+} \) & 4 & 6 & 7\\
        \(+\) & 3 & \( (1,1)_{-} \) & 5 & 6 & 7\\
        \(+\) & 3 & \( (3,0)_{c} \) & 5 & 6 & 7\\
        \(+\) & 4 & \( (0,0)_{+} \) & 6 & 8 & 9\\
        \(+\) & 4 & \( (1,1)_{+} \) & 6 & 7 & 8\\
        \(+\) & 4 & \( (2,2)_{+} \) & 6 & 7 & 8\\
        \(+\) & 5 & \( (1,1)_{-} \) & 7 & 8 & 9\\
        \(+\) & 5 & \( (3,0)_{c} \) & 7 & 8 & 9\\
        \bottomrule
        \CodeAfter
    \end{NiceTabular}
    \hspace{4em}
    \begin{NiceTabular}[t]{ccl|ccc}
        \CodeBefore
        \Body
        \toprule
        \( p \) & \( \ell \) & \( (\lambda_{1}, \lambda_{2})_{(c,\pm)} \) & \( \Delta \)  \\ \midrule
        \(-\) & 0 & \( (0,0)_{+} \) & 5 & 6 & 7\\
        \(-\) & 0 & \( (1,1)_{+} \) & 5 & 6 & 7\\
        \(-\) & 0 & \( (2,2)_{+} \) & 5 & 6 & 7\\
        \(-\) & 1 & \( (1,1)_{-} \) & 4 & 6 & 7\\
        \(-\) & 1 & \( (3,0)_{c} \) & 4 & 6 & 7\\
        \(-\) & 2 & \( (0,0)_{+} \) & 5 & 7 & 8\\
        \(-\) & 2 & \( (1,1)_{+} \) & 5 & 6 & 7\\
        \(-\) & 2 & \( (1,1)_{-} \) & 5 & 6 & 7\\
        \(-\) & 2 & \( (3,0)_{c} \) & 5 & 6 & 7\\
        \(-\) & 2 & \( (2,2)_{+} \) & 5 & 6 & 7\\
        \(-\) & 3 & \( (0,0)_{+} \) & 6 & 7 & 8\\
        \(-\) & 3 & \( (1,1)_{+} \) & 6 & 7 & 8\\
        \(-\) & 3 & \( (1,1)_{-} \) & 6 & 7 & 8\\
        \(-\) & 3 & \( (3,0)_{c} \) & 6 & 7 & 8\\
        \(-\) & 3 & \( (2,2)_{+} \) & 6 & 7 & 8\\
        \(-\) & 4 & \( (0,0)_{+} \) & 7 & 8 & 9\\
        \(-\) & 4 & \( (1,1)_{+} \) & 7 & 8 & 9\\
        \(-\) & 4 & \( (1,1)_{-} \) & 7 & 8 & 9\\
        \(-\) & 4 & \( (3,0)_{c} \) & 7 & 8 & 9\\
        \(-\) & 4 & \( (2,2)_{+} \) & 7 & 8 & 9\\
        \(-\) & 5 & \( (0,0)_{+} \) & 8 & 9 & 10\\
        \(-\) & 5 & \( (1,1)_{+} \) & 8 & 9 & 10\\
        \(-\) & 5 & \( (1,1)_{-} \) & 8 & 9 & 10\\
        \(-\) & 5 & \( (3,0)_{c} \) & 8 & 9 & 10\\
        \(-\) & 5 & \( (2,2)_{+} \) & 8 & 9 & 10 \\
        \bottomrule
        \CodeAfter
    \end{NiceTabular}
    \caption{
    \( SU(3) \) GFV  spectrum up to \( \Delta = 10 \) and \( \ell = 5 \), showing only sectors exchanged in $\braket{JJJJ}$. In the parity-even scalar sector, non-exchanged sectors are highlighted in orange.\label{tab:spectrum-SU3}}}
\end{table}

In the $N = 2$ case, the computation is much simpler since we replace $PSU(N) \rtimes \Z_2^{\mathcal{C}} \to SO(3)$. 
The group still has two disconnected components due to the action of parity, resulting in a two-dimensional vector of generating functions. 
The previous discussion can be easily adapted to this simpler case.

We report in Tables~\ref{tab:spectrum-SU2} and~\ref{tab:spectrum-SU3} the spectrum of the first primary operators in the \( SU(2) \) and $SU(3)$ GFV up to \( \Delta = 10 \), and up to $\ell = 5$ for $SU(3)$. 
For each parity, spin, and $PSU(N)$ representation, we report the scaling dimensions of up to the first 3 operators.
For self-conjugate representations in $SU(3)$, the intrinsic charge conjugation is also reported. 
For the specific $(p,\ell) = (+,0)$ sector, and only for this one, we show all operators, highlighting in orange the ones that cannot be exchanged in the correlator $\braket{JJJJ}$ in general. 
These include operators in representations different from $\mathsf{R}_s$ and $\mathsf{R}_a$ in~\eqref{eq:SUNZ2-adj-decompose-sym-alt} and~\eqref{eq:SU2-adj-decompose-sym-alt}, and those in $\mathsf{R}_s$ and $\mathsf{R}_a$ which do not satisfy the selection rules reported in table~\ref{tab:selection-rules}. 
The purpose is to show that the lightest scalars are exchanged in our bootstrap setup. 
In all other sectors, we display only the operators exchanged in the $\braket{JJJJ}$ correlator. 
We omit the label $n$ denoting the number of constituent currents. 
For $SU(2)$, $SU(3)$ and $SU(4)$ cases, the full spectrum up to \( \Delta = 10 \) ($\Delta = 8$ for $SU(4)$), including degeneracies, is available in the ancillary file \texttt{GFV-spectrum.nb}.

\section{Projectors to irreducible representations}\label{sec:projectors}
Our task is to find the explicit decomposition into irreducible representations of a tensor $O^{ab}$, where $a,b=1,\ldots N^2-1$ are adjoint indices of $SU(N)$.
The decomposition can be made by decomposing or not decomposing the adjoint indices into fundamental ones. 
We choose not to split them, using the results of~\cite{Haber:2019sgz}, as done in~\cite{Ciccone:2024guw}. 
A generic tensor operator $O^{ab}$ can be split into its symmetric and antisymmetric parts:
\begin{equation}
    O^{ab} = O^{(ab)}+O^{[ab]} = O_{s} + O_{a} \,.
\end{equation}
The two tensors have $M(M\pm 1)/2$ components, with $M=(N^2-1)$, respectively. 
They coincide with the projection of $\mathbf{adj} \otimes \mathbf{adj}$ in~\eqref{eq:SUN-adj-decompose} into the symmetric square and alternating square components 
\begin{equation}
    \mathrm{Sym}^{2}\mathbf{adj} \cong \mathbf{S} \oplus \mathbf{adj}_{s} \oplus \mathbf{S\bar{S}} \oplus \mathbf{A\bar{A}} \, , \qquad
    \mathrm{Alt}^{2}\mathbf{adj} \cong \mathbf{adj}_{a} \oplus \mathbf{S\bar{A}} \oplus \mathbf{A\bar{S}} \, .
\end{equation}
It is easy to check that the dimensions agree between the left- and right-hand side.
We write
\begin{equation}
    O^{(ab)} = \sum_{\mathbf{r} \in \mathrm{Sym}^{2}\mathbf{adj}} P^{(ab)}_{i_s,(cd)} O^{(cd)} \, , \qquad 
    O^{[ab]} = \sum_{\mathbf{r} \in \mathrm{Alt}^{2}\mathbf{adj}} P^{[ab]}_{\mathbf{r},[cd]} O^{[cd]} \, .
\end{equation}
The projectors $P_{\mathbf{r}}$ and $P_{\mathbf{r}}$ can be determined by demanding the following conditions:
\begin{equation}\label{eq:proj1}
    \begin{aligned}
        P^{(ab)}_{\mathbf{r},(ab)} =  \dim \mathbf{r} \, , 
        \qquad
        P^{[ab]}_{\mathbf{r},[ab]} &=  \dim \mathbf{r} \, , 
        \qquad
        P_{\mathbf{r}} P_{\mathbf{s}} = \delta_{\mathbf{r},\mathbf{s}} P_{\mathbf{s}} \, ,
    \end{aligned}
\end{equation}
where we have omitted the indices on the last of the three equations, as well as completeness:
\begin{equation}\label{eq:proj2}
    \sum_{\mathbf{r} \in \mathrm{Sym}^{2}\mathbf{adj}} P_{\mathbf{r}, (cd)}^{(ab)} = (\id_{s})^{ab}_{cd} \, , \qquad
    \sum_{\mathbf{r} \in \mathrm{Alt}^{2}\mathbf{adj}} P_{\mathbf{r}, [cd]}^{[ab]} = (\id_{a})^{ab}_{cd} \, ,
\end{equation}
where
\begin{equation}    \label{eq:proj3}
    (\id_{s})^{ab}_{cd} = \frac{1}{2}\left(\delta_{ac}\delta_{bd} + \delta_{ad}\delta_{bc}\right) \, , \qquad
    (\id_{a})^{ab}_{cd} = \frac{1}{2}\left(\delta_{ac}\delta_{bd} - \delta_{ad}\delta_{bc}\right) \, .
\end{equation}
The conditions~\eqref{eq:proj1},~\eqref{eq:proj2},~\eqref{eq:proj3} are overconstraining.
We only report here the final results:
\begin{align}\label{eq:proj4p}
    P_{{\bf S},cd}^{ab} & = \frac{1}{N^2-1}\delta_{ab} \delta_{cd} \,, \nn \\ 
    P_{{\bf adj}_{s},cd}^{ab} & = \frac{N}{N^2-4} d_{abz} d_{cdz} \,, \\
    P_{{\bf S\bar S},cd}^{ab} & = \frac{1}{4} \Big(\delta_{ac} \delta_{bd} + \delta_{ad} \delta_{bc} \Big) 
    + \frac{1}{2(N+1)} \delta_{ab} \delta_{cd} 
    - \frac{1}{4} \Big(f_{adz} f_{bcz} + f_{acz} f_{bdz}\Big) 
    + \frac{N}{4(N+2)} d_{abz} d_{cdz}\,, \nn \\
    P_{{\bf A\bar A},cd}^{ab} & = \frac{1}{4} \Big(\delta_{ac} \delta_{bd} + \delta_{ad} \delta_{bc} \Big)
    - \frac{1}{2(N-1)} \delta_{ab} \delta_{cd} 
    + \frac{1}{4} \Big(f_{adz} f_{bcz} + f_{acz} f_{bdz}\Big) 
    - \frac{N}{4(N-2)} d_{abz} d_{cdz} \,, \nn  
\end{align}
for the symmetric tensors, and
\begin{align}\label{eq:proj4m}
    P_{{\bf adj}_{a},cd}^{ab} & = \frac{1}{N} f_{abz} f_{cdz} \,,  \\
    P_{{\bf A\bar S},cd}^{ab} & = \frac{1}{4} \Big(\delta_{ac} \delta_{bd} - \delta_{ad} \delta_{bc} \Big) 
    - \frac{1}{2N} f_{abz} f_{cdz} 
    + \frac{i}{8} \Big(f_{acz} d_{bdz} - f_{adz} d_{bcz} - f_{bcz} d_{adz} + f_{bdz} d_{acz}\Big) \,, \nn  \\
    P_{{\bf S\bar A},cd}^{ab} &= \frac{1}{4} \Big(\delta_{ac} \delta_{bd} - \delta_{ad} \delta_{bc} \Big) 
    - \frac{1}{2N} f_{abz} f_{cdz} 
    - \frac{i}{8} \Big(f_{acz} d_{bdz} - f_{adz} d_{bcz} - f_{bcz} d_{adz} + f_{bdz} d_{acz}\Big) \,, \nn  
\end{align}
for the antisymmetric ones, where
\begin{equation}\label{eq:dabc-and-fabc}
    d^{abc} = - 2 i \tr\!\Big[\{T^{a}, T^{b}\} T^{c}\Big] \, , \qquad 
    f^{abc} = - 2 \tr\!\Big[[T^{a}, T^{b}] T^{c}\Big] \,,
\end{equation}
in terms of the anti-hermitian generators in~\eqref{eq:sungenerators}.
The same projectors can also be used to write projectors in the representations in~\eqref{eq:SUNZ2-adj-decompose-sym-alt}, which read
\begin{equation}\label{eq:Projneeded}
    \begin{gathered}
        P_{\mathbf{S}_{+}} = P_{\mathbf{S}} \, , \qquad 
        P_{\mathbf{S\bar{S}}_{+}} = P_{\mathbf{S\bar{S}}} \, , \qquad 
        P_{\mathbf{A\bar{A}}_{+}} = P_{\mathbf{A\bar{A}}} \, , \\
        P_{\mathbf{adj}_{+}} = P_{\mathbf{adj}_{s}} \, , \qquad 
        P_{\mathbf{adj}_{-}} = P_{\mathbf{adj}_{a}} \, , \qquad 
        (N \text{ odd}) \, , \\
        P_{\mathbf{adj}_{+}} = P_{\mathbf{adj}_{a}} \, , \qquad
        P_{\mathbf{adj}_{-}} = P_{\mathbf{adj}_{s}} \, , \qquad
        (N \text{ even}) \, , \\
        P_{\mathbf{S\bar{A}}_{c}} = P_{\mathbf{S\bar{A}}} + P_{\mathbf{A\bar{S}}} \, .
    \end{gathered}
\end{equation}

\section{Details on the numerical setup}\label{app:numerics}

\subsection{\texorpdfstring{$C_{J}$}{CJ} normalisation}\label{app:CJ-normalisation}

The relation between the current central charge in the theory of a Free Boson (FB) in the fundamental of $SU(N)$, $C_{J}\ap{FB}$, and the optimal objective of the SDP~\eqref{eq:SDP-CJbound-SU3} is not completely straightforward to obtain. It requires matching the normalisation conventions in~\eqref{eq:CJ-definition} with those of the conformal blocks in \texttt{blocks\_3d}~\cite{Erramilli:2020rlr} used in the numerics.

We have thus adopted a pragmatic approach. We can focus the SDP~\eqref{eq:SDP-CJbound-SU3} on FB theory by fixing $\alpha_{JJJ}=1/2$ and inputting strong assumptions on the spectrum, i.e.\ imposing that the lightest 10 operators in each sector $(\ell, p, \mathbf{r})$ are isolated and have dimensions as in the FB theory.
The primary spectrum of the FB theory can be computed at low scaling dimensions by adapting the character technique described in Appendix~\ref{app:characters-GFV}, and then inferring integer gaps between the heavier operators.
Then we computed upper and lower bounds on $C_J$. 
The result was an interval of values that allowed us to determine $C_J\ap{FB}$ within a precision of $10^{-3}$. 
We repeated it for several results of $N$. 

Our findings are consistent with the normalisation-independent relation 
\begin{equation}
    \frac{C_{J}}{C_{J}\ap{FB}} \geq -\frac{4N}{\sqrt{N^2-1}} \frac{1}{P} \, ,
\end{equation}
whose $N$-dependence can be inferred by comparing~\eqref{eq:CJ-definition} with the symmetry tensor structures~\eqref{eq:Trabcd} used in~\eqref{eq:JJJJ-gfunctions}.

\subsection{Gap maximisation bounds}

In what follows  we denote by \( \vec{0} \) a vector of identical zero components of length 
\begin{equation}
    \dim(\vec 0) = (d\ped{even}+4d\ped{odd}+1)r_{s} + (4d\ped{even}+d\ped{odd}+1)r_{a} \, , 
\end{equation}
see~\eqref{eq:counting-parameters} for the definition.

The SDP used to obtain the $SU(2)$ bounds is
\begin{equation}\label{eq:SDP-gap-maximisation-SU2}
    \begin{aligned}
        \text{objective} \colon \quad
        & \alpha(\vec{0}) = 0 \\
        \text{normalisation} \colon  \quad
        & \alpha(\vec{\mathcal{V}}_{\id}) \\
        \text{conditions} \colon \quad
        & \alpha(\vec{\mathcal{V}}_{\Delta,0,+,\mathbf{S}}) \succeq 0 \qquad && \Delta \geq  \Delta_{s} \\
        & \alpha(\vec{\mathcal{V}}_{\Delta,0,+,\mathbf{S\bar{S}}}) \succeq 0 \qquad && \Delta \geq  \Delta_{s\bar{s}} \\
        & \alpha(\vec{\mathcal{V}}_{\Delta, \ell, +, \mathbf{r}}) \succeq 0 \qquad && \Delta \geq  \Delta\ped{UB}(\ell) + \tau && \mathbf{r} \in\mathsf{R}_{s} && \ell = 2, 4,6, \ldots  \\
        & \alpha(\vec{\mathcal{V}}_{\Delta, \ell, +, \mathbf{r}}) \succeq 0 \qquad && \Delta \geq  \Delta\ped{UB}(\ell) + \tau && \mathbf{r} \in \mathsf{R}_{a} && \ell = 1,3,5, \ldots  \\
        & \alpha(\vec{\mathcal{V}}_{\Delta, \ell, -, \mathbf{r}}) \succeq 0 \qquad && \Delta \geq  \Delta\ped{UB}(\ell) + \tau && \mathbf{r} \in\mathsf{R}_{s} && \ell = 0,2,3 \ldots  \\
        & \alpha(\vec{\mathcal{V}}_{\Delta, \ell, -, \mathbf{r}}) \succeq 0 \qquad && \Delta \geq  \Delta\ped{UB}(\ell) + \tau && \mathbf{r} \in \mathsf{R}_{a} && \ell = 1,2,3 \ldots  \\
        & \alpha(\vec{\mathcal{V}}_{J}) \succeq 0
    \end{aligned}
\end{equation}
where we imposed a small twist gap $\tau = 10^{-6}$, which improves numerical stability but does not have a noticeable impact on the bounds.

The SDP used to obtain the $SU(3)$ bounds is
\begin{equation}\label{eq:SDP-gap-maximisation-SU3}
    \begin{aligned}
        \text{objective} \colon \quad
        & \alpha(\vec{0}) = 0 \\
        \text{normalisation} \colon  \quad
        & \alpha(\vec{\mathcal{V}}_{\id}) \\
        \text{conditions} \colon \quad
        & \alpha(\vec{\mathcal{V}}_{\Delta,0,+,\mathbf{S}_{+}}) \succeq 0 \qquad && \Delta \geq  \Delta_{s} \\
        & \alpha(\vec{\mathcal{V}}_{\Delta,+,0,\mathbf{adj}_{+}}) \succeq 0 \qquad && \Delta \geq  \Delta_{a} \\
        & \alpha(\vec{\mathcal{V}}_{\Delta,0,+,\mathbf{S\bar{S}}_{+}}) \succeq 0 \qquad && \Delta \geq  \Delta_{s\bar{s}} \\
        & \alpha(\vec{\mathcal{V}}_{\Delta, \ell, +, \mathbf{r}}) \succeq 0 \qquad && \Delta \geq  \Delta\ped{UB}(\ell) + \tau && \mathbf{r} \in\mathsf{R}_{s} && \ell = 2, 4,6, \ldots  \\
        & \alpha(\vec{\mathcal{V}}_{\Delta, \ell, +, \mathbf{r}}) \succeq 0 \qquad && \Delta \geq  \Delta\ped{UB}(\ell) + \tau && \mathbf{r} \in \mathsf{R}_{a} && \ell = 1,3,5, \ldots  \\
        & \alpha(\vec{\mathcal{V}}_{\Delta, \ell, -, \mathbf{r}}) \succeq 0 \qquad && \Delta \geq  \Delta\ped{UB}(\ell) + \tau && \mathbf{r} \in\mathsf{R}_{s} && \ell = 0,2,3 \ldots  \\
        & \alpha(\vec{\mathcal{V}}_{\Delta, \ell, -, \mathbf{r}}) \succeq 0 \qquad && \Delta \geq  \Delta\ped{UB}(\ell) + \tau && \mathbf{r} \in \mathsf{R}_{a} && \ell = 1,2,3 \ldots  \\
        & \alpha(\vec{\mathcal{V}}_{J}) \succeq 0
    \end{aligned}
\end{equation}
where $\tau = 10^{-6}$.

\subsection{Allowed region with strong assumptions}
The bound was obtained using the following SDP
\begin{equation}\label{eq:SDP-strong-gaps}
    \begin{aligned}
        \text{objective} \colon \quad
        & \alpha(\vec{0}) = 0 \\
        \text{normalisation} \colon  \quad
        & \alpha(\vec{\mathcal{V}}_{\id}) \\
        \text{conditions} \colon \quad
        & \alpha(\vec{\mathcal{V}}_{\Delta_{s},0,+,\mathbf{S}_{+}}) \succeq 0 \\
        & \alpha(\vec{\mathcal{V}}_{\Delta_{a},+,0,\mathbf{adj}_{+}}) \succeq 0 \\
        & \alpha(\vec{\mathcal{V}}_{\Delta,0,+,\mathbf{S}_{+}}) \succeq 0 \qquad && \Delta \geq  6 \\
        & \alpha(\vec{\mathcal{V}}_{\Delta,+,0,\mathbf{adj}_{+}}) \succeq 0 \qquad && \Delta \geq  6 \\
        & \alpha(\vec{\mathcal{V}}_{\Delta,2,+,\mathbf{S}_{+}}) \succeq 0 \qquad && \Delta \geq  4 \\
        & \alpha(\vec{\mathcal{V}}_{\Delta,0,+,\mathbf{S\bar{S}}_{+}}) \succeq 0 \qquad && \Delta \geq  \Delta\ped{UB}(0)  \\
        & \alpha(\vec{\mathcal{V}}_{\Delta, \ell, +, \mathbf{r}}) \succeq 0 \qquad && \Delta \geq  \Delta\ped{UB}(\ell) + \tau && \mathbf{r} \in\mathsf{R}_{s} && \ell = 4,6, \ldots  \\
        & \alpha(\vec{\mathcal{V}}_{\Delta, \ell, +, \mathbf{r}}) \succeq 0 \qquad && \Delta \geq  \Delta\ped{UB}(\ell) + \tau && \mathbf{r} \in \mathsf{R}_{a} && \ell = 1,3,5, \ldots  \\
        & \alpha(\vec{\mathcal{V}}_{\Delta, \ell, -, \mathbf{r}}) \succeq 0 \qquad && \Delta \geq  \Delta\ped{UB}(\ell) + \tau && \mathbf{r} \in\mathsf{R}_{s} && \ell = 0,2,3 \ldots  \\
        & \alpha(\vec{\mathcal{V}}_{\Delta, \ell, -, \mathbf{r}}) \succeq 0 \qquad && \Delta \geq  \Delta\ped{UB}(\ell) + \tau && \mathbf{r} \in \mathsf{R}_{a} && \ell = 1,2,3 \ldots  \\
        & \alpha(\vec{\mathcal{V}}_{J}) \succeq 0
    \end{aligned}
\end{equation}
where $\tau = 10^{-6}$.

\subsection{Allowed region with moderate assumptions}

The bound was obtained using the following SDP
\begin{equation}\label{eq:sdp-gfv-medium-gaps}
    \begin{aligned}
        \text{objective} \colon \quad
        & \alpha(\vec{0}) = 0 \\
        \text{normalisation} \colon  \quad
        & \alpha(\vec{\mathcal{V}}_{\id}) \\
        \text{conditions} \colon \quad
        & \alpha(\vec{\mathcal{V}}_{\Delta_{s},0,+,\mathbf{S}_{+}}) \succeq 0 \\
        & \alpha(\vec{\mathcal{V}}_{\Delta_{a},+,0,\mathbf{adj}_{+}}) \succeq 0 \\
        & \alpha(\vec{\mathcal{V}}_{\Delta, \ell, +, \mathbf{r}}) \succeq 0 \qquad && \Delta \geq \Delta\ped{\ell, +, \mathbf{r}}\ap{gap} && \mathbf{r} \in\mathsf{R}_{s} && \ell = 0,2,4, \ldots  \\
        & \alpha(\vec{\mathcal{V}}_{\Delta, \ell, +, \mathbf{r}}) \succeq 0 \qquad && \Delta \geq \Delta_{\ell, +, \mathbf{r}}\ap{gap} && \mathbf{r} \in \mathsf{R}_{a} && \ell = 1,3,5, \ldots  \\
        & \alpha(\vec{\mathcal{V}}_{\Delta, \ell, -, \mathbf{r}}) \succeq 0 \qquad && \Delta \geq \Delta_{\ell, -, \mathbf{r}}\ap{gap} && \mathbf{r} \in\mathsf{R}_{s} && \ell = 0,2,3, \ldots  \\
        & \alpha(\vec{\mathcal{V}}_{\Delta, \ell, -, \mathbf{r}}) \succeq 0 \qquad && \Delta \geq \Delta_{\ell, -, \mathbf{r}}\ap{gap} && \mathbf{r} \in \mathsf{R}_{a} && \ell = 1,2,3 \ldots  \\
        & \alpha(\vec{\mathcal{V}}_{J}) \succeq 0
    \end{aligned}
\end{equation}
with $\Delta_{\ell, p, \mathbf{r}}\ap{gap}$ given in Table~\ref{tab:gfv-medium-gaps}.

\begin{table}[h!]
    \centering
    \begin{NiceTabular}[t]{cccc}
        \toprule
        parity & rep & spin & \( \Delta_{\ell, p, \mathbf{r}}\ap{gap} \)  \\ \midrule
        \( + \) & \( \mathbf{S}_{+}, \mathbf{adj}_{+} \) & \( \ell = 0 \) & 5 \\
        \( + \) & \( \mathbf{S\bar{S}}_{+} \) & \( \ell = 0 \) & 3 \\
        \( + \) & \(\mathsf{R}_{s} \) & \( \ell = 2 \) & 3.5 \\
        \( + \) & \(\mathsf{R}_{s} \) & \( \ell = 4 \) & 4.5 \\
        \( + \) & \(\mathsf{R}_{s} \) & \( \ell \geq 6 \) & \( \Delta\ped{UB}(\ell) + \tau \) \\
        \( + \) & \( \mathsf{R}_{a} \) & \( \ell = 1 \) & 4 \\
        \( + \) & \( \mathsf{R}_{a} \) & \( \ell = 3 \) & 4.5 \\
        \( + \) & \( \mathsf{R}_{a} \) & \( \ell = 5 \) & 6.5 \\
        \( + \) & \( \mathsf{R}_{a} \) & \( \ell \geq 7 \) & \( \Delta\ped{UB}(\ell) + \tau \) \\ \bottomrule
    \end{NiceTabular}
    \hspace{2em}
    \begin{NiceTabular}[t]{cccc}
        \toprule
        parity & rep & spin & \( \Delta_{\ell, p, \mathbf{r}}\ap{gap} \)  \\ \midrule
        \( - \) & \(\mathsf{R}_{s} \) & \( \ell = 0 \) & 3 \\
        \( - \) & \(\mathsf{R}_{s} \) & \( \ell = 2 \) & 4 \\
        \( - \) & \(\mathsf{R}_{s} \) & \( \ell = 3 \) & 5 \\
        \( - \) & \(\mathsf{R}_{s} \) & \( \ell = 4 \) & 6 \\
        \( - \) & \(\mathsf{R}_{s} \) & \( \ell \geq 5 \) &  \( \Delta\ped{UB}(\ell) + \tau \) \\
        \( - \) & \( \mathsf{R}_{a} \) & \( \ell = 1 \) & 3 \\
        \( - \) & \( \mathsf{R}_{a} \) & \( \ell = 2 \) & 4 \\
        \( - \) & \( \mathsf{R}_{a} \) & \( \ell = 3 \) & 5 \\
        \( - \) & \( \mathsf{R}_{a} \) & \( \ell = 4 \) & 6 \\
        \( - \) & \( \mathsf{R}_{a} \) & \( \ell \geq 5 \) &  \( \Delta\ped{UB}(\ell) +\tau \) \\ \bottomrule
    \end{NiceTabular}
    \caption{Gaps imposed in the SDP~\eqref{eq:sdp-gfv-medium-gaps}. Here $\tau = 10^{-6}$.\label{tab:gfv-medium-gaps}}
\end{table}

\subsection{Allowed region with parametric assumptions}
The bounds in the case of $SU(3)$ are obtained with the following SDP
\begin{equation}\label{eq:sdp-gfv-delta-gaps}
    \begin{aligned}
        \text{objective} \colon \quad
        & \alpha(\vec{\mathcal{V}}_{\id}) \\
        \text{normalisation} \colon  \quad
        & \alpha(\vec{\Sigma}) \\
        \text{conditions} \colon \quad
        & \alpha(\vec{\mathcal{V}}_{\Delta_{s},0,+,\mathbf{S}_{+}}) \succeq 0 \\
        & \alpha(\vec{\mathcal{V}}_{\Delta_{a},0,+,\mathbf{adj}_{+}}) \succeq 0 \\
        & \alpha(\vec{\mathcal{V}}_{\Delta_{s\bar{s}},0,+,\mathbf{S\bar{S}}_{+}}) \succeq 0 \\
        & \alpha(\vec{\mathcal{V}}_{\Delta, \ell, +, \mathbf{r}}) \succeq 0 \qquad && \Delta \in \Delta\ped{\ell, +, \mathbf{r}}(\delta) && \mathbf{r} \in\mathsf{R}_{s} && \ell = 0,2,4, \ldots  \\
        & \alpha(\vec{\mathcal{V}}_{\Delta, \ell, +, \mathbf{r}}) \succeq 0 \qquad && \Delta \in \Delta\ped{\ell, +, \mathbf{r}}(\delta) && \mathbf{r} \in \mathsf{R}_{a} && \ell = 1,3,5, \ldots  \\
        & \alpha(\vec{\mathcal{V}}_{\Delta, \ell, -, \mathbf{r}}) \succeq 0 \qquad && \Delta \in \Delta\ped{\ell, -, \mathbf{r}}(\delta) && \mathbf{r} \in\mathsf{R}_{s} && \ell = 0,2,3, \ldots  \\
        & \alpha(\vec{\mathcal{V}}_{\Delta, \ell, -, \mathbf{r}}) \succeq 0 \qquad && \Delta \in \Delta\ped{\ell, -, \mathbf{r}}(\delta) && \mathbf{r} \in \mathsf{R}_{a} && \ell = 1,2,3 \ldots  \\
        & \alpha(\theta^{\mathsf{T}}\vec{\mathcal{V}}_{J} \theta) \succeq 0
    \end{aligned}
\end{equation}
The normalisation used is the so-called ``\( \Sigma \)-navigator'', see~\cite[Sec.\ 2.1.2]{Reehorst:2021ykw}. The gap assumptions are defined as
\begin{equation}
    \begin{aligned}
        \Delta_{0, +, \mathbf{r}}(\delta) &= \left[\Delta_{0, +, \mathbf{r}}' - \delta, \infty\right) \, , \\
        \Delta_{0, -, \mathbf{r}}(\delta) &= \left[\Delta_{0, -, \mathbf{r}} - \delta, \Delta_{0, -, \mathbf{r}}\phantom{'} - \delta \right] \cup \left[\Delta_{0, -, \mathbf{r}}' - \delta, \infty\right) \, , \\
        \Delta_{\ell, p, \mathbf{r}}(\delta) &= \left[\Delta_{\ell, p, \mathbf{r}} - \frac{\delta}{\ell}, \Delta_{\ell, p, \mathbf{r}} + \frac{\delta}{\ell}\right] \cup \left[\Delta_{\ell, p, \mathbf{r}}' - \delta, \infty\right) \, , && \ell \geq 1 \, , \ \mathbf{r} \neq \mathbf{S\bar{A}} \, , \\
        \Delta_{\ell, p, \mathbf{S\bar{A}}_{c}}(\delta) &= \left[\Delta_{\ell,p,\mathbf{S\bar{A}}_{c}} - \frac{\delta}{\ell^{2}}, \Delta_{\mathbf{S\bar{A}}_{c},p,\ell} + \frac{\delta}{\ell^{2}}\right] \cup \left[\Delta_{\ell,p,\mathbf{S\bar{A}}_{c}}' - \delta, \infty\right) \, , && \ell \geq 1 \, , \\
    \end{aligned}
\end{equation}
where \( \Delta_{\ell, p, \mathbf{r}} \) and \( \Delta_{\ell, p, \mathbf{r}}' \) are the dimensions of the leading and sub-leading operators in the GFV in the \( (\ell, p, \mathbf{r}) \) sector.\footnote{
    Ignoring the identity and the current $J$ itself.
} 
These are
\begin{equation}\label{eq:Deltarpl}
    \begin{aligned}
        \Delta_{\ell, +, \mathbf{r}} &= 2+\ell \, , \qquad && \ell > 4 \, , \\
        \Delta_{\ell, +, \mathbf{r}}' &= 3+\ell \ , \qquad && \ell > 4 \, , \\[1em]
        \Delta_{\ell, -, \mathbf{r}} &= 3+\ell \, , \qquad && \ell > 2 \, , \\
        \Delta_{\ell, -, \mathbf{r}}' &= 4+\ell \, , \qquad && \ell > 2 \, , \\
    \end{aligned}
\end{equation}
as discussed in Section~\eqref{sec:spectrum-GFV}. 
For $\ell \leq 4$, in the case of $SU(3)$ we have 
\begin{equation}\label{eq:Deltarpl-exceptions}
    \begin{array}{l|l}
    \begin{aligned}
        \Delta_{\ell, +, \mathbf{r}} &= 4 \qquad && \ell = 0,2 \text{ and } \mathbf{r} \in \mathsf{R}_{s} \\
        \Delta_{\ell, +, \mathbf{r}}' &= 6 && \ell = 0,2 \text{ and } \mathbf{r} \in\mathsf{R}_{s}\\[1em]
        \Delta_{\ell, +, \mathbf{r}} &= 5 \qquad && \ell = 1,3 \text{ and }\mathbf{r} \in \mathsf{R}_{a} \\
        \Delta_{\ell, +, \mathbf{r}}' &= 6 && \ell = 1,3 \text{ and } \mathbf{r} \in \mathsf{R}_{a}\\[1em]
        \Delta_{4,+,\mathbf{r}} &= 6 \qquad && \mathbf{r} \in\mathsf{R}_{s}\\
        \Delta_{4,+,\mathbf{r}}' &= 7 && \mathbf{r} \in \{\mathbf{adj}_{+},\mathbf{S\bar{S}}_{+}\} \\
        \Delta_{4,+,\mathbf{S}_{+}}' &= 8
    \end{aligned}
    \quad
    &
    \quad
    \begin{aligned}
        \Delta_{0, -, \mathbf{r}} &= 5 \qquad  && \mathbf{r} \in \mathsf{R}_{s}\\
        \Delta_{0, -, \mathbf{r}}' &= 6 && \mathbf{r} \in \mathsf{R}_{s}\\[1em]
        \Delta_{1,-,\mathbf{r}} &= 4 \qquad  && \mathbf{r} \in \mathsf{R}_{a}\\
        \Delta_{1,-,\mathbf{r}}' &= 6 && \mathbf{r} \in \mathsf{R}_{a} \\[1em]
        \Delta_{2,-,\mathbf{r}}' &= 5 \qquad  && \mathbf{r} \in \mathsf{R}_{s}, \mathsf{R}_{a} \\
        \Delta_{2,-,\mathbf{r}}' &= 6 \qquad && \mathbf{r} \in\{\mathbf{adj}_{+},\mathbf{S\bar{S}}_{+}\}, \mathsf{R}_{a} \\
        \Delta_{2,-,\mathbf{S}_{+}} &= 7 
    \end{aligned}
    \end{array}
\end{equation}
See Table~\ref{tab:spectrum-SU3} and the ancillary file \texttt{GFV-spectrum.nb}.

In the $SU(2)$ case, we have a similar SDP
\begin{equation}\label{eq:sdp-gfv-delta-gaps-SU2}
    \begin{aligned}
        \text{objective} \colon \quad
        & \alpha(\vec{\mathcal{V}}_{\id}) \\
        \text{normalisation} \colon  \quad
        & \alpha(\vec{\Sigma}) \\
        \text{conditions} \colon \quad
        & \alpha(\vec{\mathcal{V}}_{\Delta_{s},0,+,\mathbf{S}}) \succeq 0 \\
        &\alpha(\vec{\mathcal{V}}_{\Delta_{s\bar{s}},0,+,\mathbf{S\bar{S}}}) \succeq 0 \\
        & \alpha(\vec{\mathcal{V}}_{\Delta, \ell, +, \mathbf{r}}) \succeq 0 \qquad && \Delta \in \Delta\ped{\ell, +, \mathbf{r}}(\delta) && \mathbf{r} \in\mathsf{R}_{s} && \ell = 0,2,4, \ldots  \\
        & \alpha(\vec{\mathcal{V}}_{\Delta, \ell, +, \mathbf{r}}) \succeq 0 \qquad && \Delta \in \Delta\ped{\ell, +, \mathbf{r}}(\delta) && \mathbf{r} \in \mathsf{R}_{a} && \ell = 1,3,5, \ldots  \\
        & \alpha(\vec{\mathcal{V}}_{\Delta, \ell, -, \mathbf{r}}) \succeq 0 \qquad && \Delta \in \Delta\ped{\ell, -, \mathbf{r}}(\delta) && \mathbf{r} \in\mathsf{R}_{s} && \ell = 0,2,3, \ldots  \\
        & \alpha(\vec{\mathcal{V}}_{\Delta, \ell, -, \mathbf{r}}) \succeq 0 \qquad && \Delta \in \Delta\ped{\ell, -, \mathbf{r}}(\delta) && \mathbf{r} \in \mathsf{R}_{a} && \ell = 1,2,3 \ldots  \\
        & \alpha(\theta^{\mathsf{T}}\vec{\mathcal{V}}_{J} \theta) \succeq 0
    \end{aligned}
\end{equation}
where $\Delta_{\ell, p, \mathbf{r}}$ are as in~\eqref{eq:Deltarpl} and~\eqref{eq:Deltarpl-exceptions}, except for the absence of the $\mathbf{adj}_{+}$ and $\mathbf{S\bar{A}}_{c}$ representations, and the following change in the $(\ell,p) = (4,+)$ sector
\begin{equation}
    \begin{aligned}
        \Delta_{4,+,\mathbf{r}} &= 6 \, , \qquad && \mathbf{r} \in \mathsf{R}_{s} \, ,\\
        \Delta_{4,+,\mathbf{r}}' &= 8 \, , && \mathbf{r} \in \mathsf{R}_{s} \, .
    \end{aligned}
\end{equation}

\subsection{Details on the numerical implementation}
The numerical bootstrap computations of this work were carried out using the \texttt{simpleboot} package~\cite{simpleboot}, based on the code used in~\cite{He:2023ewx}. 
The computation of the conformal blocks was done using \texttt{blocks\_3d}~\cite{Erramilli:2020rlr}. 
The semi-definite optimisation was done using \texttt{SDPB}~\cite{Simmons-Duffin:2015qma,Landry:2019qug}. 

For the computations of Sections~\ref{sec:bootstrap-CJ} and~\ref{sec:bootstrap-dimensions-GFV}, we used a slightly modified version of \texttt{SDPB~v3} which saves a ``middle checkpoint'' at a finite value of the $\mu$ parameter and recovers from a ``stalling'' phase by aggressively increasing the $\beta$ parameter (see~\cite[Sec.\ 2.4]{Simmons-Duffin:2015qma} for the definition of these parameters). 
This modified version was previously implemented for \texttt{SDPB~v2.4} by Ning Su, available at \url{https://github.com/suning-git/sdpb/tree/sdpb2.4.0_midck_stallingrecover}, and adapted to \texttt{SDPB~v3} by one of the authors for the present work, and is made available at \url{https://gitlab.com/apiazza134/sdpb-midck-stallingrecover}.

In our experience, \texttt{SDPB~v3} provided a significant speed-up in bootstrap computations compared to \texttt{SDPB~v2.x}, offering a much better scalability across multiple nodes. 
As a benchmark, a single \texttt{SDPB} step for a computation of Section~\ref{sec:bootstrap-dimensions-GFV} took on average $40\mathrm{s}$ with \texttt{v3} and $300\mathrm{s}$ with \texttt{v2.4} (using 6 nodes, 40 cpus and 40GB each, on the SISSA Ulysses cluster).
On the other hand, when \texttt{SDPB} is run in optimisation mode, as needed for Navigator computations, it is usually impossible to ``hot start'' a new optimisation from a previous checkpoint.
The ``middle checkpoint'' and ``stalling recover'' version of \texttt{SDPB~v2.4} was designed to overcome this problem. 
We would thus like to stress the importance of using this modified version of \texttt{SDPB~v3} for the computations presented in Section~\ref{sec:bootstrap-dimensions-GFV}.

The parameters used for the conformal blocks computations are reported in Table~\ref{tab:block-params}, where $S_{\Lambda}$ denotes the set of spins included in the positivity conditions:
\begin{equation}\label{eq:spin-sets}
    \begin{aligned}
        S_{11} &= \{0,\ldots,24\}\cup\{49,50\} \, , \\
        S_{15} &= \{0,\ldots,26\}\cup\{49,50\} \, , \\
        S_{19} &= \{0,\ldots,26\}\cup\{49,50\} \, , \\
        S_{23} &= \{0,\ldots,26\}\cup\{29,30,33,34,37,38,41,42,45,46,49,50\} \, .
    \end{aligned}
\end{equation}

\begin{table}[t]
    \centering
    \begin{tabular}{c|ccccccc}
        \toprule
        \( \Lambda \) & 11 & 15 & 19 & 23\\
        \texttt{spins} & \( S_{11} \) & \( S_{15} \) &  \( S_{19} \) & \( S_{23} \) \\
        \texttt{keptPoleOrder} & 11 & 14 & 14 & 18 \\
        \texttt{order} & 30 & 56 & 56 & 72 \\ \bottomrule
    \end{tabular}
    \caption{Parameters used for \texttt{blocks\_3d}, with $S_{\Lambda}$ given in~\eqref{eq:spin-sets}.\label{tab:block-params}}
\end{table}

\bibliography{references}

\end{document}